\documentclass[aps,pre,amsmath,nofootinbib,showpacs,twocolumn]{revtex4}

\usepackage{graphicx} \usepackage{hhline} \usepackage{amsfonts}

\font\eightln=line10 at8pt \catcode`\@=11
\def\singlearrow{\@ifnextchar [{\@singlearrow }{\@singlearrow[0]}}
\def\@singlearrow[#1]{\mathrel{\,\lower0.15ex
    \hbox{\let\@linefnt\eightln\unitlength0.6ex\begin{picture}(4,3)
        \put(0,1.5){\vector(1,0){4}}
    \end{picture}}\,}}
\catcode`\@=12
 
\font\eightln=line10 at8pt \catcode`\@=11
\def\doublearrow{\@ifnextchar [{\@doublearrow }{\@doublearrow[0]}}
\def\@doublearrow[#1]{\mathrel{\,\lower0.15ex
    \hbox{\let\@linefnt\eightln\unitlength0.6ex\begin{picture}(4,3)
        \ifcase#1\put(4,0.8){\vector(-1,0){4}}\put(0,2){\vector(1,0){4}}
        \or \put(0,0.8){\vector(1,0){4}}\put(0,2){\vector(1,0){4}}\fi
    \end{picture}}\,}} 
\catcode`\@=12

\font\eightln=line10 at8pt \catcode`\@=11
\def\triplearrow{\@ifnextchar [{\@triplearrow }{\@triplearrow[0]}}
\def\@triplearrow[#1]{\mathrel{\,\lower0.15ex
    \hbox{\let\@linefnt\eightln\unitlength0.6ex\begin{picture}(4,3)
        \ifcase#1\put(0,0.3){\vector(1,0){4}}\put(0,1.5){\vector(1,0){4}}
        \put(0,2.7){\vector(1,0){4}}
        \or\put(0,0){\vector(2,1){4}}\put(0,1){\vector(2,1){4}}
        \put(0,3){\vector(4,-3){4}}
        \or\put(0,3){\vector(2,-1){4}}\put(0,2){\vector(2,-1){4}}
        \put(0,0){\vector(4,3){4}}
        \or\put(0,0){\vector(4,3){4}}\put(0,1.5){\vector(1,0){4.5}}
        \put(0,3){\vector(4,-3){4}}
        \or\put(0,0.3){\vector(1,0){4}}\put(0,1){\vector(2,1){4}}
        \put(0,3){\vector(2,-1){4}}
        \or\put(0,0){\vector(2,1){4}}\put(0,2){\vector(2,-1){4}}
        \put(0,2.7){\vector(1,0){4}}
        \or\put(4,0.3){\vector(-1,0){4}}\put(4,1.5){\vector(-1,0){4}}
        \put(0,2.7){\vector(1,0){4}}
        \or\put(4,2){\vector(-2,-1){4}}\put(4,3){\vector(-2,-1){4}}
        \put(0,3){\vector(4,-3){4}}
        \or\put(4,0){\vector(-1,0){4}}\put(4,3){\vector(-2,-1){4}}
        \put(0,3){\vector(2,-1){4}}
        \or\put(4,0.3){\vector(-1,0){4}}\put(0,1.5){\vector(1,0){4}}
        \put(0,2.7){\vector(1,0){4}} \fi
  \end{picture}}\,}} 
\catcode`\@=12

\newcommand{\asr}{\doublearrow} \newcommand{\psr}{\doublearrow[1]}
\newcommand{\pc}{\triplearrow} \newcommand{\ac}{\triplearrow[9]}
 
\makeatletter
\newcommand{\contraction}[5][1ex]{% 
  \mathchoice
    {\contraction@\displaystyle{#2}{#3}{#4}{#5}{#1}}%
    {\contraction@\textstyle{#2}{#3}{#4}{#5}{#1}}% 
    {\contraction@\scriptstyle{#2}{#3}{#4}{#5}{#1}}% 
    {\contraction@\scriptscriptstyle{#2}{#3}{#4}{#5}{#1}}}% 
\newcommand{\contraction@}[6]{% 
  \setbox0=\hbox{$#1#2$}% 
  \setbox2=\hbox{$#1#3$}%
  \setbox4=\hbox{$#1#4$}% 
  \setbox6=\hbox{$#1#5$}% 
  \dimen0=\wd2% 
  \advance\dimen0 by \wd6% 
  \divide\dimen0 by 2% 
  \advance\dimen0 by \wd4%
  \vbox{% 
    \hbox to 0pt{% 
      \kern \wd0% 
      \kern 0.5\wd2% 
      \contraction@@{\dimen0}{#6}% 
      \hss}%
    \vskip 0.2ex% 
    \vskip\ht2}}

\newcommand{\contraction@@}[3][0.06em]{% 
  \hbox{%
    \vrule width #1 height 0pt depth #3% 
    \vrule width #2 height 0pt depth #1% 
    \vrule width #1 height 0pt depth #3% 
    \relax}}
 
\makeatother
 
\begin{document} 
 
\title{Periodic-Orbit Theory of Universality in Quantum Chaos}
 
\date{\today}
 
\author{Sebastian M{\"u}ller$^1$, Stefan Heusler$^1$, Petr Braun$^{1,2}$,
  Fritz Haake$^1$, Alexander Altland$^3$}
 
\address{$^1$Fachbereich Physik, Universit{\"a}t Duisburg-Essen,
  45117 Essen, Germany\\
  $^2$Institute of Physics, Saint-Petersburg University, 198504
  Saint-Petersburg, Russia\\
  $^3$Institut f{\"u}r Theoretische Physik, Z{\"u}lpicher Str 77, 50937 K{\"o}ln,
  Germany}
 
\begin{abstract}
  We argue semiclassically, on the basis of Gutzwiller's
  periodic-orbit theory, that full classical chaos is paralleled by
  quantum energy spectra with universal spectral statistics, in
  agreement with random-matrix theory. For dynamics from all three
  Wigner-Dyson symmetry classes, we calculate the small-time spectral
  form factor $K(\tau)$ as power series in the time $\tau$. Each term
  $\tau^n$ of that series is provided by specific families of pairs of
  periodic orbits.  The contributing pairs are classified in terms of
  close self-encounters in phase space. The frequency of occurrence of
  self-encounters is calculated by invoking ergodicity.  Combinatorial
  rules for building pairs involve non-trivial properties of
  permutations. We show our series to be equivalent to perturbative
  implementations of the non-linear sigma models for the Wigner-Dyson
  ensembles of random matrices and for disordered systems; our
  families of orbit pairs are one-to-one with Feynman diagrams known
  from the sigma model.
\end{abstract} 

 
\maketitle
 
\section{Introduction}

\subsection{Background}
 
In the semiclassical limit, fully chaotic quantum systems display
universal properties. Universal behavior has been observed for many
quantities of interest in such different areas as mesoscopic transport
or nuclear physics.  One paradigmatic example stands out and will be
the object of our investigation: According to the
Bohigas-Giannoni-Schmit (BGS) conjecture put forward about two decades
ago \cite{BGS}, highly excited energy levels of generic fully chaotic
systems have universal spectral statistics. This conjecture is
supported by broad experimental and numerical evidence
\cite{Stoeckmann,Haake}.

Level statistics can be characterized by the so-called spectral form
factor. The level density $\rho(E)=\sum_i\delta(E-E_i)$ of a bounded quantum
system ($E_i$ denoting the energy levels) is split into a local
average $\overline{\rho}(E)$ and an oscillatory part $\rho_{\rm osc}(E)$
describing fluctuations around that average. The form factor is
defined as the Fourier transform of the two-point correlator
$\langle\rho_{\rm osc}(E+\frac{\epsilon}{2}) \rho_{\rm osc}(E-\frac{\epsilon}{2})\rangle$ w.r.t.
the energy difference $\epsilon$,
\begin{equation}
\label{def_formfac}
K(\tau)=\left\langle\int\!\!\! \frac{d\epsilon}{\overline{\rho}(E)}
{\rm e}^{\frac{\rm i}{\hbar}\epsilon\tau T_H}\rho_{\rm
      osc}\!\left(E+\textstyle{\frac{\epsilon}{2}}\right) \rho_{\rm
      osc}\!\left(E-\textstyle{\frac{\epsilon}{2}}\right)\!\!\right\rangle;
\end{equation}
here the time $\tau$, conjugate to the energy difference, is measured in
units of the so-called Heisenberg time
\begin{equation} T_H=2\pi\hbar\overline{\rho}(E) 
=\frac{\Omega(E)}{(2\pi\hbar)^{f-1}}\,, \end{equation} 
with $\Omega(E)$ denoting the volume of the energy shell and $f$ the
number of freedoms. Since the study of high-lying states justifies the
semiclassical limit, we may take $\hbar\to 0,T_H\to\infty$, for fixed $\tau$.  To
make $K(\tau)$ a plottable function, two averages, $\langle\ldots\rangle$ in
(\ref{def_formfac}), are necessary, like over windows of the center
energy $E$ and a small time interval $\Delta\tau\ll1$.

Given full chaos, $K(\tau)$ is found to have a universal form, as
obtained by averaging over certain {\it ensembles} of random matrices
\cite{Stoeckmann,Haake,Mehta}.  In the absence of geometric
symmetries, the prediction of random-matrix theory (RMT) only depends
on whether the system in question has no time-reversal (${\cal T}$)
invariance (unitary case), or is ${\cal T}$ invariant with either
${\cal T}^2=1$ (orthogonal case) or ${\cal T}^2=-1$ (symplectic case).
RMT yields for $0<\tau<1$
\begin{equation}\label{formfactor} 
K(\tau)=\begin{cases} 
  \tau\hspace{4cm}\mbox{unitary;}\\
  2\tau-\tau\ln(1+2\tau)=
  2\tau-2\tau^2+2\tau^3-\ldots\\\hspace{4.2cm}\mbox{orthogonal;}\\ 
  \frac{\tau}{2}-\frac{\tau}{4}\ln(1-\tau) 
  =\frac{\tau}{2}+\frac{\tau^2}{4}+\frac{\tau^3}{8}+\ldots\\ 
\hspace{4.2cm}\mbox{symplectic.} 
\end{cases}
\end{equation} 

However, a proof of the faithfulness of individual chaotic dynamics to
random-matrix theory, and even the assumptions required for a proof,
have thus far remained a challenge.  In the present paper, we take up
the challenge and derive the small-$\tau$ expansion of $K(\tau)$ for {\it
  individual} systems; we employ ergodicity and hyperbolicity of the
classical dynamics. Moreover, we require all classical relaxation
times (related to Ruelle-Pollicott resonances and Lyapunov exponents)
to be finite; we need this property to make sure that even the
shortest quantum time scale of relevance, the so-called Ehrenfest time
$T_E\sim\ln\frac{{\rm const.}}{\hbar}$ is much larger than any classical
time scale.
 
Following \cite{Berry,Argaman,SR}, we start from Gutzwiller's trace
formula \cite{Gutzwiller} which expresses the level density as a sum
over classical periodic orbits $\gamma$,
\begin{equation} 
\label{gutzwiller} 
\rho_{\rm osc}(E)\sim\frac{1}{\pi\hbar}{\rm Re}\sum_\gamma A_\gamma {\rm e}^{{\rm i}
S_\gamma/\hbar}\,,
\end{equation} 
wherein $A_\gamma$ is the stability amplitude (including the Maslov phase)
and $S_\gamma$ the action of the $\gamma$th orbit.  By (\ref{gutzwiller}), the
form factor becomes a double sum over orbits,
\begin{equation} 
\label{doublesum} 
K(\tau)=\frac{1}{T_H}\!\!\left\langle\sum_{\gamma,\gamma'}\!A_\gamma
A_{\gamma'}^*{\rm e}^{\frac{\rm i}{\hbar}(S_\gamma-S_{\gamma'})}
\delta\!\left(\!\!\tau T_H-\textstyle{\frac{T_\gamma+T_{\gamma'}}{2}}\!\!\right)\!\!\!
\right\rangle\!;
\end{equation} 
$T_\gamma$ is the period of $\gamma$.  For $\hbar\to 0$, only families of orbit
pairs with small action difference $|S_\gamma-S_{\gamma'}|\sim\hbar$ can give a
systematic contribution to the form factor. For all others, the phase
in (\ref{doublesum}) oscillates rapidly, and the contribution is
killed by the averages indicated.  Fluctuations in {\it quantum}
spectra are thus related to {\it classical} correlations among orbit
actions \cite{Argaman}.  The first periodic-orbit approach to $K(\tau)$
was taken by Berry \cite{Berry}, who derived the leading term in
(\ref{formfactor}) using ``diagonal'' pairs of coinciding ($\gamma'=\gamma$)
and, for time $\cal T$-invariant dynamics, mutually time-reversed
($\gamma'={\cal T}\gamma$) orbits, which obviously are identical in action.
Starting with Argaman et al.  \cite{Argaman}, off-diagonal orbit pairs
were studied in \cite{Cohen,Primack,Smilansky}. The potential
importance of close self-encounters in orbit pairs was first spelled
out in work on electronic transport \cite{Aleiner} and qualitatively
discussed for spectral fluctuations in \cite{Disorder}.

\begin{figure} 
\begin{center}
  \includegraphics[scale=0.36]{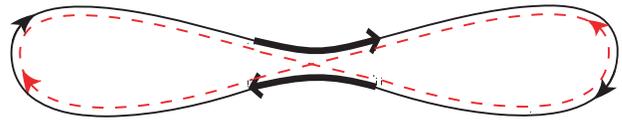}
\end{center} 
\caption{Sketch of a Sieber/Richter pair in configuration space: 
  The partner orbits, depicted by solid and dashed lines, differ
  noticeably only inside an encounter of two orbit stretches (marked
  by antiparallel arrows, indicating the direction of motion). The
  sketch greatly exaggerates the difference between the two partner
  orbits in the loops outside the encounter and depicts the loops
  disproportionally short; similar remarks apply to all subsequent
  sketches of orbit pairs.}
\label{fig:SR} 
\end{figure}

The family of orbit pairs responsible for the next-to-leading order
was definitely identified in Sieber's and Richter's seminal papers
\cite{SR} for a homogeneously hyperbolic system, the
Hadamard-Gutzwiller model (geodesic motion on a tesselated surface of
negative curvature with genus 2).  Their original formulation was
based on small-angle self-crossings of periodic orbits in
configuration space.  In each pair, the partner $\gamma'$ differs from
$\gamma$ only by narrowly avoiding one of its many self-crossings. The two
orbits almost coincide in one of the two parts separated by the
crossing, while they are nearly time-reversed in the other part.  In
phase space, both orbits contain an ``encounter" of two almost
time-reversed orbit stretches. They differ only by their connections
inside that encounter; see Fig \ref{fig:SR}.
 
As shown in \cite{Mueller}, Sieber's and Richter's reasoning can be
extended to general fully chaotic two-freedom systems. One partner
orbit $\gamma'$ arises for each encounter.  The action difference within
each orbit pair \cite{Mueller,Spehner,Turek} can be derived using the
geometry of the invariant manifolds \cite{Braun}.  It thus turned out
helpful to reformulate the treatment in terms of phase-space
coordinates \cite{Spehner,Turek}, which may also be applied to systems
with more than two freedoms \cite{Higherdim}.
 
In \cite{Tau3}, we showed that the $\tau^3$-contribution to the form
factor originates from pairs of orbits which differ either in two
encounters of the above kind, or in one encounter that involves {\it
  three} orbit stretches.
 
In the present paper we demonstrate how the whole series expansion of
$K(\tau)$ is obtained from periodic orbits. Beyond furnishing details
left out in our previous letter \cite{Letter} we here cover systems
with more than two freedoms and from all three Wigner-Dyson symmetry
classes. For the symplectic case we employ ideas presented in
\cite{Heusler,BolteHarrison}. For related work on quantum graphs, see
\cite{Berkolaiko} for the first three orders of $K(\tau)$ and
\cite{GnutzAlt} for a complete treatment.
 
\subsection{Overview}
 
\label{sec:overview}

\begin{figure}
\begin{center}
  \includegraphics[scale=0.25]{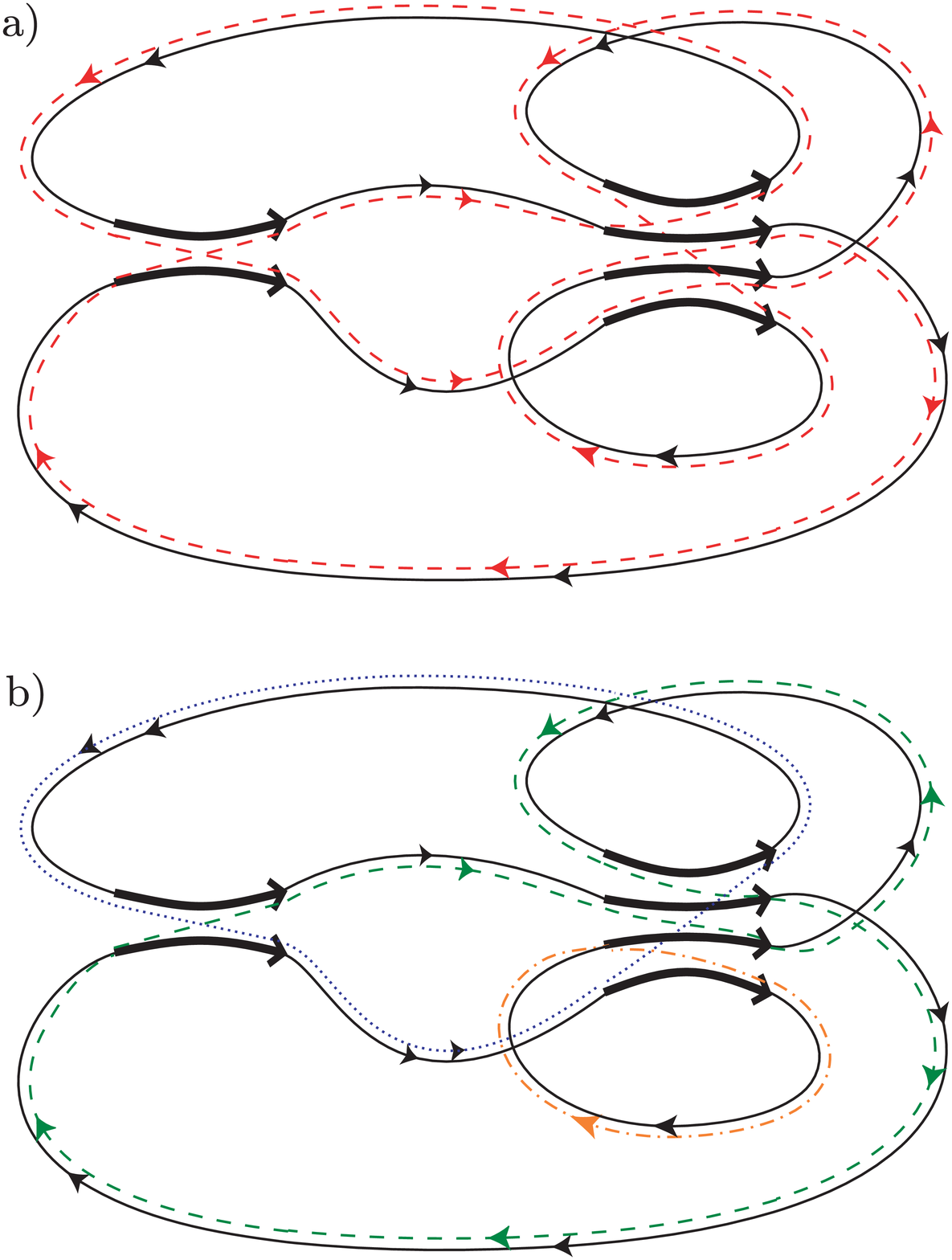}
\end{center}
\caption{a) Solid line: Periodic orbit $\gamma$ with one 4-encounter
  and one 2-encounter highlighted by bold arrows. Dotted line: Partner
  $\gamma'$ differing from $\gamma$ by connections in the encounters.  b)
  Other reconnections yield a ``pseudo-orbit" decomposing into three
  periodic orbits (dashed, dotted, and dash-dotted).}
\label{fig:24}
\end{figure}

We set out to identify the families of orbit pairs responsible for all
orders of the $\tau$-expansion. The key point is that long orbits have a
huge number of close self-encounters which may involve arbitrarily
many orbit stretches. We speak of an $l$-encounter whenever $l$
stretches of an orbit get ``close" in phase space. ``Closeness" will
be quantified below such that we may speak of the beginning, the end,
and the duration of an encounter.  Fig. \ref{fig:24}a highlights two
such encounters inside a periodic orbit, one 2-encounter and one
4-encounter.  Here, as always, we sketch orbit pairs in configuration
space, with arrows $\psr$ indicating the direction of motion inside
the encounter stretches.  The relevant encounters will turn out to
have durations of the order of the Ehrenfest time
$T_E\sim\ln\frac{\rm{const.}}{\hbar}$; even though logarithmically
divergent in the semiclassical limit (and thus larger than all
classical time scales), these encounter durations are vanishingly
small compared to the orbit periods, which are of the order of the
Heisenberg time, $T_\gamma\sim T_H\sim\hbar^{-f+1}$. In between different
self-encounters an orbit goes through ``loops", represented by thin
full lines.

Self-encounters are of interest since they lead us from a periodic
orbit $\gamma$ to partners $\gamma'$ which differ from $\gamma$ noticeably only
inside a set of encounters (see the dashed orbit in \ref{fig:24}a). In
contrast, the orbit loops in between encounters are almost identical.
The almost coinciding loops of $\gamma$ and $\gamma'$ are differently
connected inside the encounters.
 
Not all reshufflings of connections inside an encounter yield a
partner orbit. For example, reconnections as in Fig.  \ref{fig:24}b
give rise to a ``pseudo-orbit" decomposing into three separate
periodic orbits. Pseudo-orbits are not admitted in the Gutzwiller
trace formula.

\begin{figure}
\begin{center}
  \includegraphics[scale=0.18]{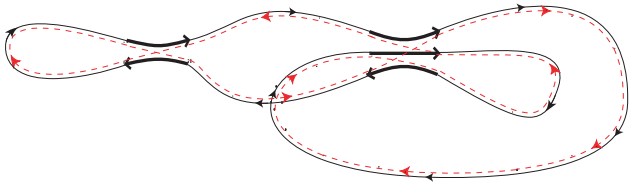}
\end{center}
\caption{Periodic orbit $\gamma$  with one 2- and one 3-encounter
  highlighted, and a partner $\gamma'$ obtained by reconnection; the
  encounters depicted only exist in ${\cal T}$-invariant systems.}
\label{fig:23}
\end{figure}
 
For ${\cal T}$-invariant dynamics, we also must account for encounters
whose stretches only get close {\it up to time reversal} as in $\ac$
and $\asr$; see Fig.  \ref{fig:23}.  Correspondingly, loops inside
mutual partner orbits may be related by time reversal.

We thus obtain a natural extension of Berry's diagonal approximation.
Instead of considering only pairs of orbits which exactly coincide (or
are mutually time-reversed), we employ all pairs whose members are
composed of similar (up to time reversal) loops.
 
We proceed to classify these orbit pairs.  Partner orbits may differ
in a {\it number $v_l$ of $l$-encounters}; we shall assemble these
numbers to a ``vector'' $\vec{v}=(v_2,v_3,\ldots)$.  The total number of
encounters is given by $V=\sum_{l\geq 2}v_l$.  The number of orbit
stretches involved in encounters, coinciding with the number of
intervening loops, reads $L=\sum_{l\geq 2}lv_l$.

The orbit pairs $(\gamma,\gamma')$ related to a fixed vector $\vec v$ may have
various {\it structures}.  Each structure corresponds to a different
ordering (and, given ${\cal T}$-invariance, different sense of
traversal) of the loops of $\gamma$ inside the partner orbit $\gamma'$.  When
drawing orbit pairs as in Figs. \ref{fig:24} and \ref{fig:23}, these
structures differ by the order in which encounters are visited in the
original orbit $\gamma$, and by the relative directions of the stretches
within each encounter (i.e. $\psr$ vs. $\asr$, or $\pc$ vs. $\ac$ in
time-reversal invariant systems).  Moreover, different reconnections
inside the same encounter may give rise to different partners, and
hence different structures.  We will see that structures have a
one-to-one correspondence to permutations, which will be used in
Section \ref{sec:combinatorics} to determine the number $N(\vec v)$ of
structures related to the same $\vec{v}$.
 
Orbit pairs sharing the same $\vec{v}$ and the same structure may
still differ in the {\it phase-space separations} between the
encounter stretches.  We shall parametrize those separations by
suitable variables $s$, $u$ and determine their density $w_T(s,u)$
inside orbits of period $T$. The double sum (\ref{doublesum}) over
orbits defining the spectral form factor will be written as a sum of
contributions from families of orbit pairs, with the family weight $\propto
N(\vec{v})w_T(s,u)$.

This paper is organized as follows.  To free the presentation of
unnecessary details we mostly disregard complications due to $f>2$ and
``non-homogeneous" hyperbolicity (i.e. Lyapunov exponents different
for different periodic orbits).  In Section \ref{sec:geometry}, we
will study the phase-space geometry of encounters and derive the
density $w_T(s,u)$.  The purely combinatorial task of determining the
number of structures $N(\vec{v})$ is attacked in Section
\ref{sec:combinatorics} with the help of the theory of permutations.
We thus obtain series expansions of $K(\tau)$ for individual chaotic
systems with and without ${\cal T}$-invariance; those series fully
coincide with the RMT predictions for the Gaussian orthogonal ensemble
(GOE) and the Gaussian unitary ensemble (GUE), respectively.  Section
\ref{sec:symplectic} generalizes these results to systems where spin
dynamics accompanies translational motion; in particular, we find
agreement with the Gaussian Symplectic Ensemble (GSE) given ${\cal
  T}$-invariance with ${\cal T}^2=-1$.  In Section \ref{sec:sigma}, we
show that our semiclassical procedure bears a close analogy to quantum
field theory.  In fact, our families of orbit pairs are equivalent to
Feynman diagrams met within the theory of disordered systems and the
perturbative implementation of the so-called nonlinear sigma model.
Finally, we present conclusions in Section \ref{sec:conclusions}.
Further details, including a generalization to $f>2$ and
non-homogeneous hyperbolicity, and remarks on the action correlation
function of \cite{Argaman}, are given in Appendices.

\section{Phase-space geometry of encounters}

\label{sec:geometry}

\subsection{Fully chaotic dynamics}

\label{sec:chaos}

At issue are fully chaotic, i.e., hyperbolic and ergodic Hamiltonian
flows without geometric symmetries with $f=2$ ``classical'' freedoms.
In the orthogonal case, the Hamiltonian is assumed to be ${\cal T}$
invariant, ${\cal T} H{\cal T}^{-1}=H$, with an anti-unitary
time-reversal operator ${\cal T}$ squaring to unity.  For convenience,
we assume ${\cal T}$ to be the conventional time-reversal operator
${\cal T}({\bf q},{\bf p})=({\bf q},-{\bf p})$; that assumption does
not restrict generality, since all Hamiltonians with non-conventional
time-reversal invariance can be brought to conventionally
time-reversal invariant form by a suitable canonical transformation
\cite{Braun}.

For each phase-space point ${\bf x}=({\bf q}, {\bf p})$ it is possible
to define a Poincar{\'e} surface of section ${\cal P}$ orthogonal to the
trajectory passing through ${\bf x}$.  Assuming a Cartesian
configuration space (and thus a Cartesian momentum space), ${\cal P}$
consists of all points ${\bf x}+\delta{\bf x}=({\bf q}+ \delta{\bf q}, {\bf
  p}+\delta{\bf p})$ in the same energy shell as ${\bf x}$ whose
configuration-space displacement $\delta{\bf q}$ is orthogonal to ${\bf
  p}$.  For $f=2$, ${\cal P}$ is a $2$-dimensional surface within the
$3$-dimensional energy shell.  Given hyperbolicity, ${\cal P}$ is
spanned by one stable direction ${\bf e}^s({\bf x})$ and one unstable
direction ${\bf e}^u({\bf x})$ \cite{Gaspard}.  We may thus decompose
$\delta{\bf x}$ as
\begin{equation} 
  \label{decompose} 
  \delta{\bf x}=\hat{s}{\bf e}^s({\bf x})+\hat{u}{\bf e}^u 
({\bf x})\,. 
\end{equation}

As long as two trajectories passing respectively through ${\bf x}$ and
${\bf x}+\delta{\bf x}$ remain sufficiently close, we may follow their
separation by linearizing the equations of motion around one
trajectory,
\begin{eqnarray} 
\label{linearize} 
\hat{s}(t)&=&\Lambda({\bf x},t)^{-1}\hat{s}(0)\nonumber\\ 
\hat{u}(t)&=&\Lambda({\bf x},t)\hat{u}(0)\,. 
\end{eqnarray}
Here, $\hat{s}(t)$ and $\hat{u}(t)$ denote stable and unstable
components in a {\it co-moving} Poincar{\'e} section at ${\bf x}(t)$, the
image of ${\bf x}={\bf x}(0)$ under time evolution over time $t$.  In
the long-time limit, the fate of the stretching factor $\Lambda({\bf x},t)$
and thus of the stable and unstable components is governed by the
(local) Lyapunov exponent $\lambda({\bf x})>0$
\begin{equation} 
\label{asymptotics} 
\Lambda({\bf x},t)\sim{\rm e}^{\lambda({\bf x})t}\,. 
\end{equation}
The ${\bf x}$-dependence of $\lambda$ and $\Lambda$ will be relevant only in
Appendix \ref{sec:maths}, when we treat non-homogeneous hyperbolicity;
until then, we may think of these quantities as constants.  As in
\cite{Spehner,Turek,Higherdim}, the directions ${\bf e}^s({\bf x})$
and ${\bf e}^u({\bf x})$ are mutually normalized by fixing their
symplectic product as
\begin{equation} 
\label{norm} 
{\bf e}^u({\bf x})\land{\bf e}^s({\bf x})={\bf e}^u({\bf x})^T
\left(\begin{array}{cc}0&-1\\1&0\end{array}\right)
{\bf e}^s({\bf
  x})=1\,. \end{equation} 

In ergodic systems, almost all trajectories fill the corresponding
energy shell uniformly. The time average of any observable along such
a trajectory coincides with an energy-shell average.
 
Periodic orbits are exceptional in the sense that they cannot visit
the whole energy shell. However, {\it long} periodic orbits still
behave ergodically: According to the equidistribution theorem
\cite{Equidistribution} (see also Appendix \ref{sec:maths}), a time
average over an orbit $\gamma$ augmented by an average over all $\gamma$ from
a small time window, with the squared stability coefficient as a
weight, equals the energy-shell average with the Liouville measure. A
special case is the sum rule of Hannay and Ozorio de Almeida
\cite{HOdA}
\begin{equation}\label{HOdA_sum}
\left\langle\sum_\gamma|A_\gamma|^2\delta(T-T_\gamma)\right\rangle_{\Delta T}=T\,.
\end{equation}
Ergodicity makes, in the limit of long times, for a uniform return
probability: A trajectory starting at ${\bf x}$ again pierces through
${\cal P}$ in a time interval $(t, t+dt)$ with stable and unstable
components of ${\bf x}(t)-{\bf x}(0)$ (or ${\cal T}{\bf x}(t)-{\bf
  x}(0)$) lying in intervals $(\hat{s}, \hat{s}+d\hat{s})$, $(\hat{u},
\hat{u}+d\hat{u})$ with uniform probability $\frac{1}{\Omega}d\hat{s}
d\hat{u} dt$.

\subsection{Encounters}

\label{sec:encounter}

To parametrize an $l$-encounter, we introduce a Poincar{\'e} surface of
section ${\cal P}$ transversal to the orbit at an arbitrary
phase-space point ${\bf x}_{1}$ (passed at time $t_1$) inside one of
the encounter stretches.  The exact location of ${\cal P}$ inside the
encounter is not important.  The remaining stretches pierce through
${\cal P}$ at times $t_j$ ($j=2,\ldots,l$) in points ${\bf x}_{j}$. If the
$j$-th encounter stretch is close to the first one in phase space, we
must have ${\bf x}_{j}\approx{\bf x}_{1}$; if it is almost time-reversed
with respect to the first one, we have ${\cal T}{\bf x}_{j}\approx{\bf
  x}_{1}$.  In the sequel, we shall shorthand as ${\bf y}_{j}\approx{\bf
  x}_{1}$ with ${\bf y}_{j}$ either ${\bf x}_{j}$ or ${\cal T}{\bf
  x}_{j}$.

The small difference ${\bf y}_{j}-{\bf x}_{1}$ can be decomposed in
terms of the stable and unstable directions at ${\bf x}_{1}$,
\begin{equation}
{\bf y}_{j}-{\bf x}_{1}=\hat{s}_{j}{\bf e}^s({\bf x}_{1})
+\hat{u}_{j}{\bf e}^u({\bf x}_{1})\,;
\end{equation}
the stable and unstable components $\hat{s}_{j}$, $\hat{u}_{j}$ depend
on the location ${\bf x}_{1}$ of the Poincar{\'e} section ${\cal P}$
chosen within the encounter. If we shift ${\cal P}$ through the
encounter, the stable components will asymptotically decrease and the
unstable components will asymptotically increase with growing $t_1$,
according to (\ref{linearize},\ref{asymptotics}).

We can now give a more precise definition of an $l$-encounter.  To
guarantee that all $l$ stretches are mutually close, we demand the
stable and unstable differences $|\hat{s}_{j}|$, $|\hat{u}_{j}|$ of
all stretches from the first one to be smaller than a constant $c$.
The bound $c$ must be chosen small enough for the motion around the
$l$ orbit stretches to allow for the mutually linearized treatment
(\ref{linearize}); however, the exact value of $c$ is irrelevant.

The stable and unstable coordinates determine the {\it duration}
$t_{\rm enc}$ of an encounter. We have to sum the durations of the
``head" of the encounter (i.e. the time $t_u$ until the end of the
encounter, when the first of the unstable components $|\hat{u}_{j}|$
reaches $c$) and its ``tail" (i.e. the time $t_s$ passed since the
beginning of the encounter, when the last of the stable components
$|\hat{s}_{j}|$ has fallen below $c$). Using the exponential
divergence of the unstable phase-space separations
(\ref{asymptotics}), we see that the coordinates $|\hat{u}_{j}|$
approximately need the time $\frac{1}{\lambda}\ln\frac{c}{|\hat{u}_{j}|}$
to reach $c$; similarly, the stable coordinates need
$\frac{1}{\lambda}\ln\frac{c}{|\hat{s}_{j}|}$.  We thus obtain
\begin{eqnarray}
\label{tu}
t_u&=&\min_{j}\left\{\frac{1}{\lambda}\ln\frac{c}{|\hat{u}_{j}|}\right\}\,,\quad
t_s=\min_{j}\left\{\frac{1}{\lambda}\ln\frac{c}{|\hat{s}_{j}|}\right\}\,,\nonumber\\
t_{\rm enc}&=&t_s+t_u=\frac{1}{\lambda}\ln\frac{c^2}{\max_i\{|\hat{s}_i|\}
\max_j\{|\hat{u}_j|\}}\,;
\end{eqnarray}
in view of (\ref{linearize}), the duration $t_{\rm enc}$ remains
invariant if the Poincar{\'e} section ${\cal P}$ is shifted through the
encounter.

An $l$-encounter involves $l$ different orbit stretches whose initial
and final phase-space points will be referred to as ``entrance" and
"exit ports".  If all encounter stretches are (almost) parallel, as in
$\pc$, all entrance ports are located on the same side of the
encounter, and the exit ports are located on the opposite side.  If
the encounter involves mutually time-reversed orbit stretches as
$\ac$, this is no longer the case. Thus, it is useful to introduce the
following convention: All ports on the side where the first stretch
begins are called ``left ports", while those on the opposite side are
"right ports". For parallel encounters, ``entrance" and ``left" are
synonymous, as well as ``exit" and ``right".

\subsection{Partner orbits}

\label{sec:partner_orbits}

\begin{figure}
\begin{center}
  \includegraphics[scale=0.3]{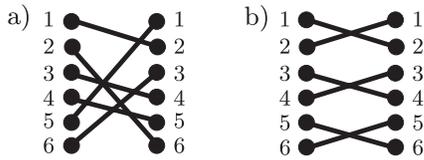}
\end{center}
\caption{Connections between left and right ports in partner orbit $\gamma'$. In b), the encounter splits into three pieces
  respectively containing the two upper, middle, and lower stretches.}
\label{fig:permutations}
\end{figure}

The partner orbits $(\gamma,\gamma')$ differ from one another only inside the
encounters, by their connections between left and right ports.  We
shall number these ports in order of traversal by $\gamma$, such that the
$i$-th encounter stretch of $\gamma$ connects left port $i$ to right port
$i$. Inside $\gamma'$, the left port $i$ is connected to a different right
port $j$; see Fig.  \ref{fig:permutations}a.

We must reshuffle connections between {\it all} stretches of a given
encounter. In contrast, Fig. \ref{fig:permutations}b shows
reconnections only between stretches 1 and 2, stretches 3 and 4, and
between stretches 5 and 6 of a 6-encounter which therefore decomposes
into three 2-encounters.

\subsubsection{Piercing points}

\begin{figure}
\begin{center}
  \includegraphics[scale=0.34]{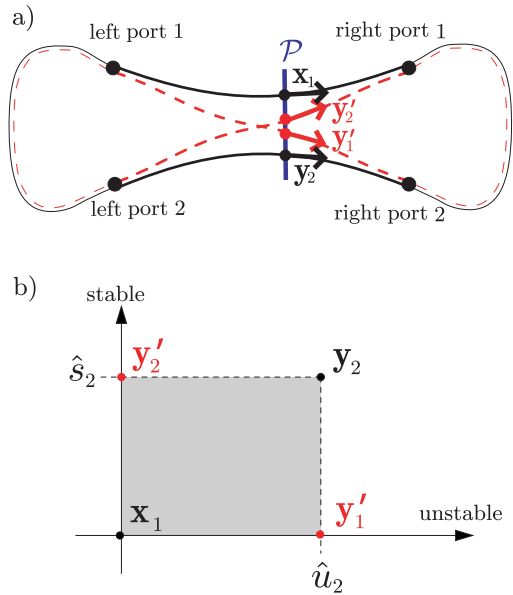}
\end{center}
\caption{Piercings ${\bf x}_1$, ${\bf y}_2$ of $\gamma$ (full line)
  and piercings ${\bf y}_1'$, ${\bf y}_2'$ of $\gamma'$ (dashed line) for
  a Sieber/Richter pair, depicted a) in configuration space, with
  arrows indicating the momentum of the above phase-space points, and
  b) in the Poincar{\'e} section ${\cal P}$ parametrized by stable and
  unstable coordinates.  The symplectic area of the rectangle is the
  action difference $\Delta S$.}
\label{fig:piercings2}
\end{figure}

\begin{figure}
\begin{center}
  \includegraphics[scale=0.34]{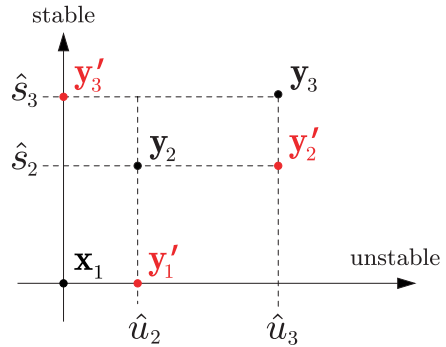}
\end{center}
\caption{Piercing points of $\gamma$,$\gamma'$ differing in a 3-encounter;
  inside $\gamma'$, left ports 1,2,3 are connected to right ports 2,3,1,
  respectively.  }
\label{fig:piercings3}
\end{figure}

\label{sec:manifolds}

The partner $\gamma'$ also pierces through our Poincar{\'e} section ${\cal
  P}$.  The corresponding piercing points are determined by those of
$\gamma$.  In particular, the unstable coordinates of a piercing point
depend on the following right port. If two stretches of $\gamma$ and $\gamma'$
lead to the same right port, they have to approach each other for a
long time - at least until the end of the encounter (which has a
duration $\sim T_E$) and half-way through the subsequent loop. Hence,
their difference must be close to the stable manifold, and the
unstable coordinates almost coincide.  Similarly, the stable
coordinates are determined by the previous left port, since stretches
with the same left port approach for large negative times.  If a
stretch of $\gamma'$ connects left port $i$ to right port $j$, it thus
pierces through our Poincar{\'e} section with stable and unstable
coordinates
\begin{equation} \hat{s}_i'\approx\hat{s}_i,\ \ \ \
  \hat{u}_i'\approx\hat{u}_j.  \end{equation}

For instance, if $\gamma$ and $\gamma'$ differ in a 2-encounter, $\gamma'$
connects left port 1 to right port 2, and left port 2 to right port 1;
see Fig. \ref{fig:piercings2}a.  Thus, the encounter stretches of
$\gamma'$ pierce through ${\cal P}$ in ${\bf y}'_1$ with $\hat{s}'_{1}\approx
\hat{s}_{1}=0$, $\hat{u}'_{1}\approx \hat{u}_{2}$, and in ${\bf y}'_2$ with
$\hat{s}'_{2}\approx \hat{s}_{2}$, $u'_{2}\approx \hat{u}_{1}= 0$, which
together with the piercings of $\gamma$ span a parallelogram in phase
space \cite{Braun} (a rectangle in Fig.  \ref{fig:piercings2}b, by
artist's license).  In Fig.  \ref{fig:piercings3}, we visualize the
locations of ${\bf y}'_j$ inside ${\cal P}$ for a 3-encounter.

\subsubsection{Action difference}

\label{sec:action_difference}

We can now determine the difference between the actions of the two
partner orbits, first for $\gamma$, $\gamma'$ only differing in one
2-encounter. Generalizing the results for configuration-space
crossings in \cite{SR,Mueller}, we will show that the action
difference is just the symplectic area of the rectangle in Fig.
\ref{fig:piercings2}b \cite{Spehner,Turek}.  Consider two segments of
the encounter stretches in Fig. \ref{fig:piercings2}a, leading from
the first left port to the piercing point ${\bf x}_1$ of $\gamma$, and to
the piercing point ${\bf y}_1'$ of $\gamma'$, respectively. Since the
action variation brought about by a shift $d{\bf q}$ of the final
coordinate is ${\bf p}\cdot d{\bf q}$, the action difference between the
two segments will be given by $\Delta S^{(1)}=\int_{{\bf y}_1'}^{{\bf
    x}_1}{\bf p}\cdot d{\bf q}$.  The integration line may be chosen to
lie in the Poincar{\'e} section; then it coincides with the unstable axis.
Repeating the same reasoning for the remaining segments, we obtain the
overall action difference $\Delta S\equiv S_\gamma-S_{\gamma'}$ as the line integral
$\Delta S=\oint{\bf p}\cdot d{\bf q}$ along the contour of the parallelogram
${\bf y}'_1\to{\bf x}_1\to{\bf y}'_2\to{\bf y}_2\to{\bf y}'_1$, spanned by
${\bf y}_1'-{\bf x}_1=\hat{u}_2{\bf e}^u({\bf x}_1)$ and ${\bf
  y}_2'-{\bf x}_1=\hat{s}_2{\bf e}^s({\bf x}_1)$.  This integral
indeed gives the symplectic area
\begin{equation}
\label{DS2}
\Delta S= \hat{u}_{2}{\bf e}^u({\bf x}_1)\land\hat{s}_{2}{\bf e}^s({\bf x}_1)
=\hat{s}_{2}\hat{u}_{2}\,. \end{equation}

\begin{figure}
\begin{center}
  \includegraphics[scale=0.28]{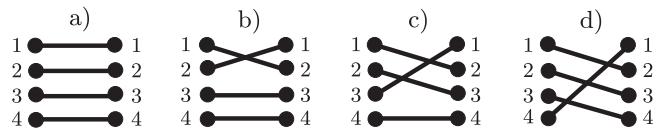}
\end{center}
\caption{Steps from connections in $\gamma$ (depicted in a) to those in $\gamma'$
  (shown in d), in each step interchanging right ports of two
  encounter stretches.}
\label{fig:steps}
\end{figure}

To generalize to arbitrary $l$-encounters, we imagine a partner orbit
$\gamma'$ constructed out of $\gamma$ by $l-1$ successive steps, as
illustrated for a special example in Fig.~\ref{fig:steps}. Each step
interchanges the right ports of two encounter stretches and
contributes to the action difference an amount given by (\ref{DS2}).
At the same time, the two piercings points change their position as
discussed in \ref{sec:manifolds}.  This step-by-step process suggests
a useful transformation of coordinates.  Let $s_{j}$, $u_{j}$ denote
the stable and unstable differences between the two stretches affected
by the $j$-th step.  Note that in contrast to $\hat{s}_{j}$,
$\hat{u}_{j}$ the index $j$ no longer represents encounter stretches
$2,\ldots,l$ but steps $1,\ldots,(l-1)$.  Now, the change of action in each
step is simply given by $ s_{j}u_{j}$.  Summing over all steps, we
obtain a total action difference
\begin{equation}
\label{DS}
\Delta S=\sum_{j=1}^{l-1}s_{j}u_{j}\,.
\end{equation}
The transformation leading from $\hat{s}_{j}$, $\hat{u}_{j}$ to
$s_{j}$, $u_{j}$ is linear and volume-preserving.\footnote{First,
  consider reconnections as depicted in Fig. \ref{fig:steps}d for
  $l=4$.  We proceed from \ref{fig:steps}a to \ref{fig:steps}d in
  $l-1=3$ steps.  In the $j$-th step, we change connections between
  left ports $j$ and $j+1$, and right ports 1 and $j+1$.  Recall that
  stable and unstable coordinates of piercing points are determined by
  the left and right ports, respectively.  Thus, the separation
  between the stretches affected has a stable component
  $s_j=\hat{s}_{j+1}-\hat{s}_j$ and an unstable component
  $u_j=\hat{u}_{j+1}-\hat{u}_{1}$.  The Jacobian of the transformation
  $\hat{s},\hat{u}\to s,u$ is equal to 1.  All other permissible
  reconnections can be brought to a form similar to Fig.
  \ref{fig:steps}d (albeit with different $l$), by appropriately
  changing the numbering of stretches; hence they allow for the same
  step-by-step procedure.  } Due to the elegant form of (\ref{DS}), it
will be convenient to use $s_{j}$, $u_{j}$ rather than $\hat{s}_{j}$,
$\hat{u}_{j}$ in defining the encounter regions, demanding all
$|s_{j}|$, $|u_{j}|$ to be smaller than our bound $c$.  Employing
(\ref{linearize}), one easily shows that $\Delta S$ remains invariant if
the Poincar{\'e} section ${\cal P}$ is shifted through the encounter.
Moreover, if the orbits $\gamma$ and $\gamma'$ differ in several encounters,
the total action difference is additive in their contributions, and
each is given by (\ref{DS}).

At this point, we can finally appreciate that the encounters relevant
for spectral universality have a duration of the order of the
Ehrenfest time. The form factor is determined by orbit pairs with an
action difference $\Delta S$ of order $\hbar$. According to our expression
(\ref{DS2}) for $\Delta S$, the relevant stable and unstable coordinates
are of the order $\sqrt{\hbar}$. The encounter duration, logarithmic in
$s$ and $u$, must consequently be of the order of the Ehrenfest time.

\subsection{Structures}

\label{sec:structures}

We want to define more precisely the notion of ``structures" of orbit
pairs $\gamma$, $\gamma'$.

(i) First of all, these structures are characterized by {\it the order
  in which encounters are traversed in $\gamma$}.  We enumerate the
encounter stretches of $\gamma$ in their order of traversal, starting from
some arbitrarily chosen stretch, and assemble the labels $1\ldots L(\vec
v)$ in $V=V(\vec v)$ groups according to the encounters they belong
to. Such a division uniquely defines the order in which the encounters
are visited. For example, in an orbit pair differing in two
2-encounters the four stretches can be distributed among the
encounters as (1,2)(3,4) or (1,3)(2,4) or (1,4)(2,3); each of these
three possibilities determines a different structure.

Some structures refer to the same orbit pair. Indeed, a different
choice of the initial stretch in the same $\gamma$ would lead to a cyclic
shift in the enumeration of stretches, and that shift may change the
structure associated with $\gamma$.  In the example of two 2-encounters,
cyclic shifts may either leave the structure (1,2)(3,4) invariant or
turn it into (1,4)(2,3), such that the structures (1,2)(3,4) and
(1,4)(2,3) are physically equivalent.

Moreover, structures are characterized (ii) by the {\it relative
  directions} of the encounter stretches (i.e. $\psr$ or $\asr$ for
2-encounters, and $\pc$ or $\ac$ for 3-encounters in ${\cal T}$
invariant systems), and (iii) by the {\it reconnections} leading from
$\gamma$ to $\gamma'$; the latter distinction is important if there exist
several such reconnections inside the same encounter set, each leading
to a different connected partner.

\subsection{Statistics of encounter sets}

\label{sec:statistics}

\begin{figure}
\begin{center}
  \includegraphics[scale=0.2]{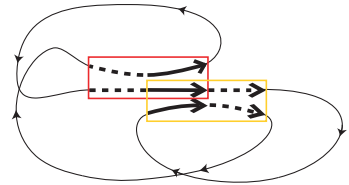}
\end{center}
\caption{Two 2-encounters (marked by boxes) overlap in one stretch and
  thus merge to a 3-encounter (solid bold arrows)}
\label{fig:overlap}
\end{figure}

The statistics of close self-encounters inside periodic orbits can be
established using the {\it ergodicity} of the classical motion.  As a
second ingredient, it is important to only consider sets of encounters
whose stretches are {\it separated by non-vanishing loops, i.e. do not
  overlap}.  For example, if two stretches of {\it different
  encounters} overlap, the two encounters effectively merge, leaving
one larger encounter with more internal stretches, see Fig.
\ref{fig:overlap}.  The partners are thus seen as differing in one
larger encounter, rather than in two smaller ones.  For the more
involved case of stretches belonging to the same encounter, see
Appendix \ref{sec:overlap_appendix}.

In the following, we will consider encounter sets within orbit pairs
$(\gamma,\gamma')$ with fixed $\vec{v}$ and fixed structure.  Each of the $V$
encounters of $\gamma$ is parametrized with the help of a Poincar{\'e} section
${\cal P}_\alpha$ ($\alpha=1,\ldots,V$) crossing the orbit at an arbitrary
phase-space point inside the encounter, traversed at time $t_{\alpha1}$.
The orbit again pierces through these sections at {\it times $t_{\alpha
    j}$} with $j=2,\ldots,l_\alpha$ numbering the remaining stretches of the
$\alpha$-th encounter.  The first piercing may occur anywhere inside the
orbit at a time $0<t_{11}<T$, $T$ denoting the period.  The remaining
$t_{\alpha j}$ follow in an order fixed by the structure at times
$t_{11}<t_{\alpha j}<T+t_{11}$.  Each of the $v_l$ $l$-encounters is
characterized by $l-1$ {\it stable and unstable coordinates} $s_{\alpha
  j}$, $u_{\alpha j}$ ($j=1,\ldots,l-1$), which in total make for $2\sum_{l\geq
  2}(l-1)v_l=2(L-V)$ components.  If ${\cal P}_\alpha$ is shifted through
the encounter, the stable and unstable coordinates change while the
contributions to the action difference $\Delta S_{\alpha j}=s_{\alpha j}u_{\alpha j}$
remain invariant.

We proceed to derive a density $w_T(s,u)$ of phase-space separations
$s$, $u$.  To understand the normalization of $w_T(s,u)$, assume that
it is multiplied with $\prod_{\alpha j}\delta(\Delta S_{\alpha j}-s_{\alpha j}u_{\alpha j})$ and
integrated over all $s$, $u$.  The result will be the average density
of partners $\gamma'$ per one orbit $\gamma$ such that the pair $(\gamma,\gamma')$ has
the given structure and action difference components $\Delta S_{\alpha j}$.
Averaging will be carried out over the ensemble of all periodic orbits
$\gamma$ with period $T$ in a given time window, assuming that the
contribution of each orbit is weighted with the square of its
stability amplitude. According to the equidistribution theorem this
ensemble is ergodic yielding the same averages as integrating over the
energy shell with the Liouville measure.

We need to count the piercings through {\it Poincar{\'e} sections}
parametrized by stable and unstable coordinates.  Due to ergodicity,
the expected number of such piercings through a given section in the
time interval $(t,t+dt)$ with stable and unstable components in
$(\hat{s},\hat{s}+d\hat{s}) \times (\hat{u},\hat{u}+d\hat{u})$ is equal to
$d\hat{s}\;d\hat{u}\;dt/\Omega$, i.e. corresponds to the uniform Liouville
density $1/\Omega$.

In fact, we need the number of {\it sets} of $L-V$ piercings through
our sections ${\cal P}_\alpha$ occurring in time intervals $(t_{\alpha
  j},t_{\alpha j}+dt_{\alpha j})$, $j=2,\ldots,l_\alpha$, with stable and unstable
coordinates inside $(s_{\alpha j},s_{\alpha j}+ds_{\alpha j})$, $(u_{\alpha j},u_{\alpha
  j}+du_{\alpha j})$, $j=1,\ldots,l_\alpha-1$; that number will be denoted by
$\rho_T(s,u)\,d^{L-V}s\,d^{L-V}u\,d^{L-V}t$.  The uniform Liouville
density carries over to the coordinates $s_{\alpha j},u_{\alpha j}$ since the
transformation from $\hat s,\hat u$ to $s,u$ is volume-preserving; so
we may expect $\rho_T(s,u)$ equal to $1/\Omega^{L-V}$.

However, recall that we are only interested in encounters separated by
non-vanishing loops.  To implement that restriction, we employ a
suitable characteristic function $\Theta_T(s,u,t)$ which vanishes if the
piercings described by $s$, $u$ and $t$ correspond to overlapping
stretches, and otherwise equals $1$. We thus obtain
\begin{equation}
\label{rho}
\rho_T(s,u,t)=\Theta_T(s,u,t)
\frac{1}{\Omega^{L-V}}\,.  \end{equation}

(Actually the duration of the connecting loops must be not just
positive but also larger than all classical relaxation times
describing correlation decay, to guarantee the statistical
independence of the piercings.  However, that classical minimal loop
length $t_{\rm cl}$ is not worth further mention since it is
vanishingly small compared to the Ehrenfest time, the smallest time
scale of semiclassical relevance.)

Proceeding towards $w_T(s,u)$ we integrate over the $L-V$ piercing
times $t_{\alpha j}$, $j\geq 2$, still for {\it fixed Poincar{\'e} sections}
${\cal P}_\alpha$. The integral yields a density of the stable and
unstable components $s$, $u$ of $L-V$ piercings, reckoned from the $V$
reference piercings.  To finally get to $w_T$, we must keep track of
all encounters along the orbits in question. To that end we have to
consider all possible positions of Poincar{\'e} sections and thus
integrate over the times $t_{\alpha1}$ (of the reference piercings) as
well.  Doing so, we weigh each encounter with its duration $t_{\rm
  enc}^\alpha$, since we may move each Poincar{\'e} section to any position
inside the duration of the encounter. In order to count each encounter
set exactly once, we divide out the factors $t_{\rm enc}^\alpha$, and thus
arrive at the desired density
\begin{equation}
\label{wintegral}
w_T(s,u)=\frac{\int d^Lt\,\Theta_T(s,u,t)}
{\prod_\alpha t_{\rm enc}^\alpha\,\Omega^{L-V}}\,.
\end{equation}

It remains to evaluate the $L$-fold time integral in
(\ref{wintegral}).  The integration over $t_{11}$ runs from 0 to $T$;
it will be done as the last integral and then give a factor $T$. The
$L-1$ other $t_{\alpha j}$ must lie inside the interval
$(t_{11},t_{11}+T)$ and respect the ordering dictated by the encounter
structure. Moreover, subsequent encounter stretches must not overlap.
Thus, the time between the piercings of two subsequent stretches must
be so large as to contain both the head of the first stretch and,
after a non-vanishing loop, the tail of the second stretch.  (Given
${\cal T}$-invariance, we rather need to include the tail of the first
stretch, if it is time-reversed w.~r.~t. the earliest stretch of its
encounter; likewise the second stretch may also participate with its
head.)  These minimal distances sum up to the total duration of all
encounter stretches $\sum_\alpha l_\alpha t_{\rm enc}^\alpha$, since each stretch
appears in this sum once with head and tail.

The minimal distances effectively reduce the integration range, as we
may proceed to a new set of times $\tilde{t}_{\alpha j}$ obtained by
subtracting from $t_{\alpha j}$ both $t_{11}$ and the sum of minimal
distances between $t_{11}$ and $t_{\alpha j}$. The $\tilde{t}_{\alpha j}$ just
have to obey the ordering in question, and lie in an interval
$(0,T-\sum_\alpha l_\alpha t_{\rm enc}^\alpha)$, where the subtrahend is the total
sum of minimal distances. We are thus left with a trivial integral
over a constant. Perhaps surprisingly, the resulting density
\begin{equation}
\label{density}
w_T(s,u)=\frac{T(T-\sum_\alpha l_\alpha t_{\rm enc}^\alpha)^{L-1}}
{(L-1)!\prod_\alpha t_{\rm enc}^\alpha\Omega^{L-V}}
\end{equation}
depends only on $\vec{v}$ but not on the structure considered, and
that fact strongly simplifies our treatment.

The number $P_{\vec{v}}(\Delta S)d\Delta S$ of orbit pairs with given
$\vec{v}$ and action difference within $(\Delta S,\Delta S+d\Delta S)$ now reads
\begin{eqnarray} \label{numorb}
&&\!\!\!\!\!\! P_{\vec{v}}(\Delta S)d\Delta S
= d\Delta S\frac{N(\vec v)}{L}\times\nonumber\\
&&\!\!\!\!\!\!\int d^{L-V}s\,d^{L-V}u\, \delta\Big(\Delta
S-\sum_{\alpha j} s_{\alpha j} u_{\alpha j}\Big) w_T(s,u)
\end{eqnarray}
where $N(\vec v)$ is the number of structures existing for the given
$\vec v$.  Multiplication by $N(\vec v)$ is equivalent to summation
over all structures belonging to the same $\vec v$, since $w_T$ is the
same for all such structures.  The denominator $L$ prevents an
overcounting. To understand this, remember that one encounter stretch
was arbitrarily singled out as ``the first'' and assigned the piercing
time $t_{11}$. Each of the $L$ possible such choices leads to a
different parametrization by $s,u$ of the same encounter set, and may
also lead to a different structure.  The integral over $s,u$ in
(\ref{numorb}) includes the contributions of all equivalent
parametrizations, and this is why the factor $L$ must be divided out.

\subsection{Contribution of each structure}

\label{sec:contribution}

To determine the spectral form factor, we have to evaluate the double
sum over periodic orbits $\gamma$, $\gamma'$ in (\ref{doublesum}).  In doing
so, we will account for all families of orbit pairs whose members are
composed of loops similar up to time reversal, i.e.  both ``diagonal"
pairs and orbit pairs differing in encounters.  We assume that these
are the only orbit pairs to give rise to a systematic contribution (an
assumption that will be further discussed in the conclusions).  For
the pairs $\gamma,\gamma'$ related to encounters not only the action
difference $S_\gamma-S_{\gamma'}$ but also the difference of the stability
amplitudes and the difference of the periods are very small.\footnote{
  As shown in \cite{Higherdim}, the quantities determining $A_\gamma$ (the
  period, the Lyapunov exponent and the Maslov index of $\gamma$) may be
  written as integrals of time-reversal invariant quantities along the
  orbit; see also \cite{Mueller,Turek,Robbins} for the Maslov index.
  Since $\gamma'$ locally almost coincides with $\gamma$ up to time reversal,
  we have $A_\gamma\approx A_{\gamma'}$.  For the case of the Hadamard-Gutzwiller
  model, a more careful treatment of these points is given in
  \cite{HeuslerPhD,MuellerPhD}.}  Since only the action difference is
discriminated by the small quantum unit we may simplify the double sum
(\ref{doublesum}) as
\begin{equation}
\label{doublesumsimplified}
K(\tau)=\frac{1}{T_H}\left\langle\sum_{\gamma,\gamma'}|A_\gamma|^2
{\rm e}^{{\rm i}(S_\gamma-S_{\gamma'})/\hbar}
\delta\left(\tau T_H-T_\gamma\right)\right\rangle\,.
\end{equation}
The summation over $\gamma$ is evaluated using the rule of Hannay and
Ozorio de Almeida (\ref{HOdA_sum}).  The diagonal pairs contribute
$\kappa\tau$, with $\kappa=1$ in the unitary and $\kappa=2$ in the orthogonal case.
The sum over partners $\gamma'$ differing from $\gamma$ in encounters can be
performed with the help of the density $P_{\vec{v}}(\Delta S)$,
\begin{equation}
\label{ksum_DS}
K(\tau)=\kappa \tau+\kappa \tau\sum_{\vec{v}}\int d\Delta S\;
P_{\vec{v}}(\Delta S) {\rm e}^{{\rm i}\Delta S/\hbar}\,.
\end{equation}
The factor $\kappa $ in the second member is inserted since apart from the
partner orbits considered so far, time-reversal invariance demands to
also take into account their time-reversed versions; the factor $\tau$
comes from the sum rule, setting $\tau=\frac{T}{T_H}$.  Substituting
(\ref{numorb}) for $P_{\vec{v}}(\Delta S)$ we get
\begin{eqnarray}
\label{ksum}
&& K(\tau)=\kappa \tau+\kappa \tau\sum_{\vec{v}} N(\vec{v}) \times\nonumber\\
&&\int d^{L-V}s\;d^{L-V}u \;
\frac{w_T(s,u)}{L}\;{\rm e}^{\frac{\rm i}{\hbar}\sum_{\alpha j}s_{\alpha j}u_{\alpha j}}\,.
\end{eqnarray}
Here, the orbit pairs with fixed $\vec{v}$, structures, and
separations $s,u$ appear with the weight
$N(\vec{v})\frac{w_T(s,u)}{L}$.

The integral over $s$ and $u$, multiplied with $\kappa \tau$, yields the
contribution to the form factor from each structure associated to
$\vec{v}$. The integral is surprisingly simple to do. Consider the
multinomial expansion of $(T-\sum_\alpha l_\alpha t_{\rm enc}^\alpha)^{L-1}$ in our
expression (\ref{density}) for the density $w_T(s,u)$. We shall show
that only a single term of that expansion contributes, the one which
involves a product of all $t_{\rm enc}^\alpha$ and therefore cancels with
the denominator,
\begin{eqnarray}
\label{wcontr}
\frac{w_T^{\rm contr}}{L}&=&\frac{T\frac{(L-1)!}{(L-V-1)!}
T^{L-V-1}\prod_\alpha(-l_\alpha)}{L!\;\Omega^{L-V}}\nonumber\\&=& h(\vec{v})
\left(\frac{T}{\Omega}\right)^{L-V}\,,\nonumber\\
h(\vec{v}) &\equiv&\frac{(-1)^V\prod_l l^{v_l}}{L(L-V-1)!}\,.
\end{eqnarray}
Due to the cancellation of $t_{\rm enc}^\alpha$, $w_T^{\rm contr}$ does
not depend on the stable and unstable coordinates and therefore the
remaining integral over $s$ and $u$ is easily calculated,
\begin{eqnarray}
\label{contribution}
&&\kappa \tau\int d^{L-V}s\;d^{L-V}u\;\frac{w_T^{\rm contr}}{L}\;
{\rm e}^{\frac{\rm i}{\hbar}\sum_{\alpha j}s_{\alpha j}u_{\alpha j}}\nonumber\\
&=&\kappa \tau h(\vec{v}) \left(\frac{T}{\Omega}\right)^{L-V}
\int_{-c}^c d^{L-V}s\;
d^{L-V}u\;{{\rm e}}^{\frac{\rm i}{\hbar}\sum_{\alpha j}s_{\alpha j}u_{\alpha j}}\nonumber\\
&\to& \kappa  h(\vec{v}) \;\tau^{L-V+1}\,;
\end{eqnarray}
we have just met with the $(L-V)$-th power of the integral
\begin{equation}\label{simple_integral}
\int_{-c}^c ds\;du\; {\rm e}^{{\rm i} s u/\hbar}\to 2\pi\hbar
\end{equation}
and used $2\pi\hbar\frac{T}{\Omega}=\frac{T}{T_H}=\tau$.  In the semiclassical
limit, the contributions of all other terms in the multinomial
expansion vanish for one of two possible reasons:

First, consider terms in which at least one encounter duration $t_{\rm
  enc}^\alpha$ in the denominator is not compensated by a power of $t_{\rm
  enc}^\alpha$ in the numerator. The corresponding contribution to the
form factor is proportional to
\begin{equation}
\label{oscillating}
\int_{-c}^c\prod_{j}ds_{\alpha j}du_{\alpha j}\frac{1}{t_{\rm enc}^\alpha}
{\rm e}^{\frac{\rm i}{\hbar}\sum_{j}s_{\alpha j}u_{\alpha j}}\,.
\end{equation}
As shown in Appendix \ref{sec:integral}, such integrals oscillate
rapidly and effectively vanish in the semiclassical limit, as $\hbar\to0$.

Second, there are terms with, say, $p$ factors $t_{\rm enc}^\alpha$ in the
numerator left uncancelled.  To show that such terms do not contribute
we employ a scaling argument.  Obviously, the considered terms may
only involve a smaller order of $T$ than $w_T^{\rm contr}$; they are
of order $T^{L-V-p}$.  However, $\Omega$ still appears in the same order
$\frac{1}{\Omega^{L-V}}$.  To study the scaling with $\hbar$, we transform to
variables $\tilde{s}_{\alpha j}=\frac{s_{\alpha j}}{\sqrt{\hbar}}$,
$\tilde{u}_{\alpha j}=\frac{u_{\alpha j}}{\sqrt{\hbar}}$, eliminating the
$\hbar$-dependence in the phase factor of (\ref{ksum}).  The resulting
expression is proportional to $\hbar^{L-V}$ due to the Jacobian of the
foregoing transformation, and proportional to $(\ln\hbar)^p$ due to the
$p$ remaining encounter durations $\sim\ln\hbar$.  Together with the factor
$\tau$ originating from the sum rule, the corresponding contribution to
the form factor scales like
\begin{eqnarray} &&\tau T^{L-V-p}\left(\frac{\hbar}{\Omega}\right)^{L-V}\!(\ln
  \hbar)^p \propto \tau\frac{T^{L-V-p}}{T_H^{L-V}}(\ln\hbar)^p\nonumber\\ &&
\propto
  \left(\frac{\ln\hbar}{T_H}\right)^p\tau^{L-V+1-p}, \end{eqnarray}
and thus disappears as $\hbar\to 0$, $T_H\propto\hbar^{-1}\to\infty$.

Therefore, the contribution to the form factor arising from each
structure with the same $\vec{v}$ is indeed determined by
(\ref{contribution}).  Remarkably, this result is due to a subleading
term in the multinomial expansion of $w_T(s,u)$, originating only from
the small corrections due to the ban of encounter overlap.

The calculation of the form factor is now reduced to the purely
combinatorial task of determining the numbers $N(\vec{v})$ of
structures and evaluating the sum
\begin{equation}
\label{k}
K(\tau)=\kappa \tau+\kappa \sum_{\vec{v}}N(\vec{v})h(\vec{v}) \tau^{L-V+1}\,.
\end{equation}
The $n$-th term in the series $K(\tau)=\kappa \tau+\sum_{n\geq 2}K_n\tau^n$ is
exclusively determined by structures with $\nu(\vec{v})\equiv
L(\vec{v})-V(\vec{v})+1=n$.  It will be convenient to represent $K_n$
as\footnote{ We slightly depart from the notation in \cite{Letter},
  where $\tilde{N}(\vec{v})$ was defined to include the denominator
  $(n-2)!$.}
\begin{eqnarray}
\label{kn}
  K_n&=&\frac{\kappa }{(n-2)!}\sum_{\vec{v}}^{\nu(\vec{v})=n}\tilde{N}(\vec{v})\,,
\\
\tilde{N}(\vec{v})&\equiv&N(\vec{v})\frac{(-1)^V\prod_l l^{v_l}}{L(\vec{v})}\,.
\label{tilde}
\end{eqnarray}

\section{Combinatorics}

\label{sec:combinatorics}

\subsection{Unitary case}

\label{combinatorics_unitary}

\subsubsection{Structures and permutations}

To determine the combinatorial numbers $N(\vec{v})$, first for systems
without ${\cal T}$-invariance, we must relate structures of orbit
pairs to permutations.

Most importantly, we require the notion of cycles \cite{Permutations}.
We may denote a permutation of $l$ objects (say the natural numbers
$1,2,\ldots l$) by $\{a\to P(a), a=1\ldots l\}$ or $ P=\left({1\atop
    P(1)}{2\atop P(2)}{\ldots\atop } {l\atop P(l)}\right)$. An alternative
bookkeeping starts with some object $a_1$ and notes the sequence of
successors, $a_1\to a_2= P(a_1)\to a_3= P(a_2)\ldots$; if that sequence
first returns to the starting object $a_1$ after precisely $l$ steps
one says that the permutation in question is a single cycle, denotable
simply as $(a_1,a_2,\ldots a_l)$. A cycle is defined up to cyclic
permutations of its member objects. The number $l$ of objects in a
cycle is called the length of that cycle.  Obviously, not every
permutation is a cycle. A more general permutation can be decomposed
into several cycles.

We now turn to applying the notion of cycles to self-encounters of a
long periodic orbit $\gamma$ and its partner orbit(s).  We first focus on
an orbit pair differing in a single $l$-encounter.  This encounter
involves $l$ orbit stretches, whose entrance and exit ports will be
labelled by $1,2,\ldots,l$.  Inside $\gamma$ the $j$-th encounter stretch
connects the $j$-th entrance and the $j$-th exit; the permutation
defining which entrance port is connected to which exit port thus
trivially reads $P_{\rm enc}^{\gamma}=\left({1\atop 1}{2\atop 2}{\ldots\atop
    \ldots}{l\atop l}\right)$.  A partner orbit $\gamma'$ differing from $\gamma$
in the said encounter has the ports differently connected: The $j$-th
encounter stretch connects the $j$-th entrance with a different exit $
P_{\rm enc}(j)$.  This reconnection can be expressed in terms of a
different permutation $ P_{\rm enc}= \left({1\atop P_{\rm
      enc}(1)}{2\atop P_{\rm enc}(2)} {\ldots\atop \ldots}{l\atop P_{\rm
      enc}(l)}\right)$; e.g.  reconnections as in
Fig.~\ref{fig:permutations}a are described by the permutation $ P_{\rm
  enc}=\left({1\atop 2}{2\atop 6}{3\atop 4}{4\atop 5}{5\atop 1}{6\atop
    3}\right)$. Note that we refrain from indexing the latter
permutation by a superscript $\gamma'$.

A permutation $ P_{\rm enc}$ accounting for a single $l$-encounter is
a single cycle of length $l$, e.g. $(1,2,6,3,4,5)$ in the above
example.  If it were multiple-cycle, the encounter would effectively
fall into several disjoint encounters.  For example, Fig.
\ref{fig:permutations}b visualizes a permutation with three cycles
$(1,2)$, $(3,4)$, and $(5,6)$.  As already mentioned, reconnections
only take place between stretches 1 and 2, stretches 3 and 4, and
stretches 5 and 6, which thus have to be considered as three
independent encounters.

If $\gamma$ and $\gamma'$ differ in several encounters, the connections
between entrance and exit ports are reshuffled separately within these
encounters. The corresponding permutation $ P_{\rm enc}$ then has
precisely one $l$-cycle corresponding to each of the $v_l$
$l$-encounters, for all $l\geq 2$, the total number of permuted objects
being $L=\sum_{l\geq 2}lv_l$.

We also have to account for the orbit loops.  The $a$-th loop connects
the exit of the $(a-1)$-st encounter stretch with the entrance of the
$a$-th one. These connections can be associated with the permutation $
P_{\rm loop}=\left({1\atop 2}{2\atop 3}{\ldots\atop \ldots} {L\atop 1}\right)=
(1,2,\ldots,L)$ which obviously is single-cycle.  The order in which
entrance ports (and thus loops) are traversed in $\gamma$ is then given by
the product $ P^\gamma= P_{\rm loop} P_{\rm enc}^\gamma= P_{\rm loop}$.  This
product is single-cycle - as it should be, because $\gamma$ is a periodic
orbit and hence returns to the first entrance port only after
traversing all others.

Similarly, the sequence of entrance ports (or, equivalently, loops)
traversed by $\gamma'$ is represented by
\begin{equation}
\label{porb}
 P= P_{\rm loop} P_{\rm enc}.
\end{equation}
with the same $P_{\rm loop}$ as above.  We must demand $ P$ to be
single-cycle for $\gamma'$ not to decompose to a pseudo-orbit.

We shall denote by ${\cal M}(\vec{v})$ the set of permutations $
P_{\rm enc}$ (representing intra-encounter connections) which have
$v_l$ $l$-cycles, for each $l\geq 2$, and upon multiplication with $
P_{\rm loop}$ yield single-cycle permutations (\ref{porb}).  These
permutations $ P_{\rm enc}$ are in one-to-one correspondence to the
structures of orbit pairs defined in \ref{sec:structures}, i.e.
determine how the encounter stretches are ordered, and how they are
reconnected to form a partner orbit.  The number of elements of ${\cal
  M}(\vec{v})$ is thus precisely the number $N(\vec{v})$ of structures
related to $\vec{v}$.

\subsubsection{Examples}

The numbers $N(\vec{v})$ can be determined numerically, by generating
all possible permutations $ P_{\rm enc}$ with $v_l$ $l$-cycles and
counting only those for which $ P= P_{\rm loop} P_{\rm enc}$ is
single-cycle. The $ P_{\rm enc}$'s contributing to the orders $n=3$
and $5$ of the spectral form factor are shown in Table
\ref{tab:unitary}.

\begin{table}
\begin{center}
\begin{tabular}{|c|c|c|c|r|r|r|} \hline
order & $\vec{v}$ & $L$ & $V$ & $N(\vec{v})$ & $\tilde{N}(\vec{v})$ & contribution \\
\hline\hline
$\tau^3$
& $(2)^2$ &4 & 2 & 1\;\; & 1\; & $1\tau^3$\;\;\;\;\; \\ \hhline{~------}
& $(3)^1$ &3 & 1 & 1\;\; & $-1$\; & $-1\tau^3$\;\;\;\;\; \\ \hhline{~------}
\hhline{~------}
&&&&&0\;&0$\tau^3$\;\;\;\;\; \\
\hline\hline
$\tau^5$
& $(2)^4$ &8 & 4 & 21\;\; & 42\; & $7\tau^5$\;\;\;\;\; \\ \hhline{~------}
& $(2)^2(3)^1$ &7 & 3 & 49\;\; & $-84$\; & $-14\tau^5$\;\;\;\;\; \\ \hhline{~------}
& $(2)^1(4)^1$ &6 & 2 & 24\;\; & 32\; & $\frac{16}{3}\tau^5$\;\;\;\;\; \\ \hhline{~------}
& $(3)^2$ &6 & 2 & 12\;\; & 18\; & $3\tau^5$\;\;\;\;\; \\ \hhline{~------}
& $(5)^1$ &5 & 1 & 8\;\; & $-8$\; & $-\frac{4}{3}\tau^5$\;\;\;\;\; \\ \hhline{~------}
\hhline{~------}
&&&&&0\;&0$\tau^5$\;\;\;\;\; \\
\hline
\end{tabular}
\end{center}
\caption{Permutations, and thus families of orbit pairs, giving rise to orders
$\tau^3$ and $\tau^5$ of the form factor, for systems
without ${\cal T}$-invariance.
We represent $\vec{v}$ by $(2)^{v_2}(3)^{v_3}\ldots$. The order of each
contribution is given by $n=L-V+1$.
We see that contributions add up to zero for odd $n$,
whereas there are no permutations for even $n$.}
\label{tab:unitary}
\end{table}

\begin{figure}
\begin{center}
  \includegraphics[scale=0.2]{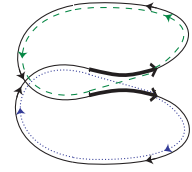}
\end{center}
\caption{Reconnections inside a parallel 2-encounter yield
  a pseudo-orbit decomposing into two separate periodic
  orbits.}\label{fig:2decompose}
\end{figure}

Interestingly, no qualifying $ P_{\rm enc}$'s exist for even
$L-V-1=n$. For example, the only candidate for $n=2$ would be $ P_{\rm
  enc}=\left({1\atop 2}{2\atop 1}\right)$, describing reconnections
inside an encounter of two parallel orbit stretches $\psr$. However,
the corresponding partner decomposes into two separate periodic orbits
(corresponding to the cycles $(1)$ and $(2)$ of $ P= P_{\rm loop}
P_{\rm enc}=\left({1\atop 1}{2\atop 2}\right)$), see Fig.
\ref{fig:2decompose}. The same happens for all other permutations with
$n$ even.\footnote{The proof is based on the parities of the
  permutations involved. A permutation is said to have parity 1 if it
  can be written as a product of an even number of transpositions, and
  to be of parity $-1$ if it is a product of an odd number of
  transpositions. Parity is given by $(-1)^{L-V}$, where $L$ is the
  number of permuted elements and $V$ the number of cycles, and the
  parity of a product of permutations equals the product of parities
  of the factors.  Since $ P$ and $ P_{\rm loop}$ both consist of one
  single cycle, they are of the same parity.  Therefore, $ P= P_{\rm
    loop} P_{\rm enc}$ implies that all allowed $ P_{\rm enc}$ need to
  have parity 1, i.e.  $n=L-V+1$ must be odd.}

\begin{figure}
\begin{center}
  \includegraphics[scale=0.2]{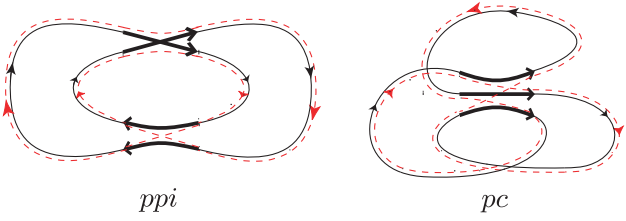}
\end{center}
\caption{Two families of pairs of orbits differing in parallel encounters;
  both exist for systems with and without ${\cal T}$-invariance; each
  contributes to $\tau^3$, but the contributions mutually cancel. For
  labels see text.}\label{fig:tau3parallel}
\end{figure}

For $n$ odd, the individual numbers $N(\vec{v})$ and
$\tilde{N}(\vec{v})$ do not vanish. However, we see in Table
\ref{tab:unitary} that the $\tilde{N}(\vec{v})$ related to the same
$n$ sum up to zero. That remarkable cancellation, a non-trivial
property of the permutation group to be discussed below, is the reason
why all off-diagonal contributions to the spectral form factor vanish
in the unitary case.  For example, the $\tau^3$ term is determined by $
P_{\rm enc}=\left({1\atop 3}{2\atop 4}{3\atop 1}{4\atop 2}\right)$
describing reconnections inside two 2-encounters of parallel stretches
$\psr$ (case {\it ppi} in \cite{Tau3}), and $ P_{\rm
  enc}=\left({1\atop 2}{2\atop 3}{3\atop 1}\right)$ describing
reconnections inside a 3-encounter of parallel orbit stretches $\pc$
(case {\it pc} in \cite{Tau3}). The respective contributions, $\tau^3$
and $-\tau^3$, mutually cancel; Fig.~\ref{fig:tau3parallel} displays the
orbits.

\subsubsection{Recursion relation for $N(\vec{v})$}

\label{sec:unitary_recursion}

We now derive a recursion formula for $N(\vec{v})$, imagining one loop
(e.g. the one with index $L$) of an orbit removed and studying the
consequences on the encounters. We shall reason with permutations but
the translation rule {\it cycle} $\to$ {\it encounter} yields an
interpretation for orbits. Readers wanting to skip the reasoning may
jump to the result (\ref{rec2_unit}).

As a preparation, let us introduce a subset ${\cal M}(\vec{v},l)$ of
${\cal M}(\vec{v})$ such that the largest of the permuted numbers,
i.e.  $L(\vec{v})=\sum_k kv_k$ belongs to a cycle of length $l$ (it is
assumed that $v_l>0$).  The full set can be obtained by applying to
this subset all $L(\vec{v})$ possible cyclic permutations.  In fact,
we thus get the set ${\cal M}(\vec{v})$ in $l v_l$ copies, since
cyclic permutations shifting the element $L(\vec{v})$ inside an
$l$-cycle or between different $l$-cycles with the same $l$ leave the
subset ${\cal M}(\vec{v},l)$ unchanged. Consequently, the sizes of
${\cal M}(\vec{v})$ and ${\cal M}(\vec{v},l)$ are related as
\begin{equation} \label{mengetomenge} N(\vec{v},l) = \frac{l
    v_l}{L(\vec{v})} N(\vec{v})\,.
\end{equation}

We need a mapping that leads from a given permutation $ P_{\rm enc}$
to a permutation $ Q_{\rm enc}$ of smaller size, with a different
cycle structure.  Recall that any permutation $ P_{\rm enc}\in{\cal
  M}(\vec{v},l)$ may be written as $ P_{\rm enc}= P_{\rm loop}^{-1}
P$, with $ P_{\rm loop}$ the single-cycle permutation $(1,2,\ldots,L)$,
and $P$ single-cycle as well.  Now suppose that the element $L$
(corresponding to the entrance port following the $L$-th orbit loop)
is deleted from the cycle representations of both $ P_{\rm loop}$ and
$ P$.  We thus obtain two new single-cycle permutations $ Q_{\rm
  loop}$ and $ Q$, acting on the numbers $1,2\ldots,L-1$. Here, $ Q_{\rm
  loop}$ is simply given by $ Q_{\rm loop}=(1,2,\ldots,L-1)$, and $ Q$
differs from $ P$ only by mapping the predecessor of $L$, i.e. $
P^{-1}(L)$, to the successor of $L$, i.e.  $ P(L)$.  Let us now define
the new ``encounter" permutation $ Q_{\rm enc} $ in analogy to
(\ref{porb}),
\begin{equation} \label{qorb}
   Q_{\rm enc}= Q_{\rm loop}^{-1}  Q \,.
\end{equation}
The $ Q_{\rm enc}$ thus obtained acts on the elements $a=1,2,\ldots,L-1$
as
\begin{equation}
\label{QPE_unitary}
 Q_{\rm enc}(a)=\begin{cases}
   P_{\rm enc}(L)&\text{if $a= P_{\rm enc}^{-1}(L-1)$}\\
  L-1&\text{if $a= P_{\rm enc}^{-1}(L)$}\\
   P_{\rm enc}(a)&\text{otherwise}\;.
\end{cases}
\end{equation}
To verify this, recall that $ Q_{\rm loop}$ differs from $ P_{\rm
  loop}$ only in the mapping of one number, same as for $ Q$ and $ P$.
Thus $ Q_{\rm enc}$ acts like $ P_{\rm enc}$ on all but two numbers
$a$.  These exceptional cases, given in the first two lines of
(\ref{QPE_unitary}), are checked by carefully applying the above
definitions of $ Q_{\rm loop}$ and $ Q$ as follows
\begin{eqnarray}
   Q_{\rm enc} P_{\rm enc}^{-1}(L-1)&=& Q_{\rm loop}^{-1} Q P^{-1} P_{\rm loop}(L-1)\nonumber\\&=&
 Q_{\rm loop}^{-1} Q P^{-1}(L)
  = Q_{\rm loop}^{-1} P(L)\nonumber\\
  &{\stackrel{(*)}{=}}& P_{\rm loop}^{-1} P(L)= P_{\rm enc}(L)\nonumber\\
   Q_{\rm enc} P_{\rm enc}^{-1}(L)&=& Q_{\rm loop}^{-1} Q P^{-1} P_{\rm loop}(L)\nonumber\\
  &=& Q_{\rm loop}^{-1} Q P^{-1}(1)
{\stackrel{(**)}{=}} Q_{\rm loop}^{-1} P P^{-1}(1)\nonumber\\
&=& Q_{\rm loop}^{-1}(1)=L-1 \,;
\end{eqnarray}
here, we used $ P(L)\neq 1$ for $(*)$, since otherwise $ P_{\rm enc}$
would have a 1-cycle (i.e.  $ P_{\rm enc}(L)= P_{\rm loop}^{-1} P(L)=
P_{\rm loop}^{-1}1=L$), and $ P(L)\neq L$, since otherwise $ P$ would
have a 1-cycle. To check $(**)$, we need $ P^{-1}(1)\neq L$ (since $
P(L)\neq 1$) and $ P^{-1}(1)\neq P^{-1}(L)$.

We need to connect the cycle structures of $ Q_{\rm enc} $ and $
P_{\rm enc}$.  Let us first consider the case that \textbf{the element
  $L-1$ of the permutation $ P_{\rm enc}$ belongs to a different cycle
  than $L$}, say a $k$-cycle. Hence, $ P_{\rm enc}$ has the form
\begin{equation}
  \label{pcyc}  P_{\rm enc}=[\ldots] (L-1,a_2,a_3,\ldots a_k)
  (L,b_2,b_3,\ldots
  b_l) \end{equation}
where the two aforementioned cycles are written in round brackets, and
$[\ldots]$ stands for all other cycles.  Then $Q_{\rm enc}$ differs from $
P_{\rm enc}$ by mapping $ P_{\rm enc}^{-1}(L-1)=a_k$ to $ P_{\rm
  enc}(L)=b_2$, and $ P_{\rm enc}^{-1}(L)=b_l$ to $L-1$. It follows
that the $k$- and $l$-cycles of $ P_{\rm enc}$ merge to a
$(k+l-1)$-cycle of
\begin{equation} \label{qcyc}  Q_{\rm enc} =[\ldots]
  (L-1,a_2,a_3,\ldots a_k,b_2,b_3,\ldots b_l)
\end{equation}
where $[\ldots]$ is the same as in (\ref{pcyc}).  Compared to $ P_{\rm
  enc}$, $ Q_{\rm enc}$ has one $k$-cycle and one $l$-cycle less, but
one additional $(k+l-1)$-cycle.  The changed cycle structure with
$v_k\to v_k-1,v_l\to v_l-1,v_{k+l-1}\to v_{k+l-1}+1$ will be denoted as
$\vec{v}^{\;[k,l\to k+l-1]}$.  In general, $\vec{v}^{[\alpha_1,\ldots,\alpha_m \to
  \beta_1,\ldots,\beta_n]}$, $m\geq0$, $n\geq0$ denotes the vector obtained from
$\vec{v}$ if we decrease all $v_{\alpha_i}$ by one, increase all
$v_{\beta_i}$ by one, and leave all other components unchanged; if no
$\beta_i$ appear on the r.~h.~s., no components of $\vec{v}$ are
increased.

The permutation $ Q_{\rm enc}$ thus belongs to the subset ${\cal
  M}(\vec{v}^{[k,l\to k+l-1]},k+l-1)$ since the largest permuted number
$L-1$ belongs to a cycle with the length $k+l-1$.  Each $ P_{\rm enc}$
with the structure (\ref{pcyc}) ($k$ and $l$ fixed) generates one
member of this subset.  Vice versa, for fixed $k$ the $ Q_{\rm enc}$
given in (\ref{qcyc}) uniquely determines one $ P_{\rm enc}$ as given
in (\ref{pcyc}).  Hence, there are
\begin{equation}
\label{no_1}
N(\vec{v}^{[k,l\to k+l-1]},k+l-1)\,
\end{equation}
members of ${\cal M}(\vec{v},l)$ structured like (\ref{pcyc}).
Physically, the present scenario corresponds to the merger of a $k$-
and an $l$-encounter into a $(k+l-1)$-encounter, by shrinking away an
intervening loop.

We now turn to the second scenario where \textbf{$L$ and $L-1$ belong
  to the same $l$-cycle of $ P_{\rm enc}$}.  If $L$ follows $L-1$
after $m$ iterations (i.e. $L= P_{\rm enc}^m(L-1)$, $1\leq m\leq
l-2$)\footnote{ Note $m=l-1$ is excluded: otherwise $ P$ would have a
  1-cycle due to $L= P_{\rm enc}^{l-1}(L-1)= P_{\rm enc}^{-1}(L-1)=
  P^{-1} P_{\rm loop}(L-1)= P^{-1}(L)$.}, the permutation $ P_{\rm
  enc}$ is of the form
\begin{equation}
  \label{pmcyc}  P_{\rm enc}=[\ldots] (L-1,a_2,a_3,\ldots a_{m},L,a_{m+2},
\ldots, a_l)\,.
\end{equation}
According to (\ref{QPE_unitary}), $ Q_{\rm enc}$ differs from $ P_{\rm
  enc}$ by mapping $ P_{\rm enc}^{-1}(L-1)=a_l$ to $ P_{\rm
  enc}=a_{m+2}$ and mapping $ P_{\rm enc}^{-1}(L)=a_{m}$ to $L-1$; $
Q_{\rm enc}$ thus reads
\begin{equation}
  \label{breakup}  Q_{\rm enc} =[\ldots] (L-1,a_2,a_3,\ldots
  a_{m})(a_{m+2},\ldots, a_l)\,; \end{equation}
the $l$-cycle of $ P_{\rm enc}$ is broken up into 2 cycles, with the
lengths $m$ and $l-m-1$. Since the largest number $L-1$ is included in
an $m$-cycle, $ Q_{\rm enc}$ belongs to ${\cal M}(\vec{v}^{[l\to
  m,l-m-1]},m)$.

In contrast to the first scenario, there are typically several $P_{\rm
  enc}$ producing the same $ Q_{\rm enc}$.  Indeed (\ref{breakup})
would not only result from (\ref{pmcyc}), but also from all $l-m-1$
permutations $P_{\rm enc}$ obtained by cyclic permutation of the last
elements $a_{m+2},\ldots,a_l$ in (\ref{pmcyc}).  Besides, $[\ldots]$ in
$P_{\rm enc}$ contains $v_{l-m-1}$ cycles of length $l-m-1$. If we
transpose the content of one of these cycles with the subsequence
$a_{m+2},\ldots,a_l$ in (\ref{pmcyc}), the resulting $P_{\rm enc}$ will
lead to the same $Q_{\rm enc}$.  Thus, for each $m$, the subset of
elements $ P_{\rm enc}\in{\cal M}(\vec{v},l)$ structured like
(\ref{pmcyc}) is $(l-m-1)(v_{l-m-1}+1)$ times larger than ${\cal
  M}(\vec{v}^{[l\to m,l-m-1]},m)$, i.e.  it has the size
\begin{equation}
  (l-m-1)(v_{l-m-1}+1)N(\vec{v}^{[l\to m,l-m-1]},m) \,.
\end{equation}

We have now decomposed ${\cal M}(\vec{v},l)$ into several subsets of
size $N(\vec{v}^{[k,l\to k+l-1]},k+l-1)$, $k\geq 2$, and further subsets
of size $(l-m-1)(v_{l-m-1}+1)N(\vec{v}^{[l\to m,l-m-1]},m)$,
$m=1,\ldots,l-1$.  The size of ${\cal M}(\vec{v},l)$ thus reads
\begin{eqnarray}
\label{NL} &&N(\vec{v},l)=\sum_{k\geq
    2}N(\vec{v}^{[k,l\to k+l-1]},k+l-1)\\
    &&+\sum_{m=1}^{l-2}(l-m-1)(v_{l-m-1}+1)N(\vec{v}^{[l\to m,l-m-1]},m)\,.\nonumber
\end{eqnarray}
We may rewrite the latter equation using $N(\vec{v},l) = \frac{l
  v_l}{L} N(\vec{v})$ and
$\tilde{N}(\vec{v})=N(\vec{v})\frac{(-1)^V\prod_l l^{v_l}}{L(\vec{v})}$,
to get
\begin{eqnarray}
\label{recur1}
&&v_l \tilde{N} (\vec{v})+ \sum_{k\geq 2}(v_{k+l-1}+1)k \tilde{N}(
\vec{v}^{[k,l\to k+l-1]})\nonumber\\
&& + \sum_{1\leq m\leq l-2}(v_{l-m-1} +1)v_{m}^{[l\to
  m,l-m-1]} \tilde{N}(\vec{v}^{[l\to m,l-m-1]})\nonumber\\
&&=0 \,;
\end{eqnarray}
Eqs.  (\ref{NL},\ref{recur1}) are the general recursion relations in
search.  (Note that $v_{k+l-1}+1=v_{k+l-1}^{[k,l\to k+l-1]}$.  Of
course, the $k$th summand vanishes if there are no $k$-cycles present,
i.e. if $v_k=0$ and thus formally $v_k^{[k,l\to k+l-1]}=-1$).

To determine the form factor for systems without time-reversal
invariance, we only need the special case $l=2$. In this case, our
recursion strongly simplifies,
\begin{equation}\label{rec2_unit}
  v_2\tilde{N}(\vec{v})+\sum_{k\geq 2}v_{k+1}^{[k,2\to k+1]}k \tilde{N}(\vec{v}^{[k,2\to k+1]})=0\,,
\end{equation}
since only the first of the two above scenarios is possible.  That is,
a 2-cycle may only merge with a $k$-cycle to form a $(k+1)$-cycle, but
not split into two separate cycles.  Recall that $\vec{v}^{[k,2\to
  k+1]}$ is obtained from $\vec{v}$ by decreasing both $v_k$ and $v_2$
by one, and increasing $v_{k+1}$ by one.

\subsubsection{Spectral form factor}

\label{sec:formfactor_unitary}

We had expressed the Taylor coefficients of the form factor as a sum
over the combinatorial numbers $\tilde{N}(\vec{v})$,
\begin{equation} K_n=\frac{1}{(n-2)!}\sum_{\vec{v}}^{\nu(\vec{v})=n}\tilde{N}(\vec{v})\,,\quad n\geq2\,,
\end{equation}
see (\ref{kn}), where the sum runs over all $\vec{v}$ with $v_1=0$
which fulfill $\nu(\vec{v})\equiv L(\vec{v})-V(\vec{v})+1=n$. Our recursion
relation for $\tilde{N}(\vec{v})$ now translates into one for $K_n$,
albeit a trivial one in the unitary case, implying that all $K_n$
except $K_1$ vanish.
(Alternatively, one may use a rather involved explicit formula for
$N(\vec{v})$ \cite{JMueller}.)

To show this, consider the recursion (\ref{rec2_unit}) for
$\tilde{N}(\vec{v})$ and sum over $\vec{v}$ as above
\begin{equation}
\label{gue_sum}
\sum_{\vec{v}}^{\nu(\vec{v})=n}\left(v_2\tilde{N}(\vec{v})+\sum_{k\geq 2}v_{k+1}^{[k,2\to
    k+1]}k\tilde{N}(\vec{v}^{[k,2\to k+1]})\right)=0\,.
\end{equation}
Each of the sums $\sum_{\vec{v}}^{\nu(\vec{v})=n}v_{k+1}^{[k,2\to
  k+1]}\tilde{N}(\vec{v}^{[k,2\to k+1]})$ may be transformed into a sum
over the argument of $\tilde{N}$, i.e.  $\vec{v}'=\vec{v}^{[k,2\to
  k+1]}$. Due to $\nu(\vec{v})=n$, we also have $\nu(\vec{v}')=n$, since
going from $\vec{v}$ to $\vec{v}'$ decreases both $L$ and $V$ by one
and thus leaves $\nu$ invariant. Given that by construction we must
have $v'_{k+1}\geq 1$, the sum over $\vec{v}'$ extends over all
$\vec{v}'$ with $\nu(\vec{v}')=n$ and $v'_{k+1}\geq 1$.  However, the
latter restriction may be dropped because due to the prefactor
$v'_{k+1}$ terms with $v'_{k+1}=0$ do not contribute.  Consequently,
the sum may be simplified as
\begin{equation} 
\label{rule_special} 
\sum_{\vec{v}}^{\nu(\vec{v})=n}v_{k+1}^{[k,2\to 
  k+1]}\tilde{N}(\vec{v}^{[k,2\to k+1]})=\sum_{\vec{v}'}^{\nu(\vec{v}')=n}v'_{k+1}\tilde{N}(\vec{v}')\,.
\end{equation} 
Applying this rule to all terms in (\ref{gue_sum}) and dropping the
primes, we obtain
\begin{equation}
\sum_{\vec{v}}^{\nu(\vec{v})=n}\Big(v_2+\sum_{k\geq2}v_{k+1}k\Big)\tilde{N}(\vec{v})=0\,. 
\end{equation} 
Since the term in brackets is just $\sum_{l\geq 2}v_l(l-1)=L-V=n-1$ we
have
\begin{equation} (n-1)\sum_{\vec{v}}^{\nu(\vec{v})=n}\tilde{N}(\vec{v})= (n-1)!\,K_n=0\,,\quad 
  n\geq2 \, . 
\end{equation} 

We see that all Taylor coefficients except $K_1$ vanish: orbit pairs
differing in encounters yield no net contribution; the diagonal
approximation exhausts the small-time form factor in full. GUE
behavior is thus ascertained.
 
\subsection{Orthogonal case}

\label{combinatorics_orthogonal} 

\subsubsection{Structures and permutations}

In $\cal T$-invariant systems the partners of an orbit $\gamma$ may
involve loops of both $\gamma$ and its time-reversed $\overline\gamma$. To
capture all partners $\gamma'$ of $\gamma$ in terms of permutations the
permuted objects must refer to both $\gamma$ and $\overline{\gamma}$ and thus
be doubled in number compared to the unitary case. Each permutation
will describe simultaneously both $\gamma'$ and $\bar \gamma'$.

\begin{figure}
\begin{center}
  \includegraphics[scale=0.31]{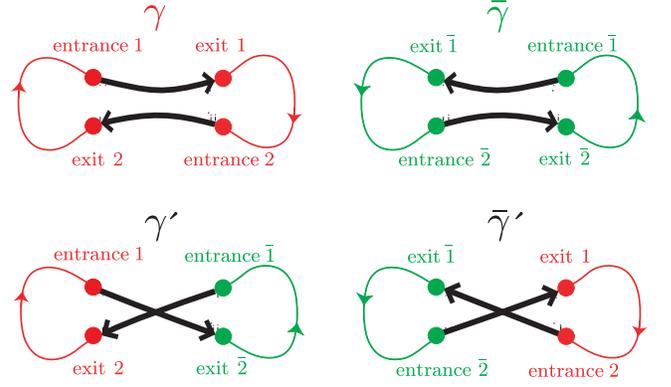}
\end{center}
\caption{Entrance-to-exit connections for a
  Sieber/Richter pair, in an orbit $\gamma$, its time-reversed $\bar{\gamma}$,
  and the partners $\gamma',\bar{\gamma'}$.}
\label{fig:2notation}
\end{figure}

We number the entrance and exit ports of self-encounters of $\gamma$ in
their order of traversal, $1,2,\ldots,L$, such that the $a$-th encounter
stretch leads from the $a$-th entrance to the $a$-th exit port; see
Fig. \ref{fig:2notation} for the example of a 2-encounter.  The
time-reversed orbit $\overline{\gamma}$ passes the same ports as $\gamma$, but
with opposite sense and entrance and exit swapped. The ports of
$\overline{\gamma}$ are labelled by $\overline{1},\overline{2},\ldots,
\overline{L}$, such that the exit port $\overline{a}$ of
$\overline{\gamma}$ is the time-reversed of the entrance port $a$ of $\gamma$,
and entrance port $\overline{a}$ of $\overline{\gamma}$ is the
time-reversed of the exit port $a$ of $\gamma$, again compare Fig.
\ref{fig:2notation}.  Consequently, inside $\overline{\gamma}$ stretch
$\overline{a}$ leads from entrance $\overline{a}$ to exit
$\overline{a}$.
 
The intra-encounter connections of $\gamma$ and $\overline{\gamma}$ are
represented by the trivial permutation $P_{\rm enc}^\gamma=\left({1\atop
    1}{2\atop 2}{\ldots\atop \ldots}{L\atop L}{\overline{1}\atop
    \overline{1}}{\overline{2}\atop \overline{2}}{\ldots\atop
    \ldots}{\overline{L} \atop \overline{L}}\right)$ which maps each
entrance (upper line) to the following exit (lower line).  The loops
are associated with $P_{\rm loop}=\left({1\atop 2}{2\atop 3} {\ldots\atop
    \ldots}{L\atop 1}{\overline{2}\atop \overline{1}}{\overline{3} \atop
    \overline{2}}{\ldots\atop \ldots}{\overline{1}\atop \overline{L}}\right)$,
since if one loop of $\gamma$ leads from the exit of the $(a-1)$-st
stretch to the entrance of the $a$-th one, its time-reversed must go
from exit $\overline{a}$ to entrance $\overline{a-1}$.  Finally,
$P^\gamma=P_{\rm loop} P_{\rm enc}^\gamma=P_{\rm loop}$ specifies the ordering
of entrance ports along $\gamma$ and $\overline{\gamma}$. That $P^\gamma$ has two
cycles $(1,2,\ldots,L)$ and
$(\overline{L},\overline{L-1},\ldots,\overline{1})$, one each for $\gamma$ and
$\overline{\gamma}$.
 
The reconnections leading to $\gamma'$, $\overline{\gamma'}$ are described by
a permutation $P_{\rm enc}$ determining the exit port connected to
each entrance.  In the example of a Sieber/Richter pair, Fig.
\ref{fig:2notation}, the partner $\gamma'$ connects the entrance 1 of $\gamma$
to the exit $\bar{2}$ of $\bar{\gamma}$, and the entrance $\bar{1}$ of
$\bar{\gamma}$ to the exit 2 of $\gamma$. Including the connections in the
time-reversed partner $\bar{\gamma'}$, we write $P_{\rm enc}=\left({1\atop
    \overline{2}}{2\atop \overline{1}}{\overline{1}\atop 2}
  {\overline{2}\atop 1}\right)$.  Note that the sequence of columns in
$P_{\rm enc}$ may be ordered arbitrarily.  We shall mostly order such
that the first lines in $P_{\rm enc}$ and $P_{\rm enc}^\gamma$ coincide;
columns describing $\gamma'$ and $\overline{\gamma'}$ may thus become mutually
interspersed.

$\cal T$-invariance imposes a restriction on $P_{\rm enc}$: If a
stretch connects entrance $a$ to exit $b$, the time-reversed stretch
must connect the entrance $\overline{b}$ of $\overline{\gamma}$ to the
exit $\overline{a}$. Thus, if $P_{\rm enc}$ maps $a$ to $b$, it has to
map $\overline{b}$ to $\overline{a}$, with $a,b$ standing for elements
out of $1,\ldots, L,\bar 1,\ldots,\bar L$ (we define
$\overline{\overline{a}}=a$).  This restriction on $P_{\rm enc}$ will
be referred to as ``$\cal T$-covariance".
 
It follows that if $(a_1,a_2,\ldots,a_{l-1},a_l)$ is a cycle of $P_{\rm
  enc}$, then so is its ``time-reversed" ,
$(\overline{a_l},\overline{a_{l-1}},\ldots,
\overline{a_2},\overline{a_1})$. These two cycles may not be
identical. Indeed a cycle coinciding with its time-reversed would have
the form $\left(a_1,\ldots,a_k,\overline{a_k},\ldots,\overline{a_1}\right),\;
k=l/2$; such cycles are not allowed, since the entrance port $a_k$ and
the exit port $\bar a_k$ coincide in configuration space and thus may
not be connected by an encounter stretch.
 
We see that given time reversal there must be a pair of twin
$l$-cycles of $P_{\rm enc}$ associated with each $l$-encounter; an
encounter associated with more than one pair of cycles would decompose
into several independent ones. In general, each cycle in a pair
describes stretches both of $\gamma'$ and of $\bar\gamma'$; only in case of
all stretches nearly parallel one cycle refers exclusively to $\gamma'$
and the other to $\overline{\gamma'}$.

The final restriction on the permutations $P_{\rm enc}$ is analogous
to the one encountered in the unitary case.  To obtain two connected
partner orbits $\gamma'$ and $\overline{\gamma'}$, we now have to demand the
permutation $P=P_{\rm loop} P_{\rm enc}$ to consist of only two
$L$-cycles, listing the entrance ports in $\gamma'$ and $\overline{\gamma'}$,
respectively.  (The second cycle can also be interpreted as the list
of exit ports of $\gamma'$, time reversed and written in reverse order.
Since an entrance $a_i$ is connected by a loop to the exit $b_i\equiv
P_{\rm loop}^{-1}(a_i)$, the two cycles of $P$ read $(a_1,a_2,\ldots,a_L)$
and $(\overline{b_L},\overline{b_{L-1}},\ldots, \overline{b_1})$.)
 
To summarize, we consider the set ${\cal M}(\vec{v})$ of permutations
$P_{\rm enc}$ acting on
$1,2,\ldots,L,\overline{1},\overline{2},\ldots,\overline{L}$ which (i) are
time-reversal covariant, (ii) have $v_l$ pairs of $l$-cycles for all
$l\geq2$, and (iii) lead to a permutation $P=P_{\rm loop} P_{\rm enc}$
consisting of two cycles as above.  Each element $P_{\rm enc}$ of the
set ${\cal M}(\vec{v})$ thus stands for one of the ``structures''
introduced in \ref{sec:structures}.  We need to calculate the number
$N(\vec{v})$ of elements of ${\cal M}(\vec{v})$.
 
\subsubsection{Examples}

\begin{table}
\begin{center}
\begin{tabular}{|c|c|c|c|r|r|r|} \hline
order & $\vec{v}$ & $L$ & $V$ & $N(\vec{v})$ & $\tilde{N}(\vec{v})$ & contribution \\
\hline\hline
$\tau^2$
& $(2)^1$ &2 & 1 & 1 & $-1$ & $-2\tau^2$\;\;\;\; \\ \hhline{~------}
\hhline{~------}
&&&&&$-1$&$-2\tau^2$\;\;\;\; \\
\hline\hline
$\tau^3$
& $(2)^2$ &4 & 2 & 5 & 5 & $10\tau^3$\;\;\;\; \\ \hhline{~------}
& $(3)^1$ &3 & 1 & 4 & $-4$ & $-8\tau^3$\;\;\;\; \\ \hhline{~------}
\hhline{~------}
&&&&&1&2$\tau^3$\;\;\;\; \\
\hline\hline
$\tau^4$
& $(2)^3$ &6 & 3 & 41 & $-\frac{164}{3}$ & $-\frac{164}{3}\tau^4$\;\;\;\; \\ \hhline{~------}
& $(2)^1(3)^1$ &5 & 2 & 60 & 72 & $72\tau^4$\;\;\;\; \\ \hhline{~------}
& $(4)^1$ &4 & 1 & 20 & $-20$ & $-20\tau^4$\;\;\;\; \\ \hhline{~------}
\hhline{~------}
&&&&&$-\frac{8}{3}$&$-\frac{8}{3}\tau^4$\;\;\;\; \\
\hline\hline
$\tau^5$
& $(2)^4$ &8 & 4 & 509 & 1018 & $\frac{1018}{3}\tau^5$\;\;\;\; \\ \hhline{~------}
& $(2)^2(3)^1$ &7 & 3 & 1092 & $-1872$ & $-624\tau^5$\;\;\;\; \\ \hhline{~------}
& $(2)^1(4)^1$ &6 & 2 & 504 & 672 & $224\tau^5$\;\;\;\; \\ \hhline{~------}
& $(3)^2$ &6 & 2 & 228 & 342 & $114\tau^5$\;\;\;\; \\ \hhline{~------}
& $(5)^1$ &5 & 1 & 148 & $-148$ & $-\frac{148}{3}\tau^5$\;\;\;\; \\ \hhline{~------}
\hhline{~------}
&&&&&12&4$\tau^5$\;\;\;\; \\
\hline
\end{tabular}
\end{center}
\caption{Permutations, and thus families of orbit pairs, giving rise to orders
$\tau^2$ to $\tau^5$ of the form factor, for systems
with ${\cal T}$-invariance; notation as in Table \ref{tab:unitary}.
The results coincide with the predictions of RMT for the GOE.}
\label{tab:orthogonal}
\end{table}

Again, the numbers $N(\vec{v})$ can be determined by numerically
counting permutations. From the results shown in Table
\ref{tab:orthogonal}, we see that indeed the form factor of the
Gaussian Orthogonal Ensemble is reproduced semiclassically.

The $\tau^2$ contribution comes from pairs of orbits differing in one
antiparallel 2-encounter \cite{SR}.  We have already shown that the
corresponding ``encounter permutation" reads $P_{\rm
  enc}=\left({1\atop \overline{2}}{2\atop
    \overline{1}}{\overline{1}\atop 2} {\overline{2}\atop 1}\right)$.

\begin{figure}
\begin{center}
  \includegraphics[scale=0.19]{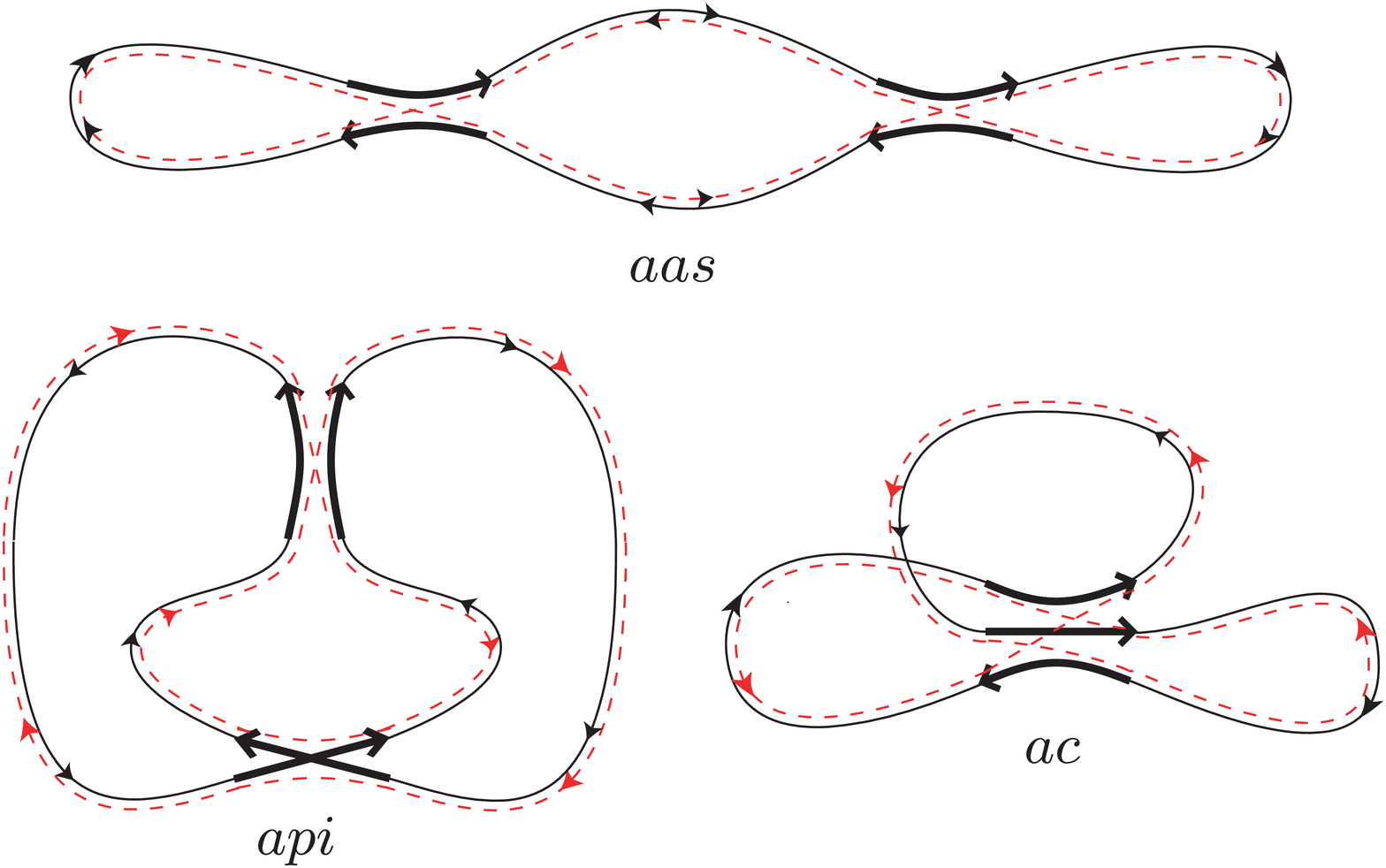}
\end{center}
\caption{Pairs of orbits  existing only
  for systems with ${\cal T}$-invariance and giving all of $\tau^3$.
  For labels see text. Two further families do not require ${\cal
    T}$-invariance, see Fig.  \ref{fig:tau3parallel}.
}\label{fig:tau3antiparallel}
\end{figure}

The $\tau^3$ contribution originates from (compare Figs.
\ref{fig:tau3parallel}, \ref{fig:tau3antiparallel} and Ref.
\cite{Tau3}) four permutations related to 3-encounters and five
permutations related to pairs of 2-encounters.  The permutation
$P_{\rm enc}=\left({1\atop 2}{2\atop 3}{3\atop 1}{\overline{1}\atop
    \overline{3}}{\overline{2}\atop
    \overline{1}}{\overline{3}\atop\overline{2}}\right)$ describes
encounters of three parallel orbit stretches $\pc$ (case {\it pc}).
For triple encounters in which one of the stretches is time-reversed
with respect to the other two $\ac$ (case {\it ac}), there are three
related permutations, $P_{\rm enc}=\left({1\atop \overline{3}}{2\atop
    3}{3\atop \overline{1}}{\overline{1}\atop 2}{\overline{2}\atop
    1}{\overline{3}\atop\overline{2}}\right)$ and its two images under
cyclic permutation of 1,2,3 as well as
$\overline{1},\overline{2},\overline{3}$; physically, the three latter
are equivalent since they differ only in which of the three stretches
is considered the first. (In Ref. \cite{Tau3} such equivalences were
taken into account by multiplicity factors ${\cal N}_{pc}$ and ${\cal
  N}_{ac}$.)

Pairs of 2-encounters may either be composed of either (i) two
parallel encounters ($P_{\rm enc}=\left({1\atop 3}{2\atop 4}{3\atop
    1}{4\atop 2}{\overline{1}\atop \overline{3}}{\overline{2}\atop
    \overline{4}}{\overline{3}\atop\overline{1}}{\overline{4}
    \atop\overline{2}}\right)$, family {\it ppi}) or (ii) one parallel
and one antiparallel encounter ($P_{\rm enc}=\left({1\atop
    \overline{3}}{2\atop 4}{3\atop \overline{1}}{4\atop
    2}{\overline{1}\atop 3}{\overline{2}\atop
    \overline{4}}{\overline{3}\atop 1}{\overline{4}\atop
    \overline{2}}\right)$ and $P_{\rm enc}=\left({1\atop 3}{2\atop
    \overline{4}}{3\atop 1}{4\atop \overline{2}}{\overline{1}\atop
    \overline{3}}{\overline{2}\atop 4}{\overline{3}\atop\overline{1}}
  {\overline{4}\atop 2}\right)$, family {\it api}), or (iii) two
antiparallel ones ($P_{\rm enc}=\left({1\atop \overline{4}}{2\atop
    \overline{3}}{3\atop \overline{2}}{4\atop
    \overline{1}}{\overline{1}\atop 4}{\overline{2}\atop
    3}{\overline{3}\atop 2}{\overline{4}\atop 1}\right)$ and $P_{\rm
  enc}=\left({1\atop \overline{2}}{2\atop \overline{1}}{3\atop
    \overline{4}}{4\atop \overline{3}}{\overline{1}\atop
    2}{\overline{2}\atop 1}{\overline{3}\atop 4}{\overline{4}\atop
    3}\right)$, family {\it aas} in \cite{Tau3}). Again, equivalent
permutations differ by cyclic permutations (of now 1,2,3,4 as well as
$\overline{1},\overline{2},\overline{3},\overline{4}$), i.e. in which
of the stretches is assigned the number 1.

\subsubsection{Recursion relation for $N(\vec{v})$}

\label{sec:orthogonal_recursion}

We are now fully equipped to establish a recursion relation for
$N(\vec{v})$ in much the same way in the unitary case. Impatient
readers may want to jump to the resulting recursion
(\ref{welcome_back}) for $K_n$.

First of all, we recover Eq. (\ref{mengetomenge}),
$N(\vec{v},l)=\frac{v_l l}{L}N(\vec{v})$ using exactly the same
arguments as in the unitary case.  From each element $P_{\rm
  enc}\in{\cal M}(\vec{v},l)$ we obtain $L$ elements of ${\cal
  M}(\vec{v})$ by the $L$ possible cyclic permutations.  Applying
these cyclic permutations to the set ${\cal M}(\vec{v},l)$ we obtain
$v_l l$ copies of ${\cal M}(\vec{v})$, since the $v_l$ pairs of twin
$l$-cycles have altogether $v_l l$ members without overbar, and
permutations shifting $L$ among these members leave ${\cal
  M}(\vec{v},l)$ invariant.

Choosing a permutation $P_{\rm enc}=P_{\rm loop}^{-1}P\in{\cal
  M}(\vec{v},l)$ we set $Q_{\rm enc}=Q_{\rm loop}^{-1}Q$ with $Q_{\rm
  loop}$ and $Q$ obtained from $P_{\rm loop}$ and $P$ by omitting $L$
and $\overline{L-1}$, and replacing $\overline{L}$ by
$\overline{L-1}$.  (This prescription can be interpreted as removing
the loop leading from exit $L-1$ to entrance $L$, and its
time-reversed leading from exit $\overline{L}$ to entrance
$\overline{L-1}$; hence the corresponding entrance ports must be
removed from $P$ and $P_{\rm loop}=P^\gamma$.)  In particular $Q_{\rm
  loop}$ will have the form $Q_{\rm
  loop}=(1,2,\ldots,L-1)(\overline{L-1},\overline{L-2},\ldots,\overline{1})$.
The two cycles of $Q$ fulfill the same relation as those of $P$ and
may thus indeed be interpreted as lists of entrance ports of two time
reversed orbits.  The resulting ``encounter permutation" $Q_{\rm enc}$
maps the remaining elements
$a=1,2,\ldots,L-1,\overline{1},\overline{2},\ldots,\overline{L-1}$ as
\begin{equation}
\label{permutation_orthogonal}
Q_{\rm enc}(a)=\begin{cases}
  P_{\rm enc}(L)&\text{if $a=P_{\rm enc}^{-1}(L-1)$}\\
  P_{\rm enc}(\overline{L-1})&\text{if $a=P_{\rm enc}^{-1}(\overline{L})$}\\
  L-1&\text{if $a=P_{\rm enc}^{-1}(L)$}\\
  P_{\rm enc}(\overline{L})&\text{if $a=\overline{L-1}$}\\
  P_{\rm enc}(a)&\text{otherwise}\,.
\end{cases}
\end{equation}
Here, the second and fourth line extend (\ref{QPE_unitary}) as
required by ${\cal T}$ invariance; the present $Q_{\rm enc}$ is indeed
${\cal T}$ covariant.

When analyzing the cycle structure of $Q_{\rm enc}$, we now have to
distinguish {\it three} cases, the first two paralleling the treatment
of \ref{sec:unitary_recursion}.  Note however a factor~2 appearing in
the second case. For each $Q_{\rm enc}\in{\cal M}(\vec{v}^{[l\to
  m,l-m-1]},m)$, there are now twice as many, namely
$2(l-m-1)(v_{l-m-1}+1)$ related $P_{\rm enc}\in {\cal M}(\vec{v},l)$
structured like (\ref{pmcyc}), since $Q_{\rm enc}$ also remains
unaffected by time reversal of $a_{m+2},\ldots,a_l$ in (\ref{pmcyc}).  The
second and fourth line in (\ref{permutation_orthogonal}) make sure
that merging or splitting of cycles is mirrored by the respective
twins.

A third possibility appears since \textbf{the cycles involving $L$ and
  $L-1$ may be twins}, and hence belong to the same encounter.  Since
the twin cycles are mutually time reversed there is one cycle
containing both $L$ and $\overline{L-1}$, and another one containing
$\overline{L}$ and $L-1$.  Assume that inside the first cycle, the
element $\overline{L-1}$ follows $L$ after $m$ iterations, i.e.
$\overline{L-1}=P_{\rm enc}^m(L)$ (with $1\leq m\leq l-1$).  Then $P_{\rm
  enc}$ can be written as
\begin{eqnarray}
  \label{PPE_orth}
  P_{\rm enc}=&&[\ldots](L,a_2,\ldots,a_{m},\overline{L-1},a_{m+2},\ldots,a_l)\nonumber\\
  &&(\overline{a_l},\ldots,\overline{a_{m+2}},L-1,
  \overline{a_{m}},\ldots,\overline{a_2},\overline{L})\,.
\end{eqnarray}
Due to (\ref{permutation_orthogonal}), $Q_{\rm enc}$ differs from
$P_{\rm enc}$ by mapping
\begin{eqnarray}
    Q_{\rm enc}(\overline{a_{m+2}})=a_2\quad\quad\quad
    Q_{\rm enc}(\overline{a_2})=a_{m+2}\nonumber\\
    Q_{\rm enc}(a_l)=L-1\quad\quad\quad
    Q_{\rm enc}(\overline{L-1})=\overline{a_l}
\end{eqnarray}
The initial pair of twin cycles of $P_{\rm enc}$ is transformed to the
following pair of twin $(l-1)$-cycles of
\begin{eqnarray}\label{QPE_orth}
    Q_{\rm enc}=&&[\ldots](a_2,\ldots,a_{m},\overline{L-1},\overline{a_l},\ldots,
    \overline{a_{m+2}})\nonumber\\
    &&(a_{m+2},\ldots,a_l,L-1,\overline{a_{m}},\ldots,\overline{a_2})\,.
\end{eqnarray}
Given that the largest number permuted by $Q_{\rm enc}$, i.e.  $L-1$,
is included in one of these cycles, we have $Q_{\rm enc}\in{\cal
  M}(\vec{v}^{[l\to l-1]},l-1)$.  Conversely, for any such $Q_{\rm
  enc}$ and each $1\leq m\leq l-1$, there is exactly one related $P_{\rm
  enc}$, since we may read off $a_2,\ldots,a_l$ from $Q_{\rm enc}$ in
(\ref{QPE_orth}) and recombine them to form a permutation $P_{\rm
  enc}$ as in (\ref{PPE_orth}).
We thus see that each of the $l-1$ subsets of ${\cal M}(\vec{v},l)$
with $\overline{L-1}=P_{\rm enc}^m(L)$ is one-to-one to ${\cal
  M}(\vec{v}^{[l\to l-1]},l-1)$ and thus has an equal number of
elements.

We have seen that ${\cal M}(\vec{v},l)$ falls into subsets similar to
the unitary case, time-reversal invariance making for the factor 2
explained above, and for $l-1$ additional subsets of size
$N(\vec{v}^{[l\to l-1]},l-1)$. The various sizes combine to the
orthogonal analogue of the recursion relation (\ref{NL}),
\begin{eqnarray}
\label{recursion_nvl}
&&N(\vec{v},l)=\sum_{k\geq 2}N(\vec{v}^{[k,l\to k+l-1]},k+l-1)\nonumber\\
&&+\sum_{m=1}^{l-2}2(l-m-1)(v_{l-m-1}+1)
N(\vec{v}^{[l\to m,l-m-1]},m)\nonumber\\
&&+(l-1)N(\vec{v}^{[l\to l-1]},l-1)\,,
\end{eqnarray}
which using (\ref{mengetomenge}) and the shorthand
$\tilde{N}(\vec{v})=N(\vec{v})\frac{(-1)^V\prod_l l^{v_l}}{L}$ may be
written as
\begin{eqnarray}
\label{recursion_orthogonal}
\label{recur2}
&&v_l\tilde{N}(\vec{v})+\sum_{k\geq 2}(v_{k+l-1}+1)k
\tilde{N}(\vec{v}^{[k,l\to k+l-1]})\nonumber\\
&&+\sum_{m=1}^{l-2}2(v_{l-m-1}+1)v_m^{[l\to m,l-m-1]}
\tilde{N}(\vec{v}^{[l\to m,l-m-1]})
\nonumber\\
&&-(l-1)(v_{l-1}+1)\tilde{N}(\vec{v}^{[l\to l-1]})=0\,;
\end{eqnarray}
recall that $v_{k+l-1}+1=v_{k+l-1}^{[k,l\to k+l-1]}$.

\subsubsection{Spectral form factor}

Similarly as in \ref{sec:formfactor_unitary} we now turn the recursion
relation for $\tilde{N}(\vec{v})$ into one for the Taylor coefficients
$K_n$.  As a preparation we generalize our rule (\ref{rule_special}).
For all similar sums over $\vec{v}$ with fixed $\nu(\vec{v})=n$ and
$v_1=0$ we find
\begin{equation}
\label{rule}
\sum_{\vec{v}}^{\nu(\vec{v})=n}\!\!f(\vec{v}^{[\alpha_1,\alpha_2,\ldots\to \beta]})\tilde{N}(\vec{v}^{[\alpha_1,\alpha_2,\ldots\to \beta]})=\!\!\!\!\sum_{\vec{v}'}^{\nu(\vec{v}')=n'}\!\!\!f(\vec{v}')\tilde{N}(\vec{v}')\,,
\end{equation}
with integers $\alpha_i\geq 2$, $\beta\geq 2$,
$n'=\nu(\vec{v}')=\nu(\vec{v}^{[\alpha_1,\alpha_2,\ldots\to \beta]})=n-\sum_i
(\alpha_i-1)+(\beta-1)$, and $f$ any function of $\vec{v}'$ vanishing for
$v'_\beta=0$. Eq. (\ref{rule}) follows in the same way as
(\ref{rule_special}), i.e. by switching to
$\vec{v}'=\vec{v}^{[\alpha_1,\alpha_2,\ldots\to \beta]}$ as the new summation variable
and dropping the restriction $v'_\beta\geq 1$ ($\vec{v}'$ with $v'_\beta=0$ do
not contribute due to the vanishing of $f$).  One similarly shows that
the foregoing rule holds even without any conditions on $f$ if $\beta$ is
removed, i.e. if no new cycles are created. It is convenient to
abbreviate the r.~h.~s. of (\ref{rule}) with the help of
\begin{equation}
\label{shorthand}
S_{n'}[f]\equiv\sum_{\vec{v}'}^{\nu(\vec{v}')=n'}f(\vec{v}')\tilde{N}(\vec{v}') \,
\end{equation}
for arbitrary $f$; we note that $K_n=\frac{2}{(n-2)!} S_{n}[1]$. Thus
equipped we turn to the three special cases $l=1,2,3$ of our recursion
relations (\ref{recursion_nvl},\ref{recursion_orthogonal}) which we
shall need below.

First, the case \underline{$l=1$} involves permutations with 1-cycles,
appearing only in intermediate steps of our calculation.  If the
element $L$ forms a 1-cycle, it may simply be removed from a
permutation without affecting the other cycles, which corresponds to a
transition $\vec{v}\to\vec{v}^{[1\to ]}$. We thus have
$N(\vec{v},1)=N(\vec{v}^{[1\to]})$ and equivalently
\begin{equation}
\label{case_one}
v_1\tilde{N}(\vec{v})+L(\vec{v}^{[1\to ]})\tilde{N}(\vec{v}^{[1\to ]})=0\,.
\end{equation}

Second, for \underline{$l=2$} (and $v_1=0$) the recursion
(\ref{recursion_orthogonal}) boils down to
\begin{eqnarray}
\label{case_two}
&& v_2\tilde{N}(\vec{v})+\sum_{k\geq 2}\left(v_{k+1}^{[k,2\to
    k+1])}k\tilde{N}(\vec{v}^{[k,2\to k+1]})\right)\nonumber\\&&
    -\tilde{N}(\vec{v}^{[2\to 1]})=0\,,
\end{eqnarray}
where only the last term is new compared to the unitary case. We bring
it to a form free from 1-cycles by invoking (\ref{case_one}) and thus
$\tilde{N}(\vec{v}^{[2\to 1]})=-L(\vec{v}^{[2\to]})
\tilde{N}(\vec{v}^{[2\to]})$, to get
\begin{eqnarray}
\label{case_two_L}
&& v_2\tilde{N}(\vec{v})+\sum_{k\geq 2}\left(v_{k+1}^{[k,2\to
    k+1])}k\tilde{N}(\vec{v}^{[k,2\to k+1]})\right)\nonumber\\
    &&+L(\vec{v}^{[2\to ]})\tilde{N}(\vec{v}^{[2\to ]})=0\,.
\end{eqnarray}
As in the unitary case, we sum over all $\vec{v}$ with $v_1=0$ and
$\nu(\vec{v})=n$. The rule (\ref{rule}) and the shorthand
(\ref{shorthand}) give
\begin{eqnarray}
\label{result_two_a}
&& S_{n}\Bigg[v_2+\sum_{k\geq
    2}v_{k+1}k\Bigg]+ S_{n-1}[L(\vec{v})]\nonumber\\
    &&=(n-1) S_{n}[1]+ S_{n-1}[L(\vec{v})]=0\,.
\end{eqnarray}
A further relation is obtained by multiplying (\ref{case_two_L}) with
$L(\vec{v})-1=L(\vec{v}^{[k,2\to k+1]})=L(\vec{v}^{[2\to]})+1$ and again
summing over $\vec{v}$ with the help of (\ref{rule}). The resulting
equation
\begin{equation}
  (n-1) S_{n}[L(\vec{v})]- S_{n}[v_2]+ S_{n-1}[L(\vec{v})(L(\vec{v})+1)]=0
\end{equation}
can be simplified by (\ref{result_two_a}) and replacing $n\to n-2$,
\begin{equation}
\label{result_two_b}
-(n-2)(n-3) S_{n-1}[1]= S_{n-2}[v_2]- S_{n-3}[L(\vec{v})(L(\vec{v})+1)]\,.
\end{equation}

Finally, we consider \underline{$l=3$} (and $v_1=0$)
\begin{eqnarray}
\label{case_three}
&& v_3\tilde{N}(\vec{v})+\sum_{k\geq 2}\left(v_{k+2}^{[k,3\to
    k+2]}k\tilde{N}(\vec{v}^{[k,3\to k+2]})\right)\nonumber\\
    &&+4\tilde{N}(\vec{v}^{[3\to 1,1]})-2v_2^{[3\to
  2]}\tilde{N}(\vec{v}^{[3\to 2]})=0\,.
\end{eqnarray}
The 1-cycles in the third term are eliminated using the identity
$\tilde{N}(\vec{v}^{[3\to 1,1]})=\frac{1}{2}L(\vec{v}^{[3\to
  ]})(L(\vec{v}^{[3\to]})+1)\tilde{N}(\vec{v}^{[3\to]})$, which follows
by twice applying (\ref{case_one}) to $\vec{v}^{[3\to 1,1]}$.  Summing
over $\vec{v}$ in (\ref{case_three}) we find
\begin{eqnarray}
&&
 S_{n}\Bigg[v_3+\sum_{k\geq
    2}v_{k+2}k\Bigg]+2 S_{n-2}[L(\vec{v})(L(\vec{v})+1)]\nonumber\\
    &&-2 S_{n-1}[v_2]=0\,.
\end{eqnarray}
This expression can be simplified using $v_3+\sum_{k\geq 2}v_{k+2}k
=\sum_{k\geq
  2}v_k(k-2)=L(\vec{v})-2V(\vec{v})=2(\nu(\vec{v})-1)-L(\vec{v})$, i.e.
\begin{eqnarray}
&& 2(n-1) S_{n}[1]- S_{n}[L(\vec{v})]+2 S_{n-2}[L(\vec{v})(L(\vec{v})+1)]\nonumber\\
&&-2 S_{n-1}[v_2]=0\,.
\end{eqnarray}
Finally applying (\ref{result_two_a}), substituting $n\to n-1$, and
dividing by 2, we proceed to
\begin{eqnarray}
\label{result_three}
&&\frac{n-1}{2} S_{n}[1]+(n-2) S_{n-1}[1]\nonumber\\
&&= S_{n-2}[v_2]- S_{n-3}[L(\vec{v})(L(\vec{v})+1)]\,.
\end{eqnarray}

Upon comparing the recursion relations (\ref{result_two_b}) and
(\ref{result_three}), obtained for the cases $l=2$ and $l=3$, we find
the coefficients $ S_{n}[1]=\frac{(n-2)!}{2}K_n$ and $
S_{n-1}[1]=\frac{(n-3)!}{2}K_{n-1}$ related as $\frac{n-1}{2}
S_{n}[1]=-(n-2)^2 S_{n-1}[1]$ or
\begin{equation}
\label{welcome_back}
(n-1)K_n=-2(n-2)K_{n-1}\,.
\end{equation}
An initial condition is provided by the Sieber/Richter result for
orbits differing in one 2-encounter, $K_2=-2$.  Thus started, our
recursion yields the Taylor coefficients
\begin{equation} K_n=\frac{(-2)^{n-1}}{n-1}
\end{equation}
coinciding with the random-matrix result. Universal behavior is thus
ascertained for the small-time form factor of fully chaotic dynamics
from the orthogonal symmetry class; the resulting series converges for
$\tau<\frac{1}{2}$.

\section{Spinning particles and the symplectic symmetry class}

\label{sec:symplectic}

We now allow for a spin with arbitrary but fixed spin quantum number
$S$. Assuming time-reversal invariance we know that for integer $S$
the time reversal operator ${\cal T}$ squares to unity, ${\cal
  T}^2=1$, whereas for half-integer spin, ${\cal T}^2=-1$; we face the
orthogonal and symplectic symmetry class, respectively. The
semiclassical theory of spinning particles is discussed in
\cite{Keppeler}; off-diagonal terms of the form factor were considered
in a preliminary version in \cite{Heusler}, and for quantum graphs in
\cite{BolteHarrison}.

The Pauli Hamiltonian reads $H=H_0+\mathbf{\mathbf{\hat B}\cdot\hat S} $
where $\mathbf{\hat S}$ is the spin operator, with $\mathbf{\hat
  S}^2=\hbar^2S(S+1)$. The vector operator $\mathbf{\hat B} (
\mathbf{\hat q}, \mathbf{\hat p})$ describes the influence of the
translational motion and external fields on the spin. The spin-orbit
interaction formally behaves like $\cal{O}(\hbar)$ and tends to zero in
the semiclassical limit; it is not a small quantum perturbation
though, since its matrix elements are infinitely larger than the
energy spacing $\Delta\sim {\cal O}(\hbar^f),\quad f>1$.

The state of the system is given by a spinor with $2S+1$ components
and the propagator by a $(2S+1)\times(2S+1)$ matrix. In the leading order
of the semiclassical approximation the propagator consists of the
scalar translational part which is the van Vleck propagator of the
spinless system, multiplied by the spin evolution matrix $d$
(belonging to the spin-$S$ representation $D^{(S)}$ of $SU(2)$); the
latter matrix has to be evaluated on the classical orbit (of the
translational motion) connecting the initial and final point. Along
such an orbit $\gamma$ the $d$-matrix is a function of the initial and
final times, $d=d_\gamma(t,t_0)$. It satisfies the equation ${{\rm
    i}}\partial_td_\gamma(t,t_0)=\mathbf{B}_\gamma(t)\cdot (\mathbf{\hat
  S}/\hbar)\;d_\gamma(t,t_0)$ where the classical time-dependent vector
$\mathbf{B}_\gamma(t)$ is obtained by substituting the classical
coordinates and momenta along $\gamma$ for the operator valued arguments
of $\mathbf{\hat B}(\mathbf{\hat q},\mathbf{\hat p})$; the initial
condition is $d_\gamma(t_0,t_0)=\mathbf{1}_{2S+1}$.  Such semiclassical
treatment keeps the translational motion unaffected by the quantum
spin. The spin, however, is driven by the translational motion. No
semiclassical approximation for the spin itself (which would require
the assumption of large $S$) is invoked.

The full quantum nature of the spin (finite $S$) notwithstanding, a
seemingly classical manner of speaking about the spin is possible, due
to the following fact: any matrix from $D^{(S)}$ can be parametrized
by three Euler angles (e.g. like $d(\theta, \phi,\psi)={\rm e}^{{{\rm i}} \psi
  \hat S_z/\hbar}{\rm e}^{{{\rm i}} \theta \hat S_x/\hbar}{\rm e}^{{{\rm i}} \phi
  \hat S_z/\hbar}$), which are time-dependent for $d=d(t)$. The angles
$\psi(t),\theta(t),\phi(t)$ may also be imagined to specify the orientation of
a fictitious rigid body in classical rotation; as done in
\cite{Keppeler,Heusler,BolteHarrison} we shall speak of that motion as
``spin rotation''.

We assume the translational motion chaotic as before and require
ergodicity of the combined spin rotation and translational motion. The
spin rotation itself is then also ergodic.  This means that time
averages of the spin dependent properties over intervals longer than a
certain relaxation time $t_{\rm cl}$ can be replaced by averages over
all $d\in D^{(S)}$; the measure to be used reads $d\mu=d\phi\,\sin \theta
d\theta\, d\psi/8\pi^2$.

\subsection{Integer spin}

The trace formula for a particle with spin \cite{Keppeler} gives the
level density as a sum over periodic orbits $\gamma$
\begin{equation}
\label{spinwiller} \rho_{\rm osc}(E) \sim \frac{1}{\pi
\hbar}{\rm Re} \sum_\gamma ({\rm tr} \ d_\gamma) A_\gamma {\rm e}^{{\rm
i}S_\gamma/\hbar}\,;
\end{equation}
beside the stability amplitude (including period and Maslov phase) and
the classical action of the $\gamma$th orbit, the factor ${\rm tr} \ 
d_{\gamma}\equiv {\rm tr} \ d_\gamma(t_0+T_\gamma,t_0)$ appears and reflects the spin
evolution over a period of the translational motion.  That ${\rm tr} \ 
d_{\gamma}$ is independent of the initial moment $t_0$: its shift leads
only to a similarity transformation of $d_\gamma$.

The form factor becomes the double sum
\begin{eqnarray}
\label{doublesum_spin}
K(\tau)&=  &\frac{1}{T_H}  \left\langle\sum_{\gamma,\gamma'} ({\rm tr} \
d_{\gamma})  ({\rm tr} \ d_{\gamma'}) A_\gamma A_{\gamma'}^*  \right. \nonumber\\
&\times&\left.{\rm e}^{{\rm i}(S_\gamma-S_{\gamma'})/\hbar}\delta\left(
\tau T_H-\frac{T_\gamma+T_{\gamma'}}{2}\right)\right\rangle\,.
\end{eqnarray}
Due to the spin, the average level density and thus the Heisenberg
time $T_H$ are increased by the factor $2S+1$.

The diagonal approximation yields the sum
\begin{equation}
\label{diagonal}
K_{{\rm diag}}(\tau) = \frac{1}{T_H}  \left\langle\sum_{\gamma}
({\rm tr} \ d_{\gamma})^2
|A_\gamma|^2
\delta\left(\tau T_H-T_\gamma\right)\right\rangle.
\end{equation}
Since the spin dynamics is ergodic and since we are averaging over an
ensemble of orbits, the equidistribution theorem
\cite{Equidistribution} allows us to treat $d_\gamma$ as random and to
integrate over all matrices of the spin-$S$ representation $D^{(S)}$
of $SU(2)$; the sum over $\gamma$ gives the usual factor $T$.  The spin
integral gives unity for any $S$, and so we have \cite{Keppeler}
\begin{equation}
 K_{{\rm diag}}(\tau)= 2 \tau \int d\mu({\cal A})
({\rm tr} {\cal A})^2=2 \tau\,.
\end{equation}

As to off-diagonal contributions from orbit pairs $\gamma$, $\gamma'$, spin
makes for two modifications compared to the previous Sections.  First,
$T_H$ contains a factor $2S+1$, such that the $L-V$ factors
$2\pi\hbar\frac{T}{\Omega}$ in (\ref{contribution}) give
$(2S+1)\frac{T}{T_H}=(2S+1)\tau$ rather than $\tau$. Second, the
contribution of each orbit pair comes with the factor $({\rm tr} \ 
d_\gamma)({\rm tr} \ d_{\gamma'})$.

To evaluate the factor $({\rm tr} \ d_\gamma)({\rm tr} \ d_{\gamma'})$, we
decompose the orbit $\gamma$ into $L$ pieces by cutting it once in each
encounter, and represent $d_\gamma$ as a product of $L$ matrices
describing the spin evolution over one orbit piece.  In the orbit
pairs contributing to the form factor the duration of each piece (the
orbit loop + segments of the preceding and following encounter
stretches) exceeds the Ehrenfest time $T_E$, and thus also the
classical relaxation time $t_{\rm cl}$.  Therefore, keeping in mind
that we are summing over an ensemble of orbits, we may invoke
ergodicity and replace the partial spin evolution matrices by matrices
randomly chosen out of $D^{(S)}$.  Numbering the orbit pieces and the
corresponding random matrices in the order of their traversal in $\gamma$
we may replace $d_\gamma$ by a product ${\cal A}_L{\cal A}_{L-1}\ldots{\cal
  A}_1$, the earliest propagator matrix written rightmost. The orbit
partner $\gamma'$ consists of practically the same pieces passed in a
different order, some of them in the opposite direction.  Hence
$d_{\gamma'}$ can be expressed in terms of the same $L$ matrices ${\cal
  A}_i$, but with the order suitably rearranged and with ${\cal A}_i\to
{\cal A}_i^{-1}$ for the time reversed pieces. The expectation value
of the trace product can now be evaluated as an integral over ${\cal
  A}_i,\;i=1,\ldots,L$.  Using the results of \cite{BolteHarrison}, one
obtains
\begin{eqnarray}\label{spi}
 &&\!\!\!({\rm tr} \ d_{\gamma})  ({\rm tr} \ d_{\gamma'})   \nonumber\\
&&\!\!\!\to  \!\int\!\!  ({\rm tr} {\cal A}_L {\cal A}_{L-1} \ldots
 {\cal A}_1)
 ({\rm tr} {\cal A}_{k_L}^{\eta_L}{\cal A}_{k_{L-1}}^{\eta_{L-1}} \ldots
 {\cal A}_{k_1}^{\eta_1})\prod_{i=1}^L d\mu({\cal
 A}_i)\nonumber\\ &&\!\!\!= \left(\frac{(-1)^{2 S}}{2 S + 1}\right)^{L-V}\,;
\end{eqnarray}
most remarkably, for an orbit pair with a given $\vec v$ the $L$-fold
integral depends only on the difference $L(\vec v)-V(\vec v) = n-1$;
in particular, it is independent of the special ordering of loops in
the partner orbit $\gamma'$ (expressed by the subscripts
$k_1,k_2,\ldots,k_L$), as well as of the senses of traversal (expressed by
the exponents $\eta_a=\pm 1$).\footnote{If desired, these indices can be
  determined from the permutation $P=P_{\rm loop} P_{\rm enc}$
  describing the partner $\gamma'$, see Section \ref{sec:combinatorics}.
  The permutation $P$ consists of two $L$-cycles relating to $\gamma'$ and
  $\overline\gamma'$; the sequence of loops $k_1\ldots k_L$ in $\gamma'$ is given
  by the appropriate cycle in which the elements with a bar
  (indicating time reversal of the loop) are modified like
  $\overline{k}\to (k+1)\mbox{ mod }L$ and simultaneously the
  associated superscripts $\eta$ are set to -1.  }

Now the two occurrences of $(2S+1)^{L-V}$ mutually cancel and the form
factor reads
\begin{equation}  \label{spinform}
K(\tau)=2\tau+2\sum_{\vec{v}}N(\vec{v})h(\vec{v})
(-1)^{2 S(L-V)}\tau^{L-V+1}.
\end{equation}
For integer spin, $(-1)^{2S}=1$ whereupon we recover the expansion
(\ref{k}) of $K(\tau)$ for the orthogonal class.

\subsection{Half-integer spin}

For half-integer spin, the minus sign in (\ref{spinform}) becomes
relevant.  Moreover, all levels become doubly degenerate {\`a} la Kramers
\cite{Haake}. With the density of levels reduced to half the density
of states we are led to the rescaling $K(\tau) \to \frac{1}{2}
K(\frac{\tau}{2})$ \cite{Keppeler}.  In this case, the form factor reads
\begin{equation}\label{Korth}
K(\tau)=\frac{\tau}{2}-\sum_{\vec{v}}  N(\vec{v})h(\vec{v})
\left(-\frac{\tau}{2}\right)^{L-V+1}
=\frac{\tau}{2}-\frac{\tau}{4}\ln(1-\tau).
\end{equation}
To understand the final step, we compare the sums over $\vec{v}$ in
(\ref{Korth}) and (\ref{k}), the latter pertinent to the orthogonal
class, and find $K(\tau)=-\frac{1}{2}K_{\rm GOE}(-\frac{\tau}{2})$.  We
have thus verified the random-matrix result for the Gaussian
symplectic ensemble (\ref{formfactor}).  As predicted in
\cite{Heusler} the sign change of the argument $\tau$, which entails the
logarithmic singularity of the symplectic form factor at $\tau=1$, comes
from the contributions of the spin integrals (\ref{spi}).

\section{Relation  to the $\sigma$-model}

\label{sec:sigma}

\subsection{Introduction}

The so-called sigma model \cite{Sigma,Efetov} is a convenient
framework for calculating averaged products of Green functions of
random Hamiltonians. Its zero dimensional variant affords, in
particular, the two-point correlator of the level density (and its
Fourier transform, the spectral form factor) for the Gaussian
ensembles of RMT; see Appendix \ref{sec:sigma_appendix} for a brief
introduction.  Perturbative implementations exist for the three
Wigner/Dyson symmetry classes and yield the respective spectral form
factors $K(\tau)$ as power series in the time $\tau$, i.e.  precisely the
series extracted from Gutzwiller's semiclassical periodic-orbit theory
in the preceding sections.

The sigma model for random-matrix theory proved of great heuristic
value for our semiclassical endeavor: We were led to the correct
combinatorics of families of orbit pairs by an analysis of the
perturbation series of the sigma model. The analogy of periodic-orbit
expansions to perturbation series might prove fruitful for future
applications of periodic-orbit theory, and that possibility motivates
the following exposition.

Before entering technicalities it is appropriate to point to some
qualitative analogies and differences between the two approaches.
Very roughly, different Feynman diagrams of the sigma model (both for
the Wigner/Dyson ensembles and disorder) correspond to different
families of (pairs of) periodic orbits, vertices to close
self-encounters, and propagator lines to orbit loops. An important
difference lies in the point character of vertices and the non-zero
duration, of the order of the Ehrenfest time $T_E\propto \ln\hbar$, of the
relevant self-encounters.  Of course, the relevant encounter durations
are vanishingly small compared to the typical loop lengths ($\sim T_H\propto
\hbar^{-f+1}$); nevertheless, we may say that self-encounters give
internal structure to vertices.

\subsection{Expansions of two-point correlator
  and form factor}
\label{sec:sigmamodel_formfactor}

The connected two-point correlator of the density of levels,
$\bar{R}=(\overline{\rho\rho}-\bar{\rho}^2)/\bar{\rho}^2$ can be obtained from
the ensemble averaged product of the retarded and advanced Green
functions $G_\pm(E)=(E\pm {\rm i}\delta-H)^{-1}$ as \cite{Haake}
\begin{eqnarray}\label{complex_correlator}
&& R(s)=\frac{{\rm
Re}\,\overline{\delta\,\mbox{tr}\, G_+(E+s/2\pi\bar
\rho)\;\delta\,\mbox{tr}\, G_-(E-s/2\pi\bar \rho)}}{2\pi^2\bar \rho^2},\quad\nonumber\\&& \delta\,
\mathrm{tr}\, G_\pm=\mathrm{tr}\; G_\pm-\overline{ \mathrm{tr }
\,G_\pm}\,.
\end{eqnarray}
Here, the overbar denotes an average over a Gaussian ensemble of
random matrices.  The argument $E$ (the average energy) is suppressed.
The energy difference is expressed in terms of the dimensionless
offset $s$.  The Fourier transform of $R$ w.r.t. the offset $s$ gives
the central object of the present paper, the spectral form factor,
\begin{equation}
K(\tau)=\frac{1}{\pi}\int_{-\infty}^\infty ds \,{\rm e}^{2{{\rm i}}s\tau}R(s).
\end{equation}

As briefly shown in an Appendix, a bosonic replica variant of the
$\sigma$-model yields the $(1/s)$-expansion of the Fourier transform of
the small-time form factor as an integral over matrices $B$,
\begin{eqnarray}\label{rs}
         R(s) &\sim& -\frac{1}{2} {\rm Re}\ {\rm lim}_{r \to 0} \frac{1}{r^2}
        \partial^2_s\,
        s^{-\kappa  r^2}\nonumber\\&&
        \times\int {dB} \,{\rm e}^{(2{{\rm i}}/\kappa ) \sum_{l=1}^{\infty}
         s^{1-l}{\rm tr}(B {B^\dagger})^l}\,,
\end{eqnarray}
where on the r.h.s. the offset $s$ must be read as $s+{\rm i}\delta$ with
$\delta\downarrow 0$; the matrices $B, {B^\dagger}$ are $r \times r$ for the GUE and $2r \times 2
r$ for the GOE; as before, the factor $\kappa $ takes the respective value
1 and 2 for the two classes. The essence of the replica ``trick'' is
to find the foregoing integral as a power series in $r$ and to isolate
the coefficient of $r^2$.

In the limit $s\to\infty $ the principal contribution to $R(s)$ comes from
the Gaussian factor $\exp [(2{{\rm i}}/\kappa )\ \mbox{tr}M]$ in the
integrand of (\ref{rs}), where $M=BB^\dagger$.  The remaining factor can be
expanded as
\begin{eqnarray}\label{expan1}
&&\exp \left(\frac{2{{\rm i}}}{\kappa }\sum_{l\geq 2} s^{1-l}{\rm tr }
M^l\right)\nonumber\\ &=&\sum_{V=0}^\infty
\frac{1}{V!}\left(\frac{2{{\rm i}}}{\kappa }\right)^V
\left(\sum_{l\geq2} s^{1-l}{\rm tr } M^l \right)^V \nonumber\\
&=&\sum_{\vec{v}} \frac{1}{\prod_{l\geq 2}v_l!}\left(\frac{2
{{\rm i}}}{\kappa }\right)^Vs^{V-L}
 \prod_{l\geq2}
(\mbox {tr}M^l)^{v_l}.
\end{eqnarray}
Here, the summation extends over integers $v_2,v_3,\ldots,v_l,\ldots$ each of
which runs from zero to infinity, and we write $\vec{v}=(v_2,v_3,\ldots)$
just like in our semiclassical treatment.  The total number of traces
in the summand $\vec v$ is $V(\vec v) =\sum_{l\geq 2} v_l$, and again we
define $L(\vec{v})=\sum_{l\geq 2}lv_l$.  The integral over $B$ and $B^\dagger$
in (\ref{rs}) may now be seen as a sum of averages like
\begin{equation}\label{sigma_quadratic_av}
  \big< f(B, {B^\dagger}) \big>         \equiv \int {dB}\,
        f(B, {B^\dagger}) \ {\rm e}^{(2{{\rm i}}/\kappa )  {\rm tr}(B {B^\dagger})}\,.
\end{equation}
We may thus write
\begin{eqnarray} 
&& R(s)\sim-\frac{1}{2}\mbox{Re}\left\{\lim_{r\to
0}\frac{1}{r^2}\partial_s^2s^{-\kappa  r^2} \times\right.\nonumber\\&&\left.\sum_{\vec{v}}
\frac{1}{\prod_{l\geq 2}v_l!}\left(\frac{2
{{\rm i}}}{\kappa }\right)^Vs^{V-L} \left\langle\prod_{l\geq2}(\mbox
{tr}M^l)^{v_l}\right\rangle\right\}.
\end{eqnarray} 

The leading term $\vec{v}=0$ corresponds to
$\left<1\right>=\left(\mathrm{const}\right)^{\kappa r^2}\stackrel{{r\to
    0}}{\longrightarrow} 1$.  The respective contribution to the two-point
correlator is $ -\frac{1}{2}\mbox{Re} \;{\rm lim}_{r \to 0}
\frac{1}{r^2} \partial^2_s (s+{{\rm i}}\delta)^{-\kappa r^2}=-\mbox{Re}\frac{\kappa
}{2(s+{{\rm i}}\delta)^2}$, and thus $\kappa |\tau|$ for the spectral form
factor, reproducing the diagonal part both in the unitary and
orthogonal cases.
 
For all other terms, the operations of taking the second derivative by
$s$ and going to the limit $r\to 0$ commute, meaning that the factor
$s^{-\kappa r^2}$ can be disregarded.  We shall show below that the
averages of the trace products with non-zero $\vec v$ have the
property
\begin{equation}\label{connum} 
 \lim_{r\to0}\frac{1}{r^2}\big<\prod_{l} \left(
\mbox{tr}M^l\right)^{v_l}\big>=\frac{\kappa^2}{(-2{\mathrm i})^{L(\vec 
 v)}}N_c(\vec v) 
\end{equation} 
in which $N_c(\vec v)$ take positive integer values; we shall in fact
come to interpreting $N_c(\vec{v})$ as the ``number of contractions'';
the traces of $M^l$ will be called $l$-traces, to stress the analogy
with the $l$-encounters of periodic orbits. The form factor $K(\tau)$ is
now obtained by Fourier transforming.  Using
$\int_{-\infty}^{\infty}\frac{ds}{\pi}{\rm e} ^{2{\rm i}s\tau}{\rm Re}\left[{\rm
    i}^{-n+1}(s+{\rm i}\delta)^{-n-1}\right]=(-2)^n|\tau|^n/n!$ one easily
shows that the Taylor coefficients of $K(\tau)=\kappa |\tau|+\sum_{n\geq
  2}K_n|\tau|^n$ are given by
\begin{equation} 
  \label{sigmak} 
  K_n=\frac{\kappa }{(n-2)!}\sum_{\vec{v}}^{L(\vec{v})-V(\vec{v})+1=n} 
  \frac{(-1)^V}{\kappa ^{V-1}\prod_{l\geq 2} v_l!}N_c(\vec{v})\,.
\end{equation} 

\subsection{Contraction rules}
 
In the following, we will derive a recursion for $N_c(\vec{v})$
similar to the recursion in our semiclassical analysis. To that end,
we calculate the averages of the products of traces in (\ref{connum})
by Wick's theorem. Each average becomes, for the GUE, a sum of
contractions of a fixed matrix $B$ in one of the traces and all
matrices $B^\dagger$; for the GOE, contractions with other matrices $B$
arise as well.  In all formulae below, $X$ and $Y$ must provide the
traces on the l.~h.~s. with an alternating sequence of $ B$'s and
$B^\dagger$'s; then the same will hold for all traces on the r.~h.~s.~.
Moreover, the term $[\ldots]$ will stand for inert traces unchanged by the
contraction.  The GUE involves two contraction rules,
\begin{subequations}
\label{sigma_con}
\begin{eqnarray}
             \big< {\rm tr} \ \contraction[2ex]{}{  {\mathbf  B}}{X{\rm tr}\,\,}
 {      {\mathbf B} } { B}
         X \ {\rm tr} { {  B^\dagger}} Y [\ldots] \big>
          &=&
        - \frac{1}{2 {{\rm i}}} \big< {\rm tr} ( X \ Y )  [\ldots]
                 \big>  \\
             \big< {\rm tr} \ \contraction[2ex]{}{  {\mathbf B}}{}{(X)
                  {\mathbf B} }
                 { B} X { { B^\dagger}} Y [\ldots] \big> &=&
                - \frac{1}{2 {{\rm i}}} \big< {\rm tr} X \ {\rm tr} Y
[\ldots] \big> \,.\ \
\end{eqnarray}
For the GOE, two more rules arise from contractions of $B$ with $B$
(and similarly $B^\dagger$ with $B^\dagger$):
\begin{eqnarray}
             \big< {\rm tr} \ \contraction[2ex]{}{
{\mathbf B}}{X{\rm tr}\,\,}{      {\mathbf B} }
        { B} X \ {\rm tr} { B} Y[\ldots] \big>
         &=&
                - \frac{1}{2 {\rm i}}\big<{\rm tr}( X \ {Y^\dagger} )
[\ldots]\big>
\\
                \big< {\rm tr} \   \contraction[2ex]{}
{ {\mathbf B}}{}{(X)      {\mathbf B} }
        { B} X { B} Y [\ldots]\big> &=&
 - \frac{1}{2 {\rm i}} \big<  {\rm tr} ( X \ {Y^\dagger} )[\ldots] \big>\,.
\end{eqnarray}
\end{subequations}
The only possible ordering of $B,{B^\dagger}$ after contraction is
alternation $B {B^\dagger} B {B^\dagger} \ldots$. We may thus express all quantities
in terms of $M = B{B^\dagger}$.  Each contraction reduces the number of
$M$'s by 1.
 
The sequences $X,Y$ in (\ref{sigma_con}b) may be absent; then they
must be replaced by the unit matrix $\mathbf 1$. In particular, if we
repeatedly invoke (\ref{sigma_con}a-d) in order to reduce the number
of $M$'s, the final step will always be
\begin{equation}\label{finstep}
  \big< {\rm tr}
                  \  \contraction[2ex]{}{  {\mathbf B}}{}{()
                  {} }
                 { B} \, B^\dagger   \big> =
                - \frac{1}{2 {{\rm i}}} \big< ({\rm tr} {\mathbf 1})^2
                \big>= - \frac{1}{2 {{\rm i}}}\kappa ^2 r^2\langle1\rangle\,.
\end{equation}
Thus, in the limit $r\to 0$ all averages of trace products vanish like
$r^2$ or faster. Since only terms $\sim r^2$ count, the contractions
between the neighboring $B$ and $B^\dagger$ in the same trace may be
disregarded, unless we are dealing with case (\ref{finstep}).  (Taking
in (\ref{sigma_con}b) $X={\mathbf 1}$, $Y$ or $[\ldots]$ not equal to
unity, the r.~h.~s.  would be $r$ times another averaged trace product
and thus vanish like $r^3$ or faster.)

\subsection{Recursion formula for the number of contractions}

To translate (\ref{sigma_con}a-d) into a recursion relation for the
numbers of contractions $N_c(\vec{v})$, let us select a trace inside
the product in (\ref{connum}), say $\mbox{tr}M^l$ (assuming $v_l>0$),
and a matrix $B$ inside. We must contract that $B$ with all other
suitable matrices in the product of traces.  Three possibilities
arise, paralleling the recursion relation for the combinatorial
numbers in \ref{sec:combinatorics}.

[i] First, we take up the contractions between our selected $B$ in
$\mbox{tr}M^l$ and all suitable matrices in some {\it other} trace
$\mbox{tr}M^k$. In the unitary case rule (\ref{sigma_con}a) implies
that one $k$-trace and one $l$-trace disappear while a $(k+l-1)$-trace
is born
\begin{equation}\label{exter}
         \big< \ {\rm tr} \contraction[2ex]
        {}{   {\mathbf B}}{(X)}{    {\mathbf B} }
          { M}^l \ {\rm tr} M^k [\ldots] \big>
          =
 - \frac{1}{2 {{\rm i}}} \big< {\rm tr }  M^{k+l-1}   [\ldots]
                  \big> \,.
\end{equation}
The contractions with all matrices $B^\dagger$ in $k$-traces $\mbox{tr}M^k$
give the same result.  We thus get $k(v_k-\delta_{kl})$ contributions like
(\ref{exter}), where $\delta_{kl}$ is subtracted to exclude contractions
between matrices inside the same trace.  In the orthogonal case we
must also invoke rule (\ref{sigma_con}c) for contractions with
$k(v_k-\delta_{kl})$ matrices $B$ in traces $\mbox{tr}M^k$ which again all
contribute identically.

Each time, one $k$-trace and one $l$-trace disappear and one
$(k+l-1)$-trace is added to the trace product.  The vector $\vec v$
thus changes to according to $v_k\to v_k-1,\,v_l\to v_l-1,\, v_{k+l-1}\to
v_{k+l-1}+1$; using the same notation as in our semiclassical analysis
we write $\vec v^{[k,l\to k+l-1]}$. The overall number of matrices $M$
is decreased by 1 such that $L\to L-1$.  Each of the above contractions
provides a contribution $N_c(\vec{v}^{[k,l\to k+l-1]})$; here, the
denominator $(-2{{\rm i}})$ in the contraction rules is compensated by
the factor $(-2{{\rm i}} )^{L(\vec v)}$ in the definition of the
contraction numbers.  Thus, the overall contribution to $N_c(\vec{v})$
reads $\kappa k(v_k-\delta_{kl}) N_c(\vec{v}^{[k,l\to k+l-1]})$.

[ii] Next, we turn to ``internal'' contractions between the selected
$B$ and all matrices $B^\dagger$ in the same trace $\mbox{tr}M^l$, apart
from those immediately preceding or following the selected $B$.  As
explained above, the latter contractions would increase the order in
$r$ and lead to a result that vanishes as $r\to 0$.  We apply rule
(\ref{sigma_con}b) and replace $\mbox{tr}M^l$ by the product of two
traces which together contain $l-1$ factors $M$.  Thus, one $l$-trace
disappears and two traces, of $M^m$ and $M^{l-m-1}$, are added, with
$m$ running 1,2,$\ldots, l-2$; the vector $\vec{v}$ then changes to $\vec
v^{[l\to m,l-m-1]}$.  From each of these contractions, $N_c(\vec{v})$
receives a contribution $N_c(\vec v^{[l\to m,l-m-1]})$.

[iii] For the orthogonal case rule (\ref{sigma_con}d) yields further
internal contractions: Pairing the selected $B$ with all other $l-1$
matrices $B$, we obtain $l-1$ times ${\rm tr}M^{l-1}$.  Each time
$\vec{v}$ is thus replaced by $\vec v^{[l\to l-1]}$.  Altogether, we
gain the contribution $(l-1)N_c(\vec v^{[l\to l-1]})$.

Summing up all contributions, we arrive at the recurrence for $N_c$,
for any $l$ with $v_l>0$
\begin{eqnarray} \label{ncrec}
N_c(\vec{v})&=&\kappa \sum_k k(v_k-\delta_{kl}) N_c(\vec{v}^{[k,l\to
k+l-1]}) \nonumber\\&+&\sum_{m=1}^{l-2}N_c(\vec v^{[l\to m,l-m-1]})  \nonumber\\
&+&(\kappa -1)(l-1)N_c(\vec v^{[l\to l-1]})\,.
\end{eqnarray}

The recurrence relation (\ref{ncrec}) reflects a single contraction
step according to the rules (\ref{sigma_con}a-d) and gives a sum of
terms each containing one matrix $M$ less than the original trace.
Repeated such contraction steps give a sum of an ever increasing
number of terms.  After $L(\vec v)-1$ steps every summand will be
reduced to $N_c(\vec{v}')$ with $v_1'=1$, $v'_l=0$ for $l\geq 2$, which
due to (\ref{connum}) and (\ref{finstep}) just equals unity.
Consequently, $N_c(\vec{v})$ gives the number of terms in the sum at
the final stage, and is thus appropriately called ``the number of
contractions".
 
Remarkably, the numbers of contractions $N_c(\vec{v})$ and the numbers
of structures $N(\vec{v})$ are related by
\begin{equation} 
\label{identify} N_c(\vec{v})=N(\vec{v})\frac{\kappa^{V-1}\prod_l
l^{v_l}v_l!}{L(\vec{v})}\,. 
\end{equation} 
With this identification, the recursion relations for both quantities,
(\ref{recur1}), (\ref{recur2}), (\ref{tilde}) for $N(\vec{v})$, and
(\ref{ncrec}) for $N_c(\vec{v})$, coincide. When comparing, note that
we may substitute $v_k-\delta_{kl}=v_k^{[k,l\to k+l-1]}+1$. A constant
proportionality factor in (\ref{identify}) was chosen to satisfy the
initial condition $N(\vec{v}')=N_c(\vec{v}')=1$.  In view of
(\ref{identify}), the series for $K(\tau)$ obtained from periodic-orbit
theory (\ref{kn}) and the $\sigma$-model (\ref{sigmak}) agree term by
term.

\section{Conclusions and outlook}
 
\label{sec:conclusions}

Within the semiclassical frame of periodic-orbit theory, we have
studied the spectral statistics of {\it individual} fully chaotic
(i.e. hyperbolic and ergodic) dynamics. Central to our work are pairs
of orbits which differ only inside close self-encounters.  These orbit
pairs yield series expansions of the spectral form factor $K(\tau)$, and
our series agree with the predictions of random-matrix theory to all
orders in $\tau$, for all three Wigner-Dyson symmetry classes.  Note
that we do not require any averaging over ensembles of systems.
Moreover, we find a close analogy between semiclassical periodic-orbit
expansions and perturbative treatments of the nonlinear sigma model.

Important questions about universal spectral fluctuations remain open.
The perhaps most urgent challenge is to go beyond the range of small
$\tau$, and treat $\tau>1$.
 
The precise conditions for a system to be faithful to random-matrix
theory remain to be established. We certainly have to demand that the
contributions of all orbit pairs unrelated to close self-encounters
mutually cancel. While one may expect such cancellation for generic
systems, there are important exceptions. For dynamics exhibiting
arithmetic chaos, strong degeneracies in the periodic-orbit spectrum
give rise to system-specific contributions to the form factor; hence
the systems in question deviate from random-matrix theory
\cite{Bogomolny}.  On the other hand, for the Sinai billiard and the
Hadamard-Gutzwiller model, system-specific families of orbit pairs
found in \cite{Primack} and \cite{HeuslerPhD}, respectively, do not
prevent universality.  In order to formulate the precise conditions
for the BGS conjecture, one has to clarify when non-universal
contributions may occur.

Moreover, a better justification is needed for neglecting the
difference between stability amplitudes and periods of the partner
orbits. So far, such a justification is only available for
Sieber/Richter pairs in the Hadamard-Gutzwiller model
\cite{HeuslerPhD,MuellerPhD}.

The study of ``correlated" orbit pairs opens a rich variety of
applications in mesoscopic physics. Recent results concern
matrix-element fluctuations \cite{Higherdim} and transport properties
such as conductance, shot noise, or delay times \cite{Transport}.  In
the latter cases, the relevant trajectories are no longer periodic,
and even e.g.  quadruples of trajectories have interesting
interpretations.  While previous work was restricted to the lowest
orders in series expansions of the quantities in question, our
machinery of encounters and permutations, together with intuition
drawn from field theory should allow to attack the full expansion.

Finally, one might wish to go beyond Wigner's and Dyson's ``threefold
way" and extend the present results to the new symmetry classes
\cite{Tenfold}, of experimental relevance for
normal-metal/superconductor heterostructures; first steps are taken in
\cite{GnutzSeif}. Further possible applications concern localization,
a clarification of open problems in the nonlinear sigma model
\cite{Jan}, and the crossover between universality classes
\cite{Nagao,Turek}.

Financial support of the Sonderforschungsbereich SFB/TR12 of the
Deutsche Forschungsgemeinschaft is gratefully acknowledged.  We have
enjoyed fruitful discussions with Gerhard Knieper, J{\"u}rgen M{\"u}ller,
Dmitry Savin, Martin Sieber, Hans-J{\"u}rgen Sommers, Dominique Spehner,
and Martin Zirnbauer.

\appendix

\section{Integrals involving $1/t_{\rm enc}$}
 
\label{sec:integral}

We want to evaluate the integral
\begin{equation} 
\label{osc_integral} 
\int_{-c}^cd^{l-1}sd^{l-1}u\frac{1}{t_{\rm 
    enc}(s,u)}{\rm e}^{{{\rm i}}\Delta S/\hbar} 
\end{equation} 
for an $l$-encounter.  The integration goes over the $2(l-1)$ stable
and unstable coordinates $s_{j}$, $u_{j}$.  These variables determine
both the duration $t_{\rm enc}(s,u)$ of the encounter in question and
its contribution to the action difference $\Delta S=\sum_{j}s_{j}u_{j}$. We
shall show that the integral may be neglected in the semiclassical
limit.

The key is the following change of picture: So far, all Poincar{\'e}
sections ${\cal P}$ inside a given encounter were integrated over; we
thus had to divide out the duration $t_{\rm enc}$.  Instead, we may
single out a section ${\cal P}^{{\rm e}}$, fixed at the end of the
encounter, and only consider the stable and unstable separations
$s_{j}^{{\rm e}}$, $u_{j}^{{\rm e}}$ therein.  For homogeneously
hyperbolic dynamics, i.e.  $\Lambda({\bf x},t)={\rm e}^{\lambda t}$ for all
${\bf x}$ and $t$, the separations inside ${\cal P}^{{\rm e}}$ are
given by $s_j^{{\rm e}}= s_j{\rm e}^{-\lambda t_u}$, $u_j^{{\rm e}}=
u_j{\rm e}^{\lambda t_u}$ with $t_u$ denoting the time difference between
${\cal P}$ and ${\cal P}^{{\rm e}}$.

We recall that the encounter ends when the first of the unstable
components, say the $J$th one, reaches $\pm c$ such that $u_J^{{\rm
    e}}=u_J{\rm e}^{\lambda t_u}=\pm c$. All $l-1$ possibilities
$J=1,2,\ldots,l-1$ and the two possibilities for the sign $u_J^{{\rm
    e}}/c=\pm 1$ give additive contributions $I_J^\pm$ to the integral
(\ref{osc_integral}).  Each of them is easily evaluated after
transforming the integration variables from $s_j$, $u_j$ to $s_j^{{\rm
    e}}$, $u_j^{{\rm e}}$ (with $j\neq J$), $s_J^{{\rm e}}$, and
$t_u=\frac{1}{\lambda}\ln\frac{c}{|u_J|}$.  The Jacobian of that
transformation equals $\lambda c$.  The new coordinates are restricted to
the ranges $\{-c<s_j^{{\rm e}}<c, \,-c<u_j^{{\rm e}}<c,\,{\mbox{
    for}}\,j\neq J\}, \, -c<s_J^{{\rm e}}<c,\,0<t_u<t_{\rm enc}$, and
determine the action difference as $\Delta S=\sum_j s_j^{{\rm e}} u_j^{{\rm
    e}}=\sum_{j\neq J} s_j^{{\rm e}} u_j^{{\rm e}}\pm s_J^{{\rm e}}c$.  We
thus obtain
\begin{eqnarray}
\label{osc_result}
I_J^\pm&=&\lambda c\int_{-c}^cd s_J^{{\rm e}}{\rm e}^{\pm {\rm i}s_J^{{\rm e}}c/\hbar}
\left(\prod_{j\neq J}\int_{-c}^cd s_j^{{\rm e}} du_j^{{\rm e}}{\rm e}
^{{\rm i}s_j^{{\rm e}}u_j^{{\rm e}}/\hbar}\right)\nonumber\\&&\times\frac{1}{t_{\rm enc}
(s^{{\rm e}},u^{{\rm e}})}
\int_0^{t_{\rm enc}(s^{{\rm e}},u^{{\rm e}})}d t_u  \nonumber\\
&\sim& \lambda (2\pi\hbar)^{l-2}2\hbar\sin\frac{c^2}{\hbar}\,;
\end{eqnarray}
note that the divisor $t_{\rm enc}$ was canceled by the
$t_u$-integral; moreover, the $2(l-2)$ integrals over $s_j^{{\rm e}}
,u_j^{{\rm e}}$, of the form already encountered in
(\ref{simple_integral}), gave the factor $(2\pi\hbar)^{l-2}$. Most
importantly, the factor $\sin\frac{c^2}{\hbar}$, provided by the integral
over $s_J^{{\rm e}}$, is a rapidly oscillating function of $c$ and
$\hbar$, annulled by averaging over these quantities; such rapidly
oscillating terms are essentially spurious and would not appear if
smooth encounter cut-offs were used (instead of our $|s|<c,|u|<c$). At
any rate, the integral (\ref{osc_integral}), just the $2(l-1)$-fold of
(\ref{osc_result}), vanishes as $\hbar\to 0$.

\section{Extension to general hyperbolicity and $f>2$}

So far, we mostly restricted ourselves to two-dimensional
homogeneously hyperbolic systems.  To generalize, we reason similarly
to Ref. \cite{Higherdim}, where only Sieber/Richter pairs were
considered.

\subsection{General hyperbolicity}

\label{sec:maths}

First, we shed the restriction to ``homogeneously hyperbolic''
dynamics, for which all phase space points ${\bf x}$ have the same
Lyapunov exponent $\lambda$ and stretching factor $\Lambda(t)={\rm e}^{\lambda t}$.
We shall now lift our reasoning to general hyperbolicity, where the
stretching factors $\Lambda({\bf x},t)$ do depend on ${\bf x}$.  In such
systems the Lyapunov exponents of {\it almost all } points still
coincide with the ${\bf x}$ independent ``Lyapunov exponent of the
system'', whereas each periodic orbit may come with its own Lyapunov
exponent \cite{Gaspard}.  Most importantly, the divergence of the
stretches involved in an encounter depends on the local stretching
factor of that encounter, rather than the Lyapunov exponent of the
system.  Our formula (\ref{tu}) for the encounter duration can only be
read as an approximation, and that approximation is now to be avoided.
We will thus allow $t_{\rm enc}^\alpha({\bf x}_{\alpha 1},s_\alpha,u_\alpha)$ to
depend not only on the stable and unstable separations $s_{\alpha j}$,
$u_{\alpha j}$, but also on the phase-space location of the piercing ${\bf
  x}_{\alpha 1}$ chosen as the origin of the respective Poincar{\'e} section.
The changes arising will be important only for showing that the
contribution arising from the $\frac{1}{t_{\rm enc}}$- integrals of
Appendix \ref{sec:integral} vanishes; recall that for the contributing
terms all occurrences of $t_{\rm enc}$ mutually cancel.

When generalizing the statistics of encounters of
\ref{sec:statistics}, we now have to discriminate between piercing
points ${\bf x}=\{{\bf x}_{1 1},{\bf x}_{2 1},\ldots,{\bf x}_{V 1}\}$ as
well.  Given that encounter stretches are separated by non-vanishing
loops, these piercing points are uncorrelated.  The analogue of $w_T$
will thus be a density w.~r.~t. ${\bf x}$, $s$ and $u$, differing from
(\ref{wintegral}) only by $t_{\rm enc}$ being a function of ${\bf x}$,
and by a normalization factor $\frac{1}{\Omega^V}$.

Preparing for a careful average over periodic orbits we first
introduce the density $\rho^\gamma({\bf x},s,u,t)$ of piercing points,
separations and piercing times {\it of one fixed orbit $\gamma$},
\begin{eqnarray*}
\label{rho_deltas}
&&\rho^\gamma({\bf x},s,u,t)=
\prod_{\alpha=1}^V\delta(\Phi_{t_{\alpha 1}}({\bf z}_0)-{\bf x}_{\alpha 1})\\
&&\times
\prod_{j=2}^{l_\alpha}\delta\left(\Phi_{t_{\alpha j}}({\bf z}_0)-{\bf x}_{\alpha 1}-\hat{s}_{\alpha j}{\bf e}^s({\bf x}_{\alpha 1})
-\hat{u}_{\alpha j}{\bf e}^u({\bf x}_{\alpha 1})\right).\nonumber
\end{eqnarray*}
Here, ${\bf z}_0$ denotes an arbitrary point of reference on $\gamma$ and
$\Phi_t({\bf z}_0)$ is the image of ${\bf z}_0$ under evolution over the
time $t$; if the $j$-th stretch of the $\alpha$-th encounter is almost
time-reversed w.r.t. the first one, we have to replace $\Phi_t({\bf
  z}_0)\to{\cal T}\Phi_t({\bf z}_0)$.  In analogy to
\ref{sec:statistics}, we integrate over the piercing times and divide
out the encounter durations, obtaining a density of piercings and
stable and unstable separations only, i.e.
\begin{equation*}
w^\gamma({\bf x},s,u)=\frac{\int d^Lt\ \rho^\gamma({\bf
      x},s,u,t)} {\prod_\alpha t_{\rm enc}^\alpha({\bf x}_{\alpha 1},s_\alpha,u_\alpha)}.
\end{equation*}
The time integrals can be split into one over $0<t_{11}<T$, and
integrals over the differences $t_{\alpha j}'=t_{\alpha j}-t_{11}$ of all
other piercing times from the first one, the latter with the same
minimal distances as in \ref{sec:statistics}.  Using the group
property $\Phi_{t_{\alpha j}}({\bf z}_0)=\Phi_{t_{11}}( \Phi_{t_{\alpha j}'}({\bf
  z}_0))$, we may thus represent $w^\gamma$ as the average of an
observable $f({\bf z})$ along $\gamma$,
\begin{equation*}
w^{\gamma}({\bf x},s,u)=\frac{1}{T}\int_0^T dt_{11} f(\Phi_{t_{11}}({\bf z}_0))\equiv
\big[f\big]_\gamma
\end{equation*}
with
\begin{eqnarray}
&& f({\bf z})=\frac{T}{\prod_\alpha t_{\rm enc}^\alpha({\bf x}_{\alpha 1},s_\alpha,u_\alpha)}\nonumber\\&&\times
\int d^{L-1}t'\prod_{\alpha=1}^V\delta(\Phi_{t_{\alpha 1}'}({\bf z})-{\bf x}_{\alpha 1})\label{fauxiliary} \\
&&\times
\prod_{j=2}^{l_\alpha}\delta\left(\Phi_{t_{\alpha j}'}({\bf z})-{\bf x}_{\alpha 1}-\hat{s}_{\alpha j}{\bf e}^s({\bf x}_{\alpha 1})
-\hat{u}_{\alpha j}{\bf e}^u({\bf x}_{\alpha 1})\right).\nonumber
\end{eqnarray}

The periodic-orbit sum for the form factor may now be written as
(compare (\ref{ksum}))
\begin{eqnarray*}
 K(\tau)=\kappa\tau&+&\frac{\kappa}{T_H}\left\langle
\sum_{\vec{v}} \frac{N(\vec{v})}{L}\right.\\&&\times
\int d^{V}\mu({\bf x})\int d^{L-V}s\;d^{L-V}u\;{\rm e}^{{\rm i}\Delta S/\hbar}\\
&&\times\left.\left\{\sum_{\gamma}|A_\gamma|^2\delta(T-T_\gamma)
\big[f\big]_\gamma\;\right\}\right\rangle,
\end{eqnarray*}
where the ${\bf x}$-integral is over $V$ points ${\bf x}_{\alpha1}$ in the
energy shell, i.e. $d^{V}\mu({\bf x})=\prod_{\alpha=1}^V d^4 x_{\alpha 1}\delta(H({\bf
  x}_{\alpha 1})-E)$.

We have not used the sum rule of Hannay and Ozorio de Almeida, except
for the diagonal part. Instead, we invoke the equidistribution theorem
\cite{Equidistribution} (recall \ref{sec:chaos}), which says that {\it
  ensembles} of periodic orbits, weighted with their stability, behave
ergodically.  More precisely, if we average an observable $f({\bf z})$
(i) along a periodic orbit $\gamma$ and subsequently (ii) over the
ensemble of all such $\gamma$ (inside a small time window and weighted
with $|A_\gamma|^2$), we obtain an energy-shell average,
\begin{equation}
\label{equidistribution}
\frac{1}{T}\!\left\langle\!\sum_\gamma |A_\gamma|^2\delta(T-T_\gamma)\big[f\big]_\gamma\!\right\rangle_{\!\!\!\Delta T}
=\int\frac{d\mu({\bf z})}{\Omega}f({\bf z})\equiv\overline{f({\bf z})}\,.
\end{equation}

For the observable given in (\ref{fauxiliary}), the energy-shell
average can be evaluated provided the dynamics is mixing
\cite{Gaspard}, i.e. if for two observables $g({\bf z})$, $h({\bf z})$
we have
\begin{equation}
\label{mixing}
\lim_{t\to \infty}\overline{g({\bf z})h(\Phi_t({\bf z}))}
=\overline{g}\;\overline{h}\,.
\end{equation}
Physically, (\ref{mixing}) implies that for sufficiently large times
$t$, we may neglect any classical correlations between ${\bf z}$ and
its time evolved $\Phi_t({\bf z})$, and hence replace $\Phi_t({\bf z})$ by
a phase-space point ${\bf z}'$ and average over all ${\bf z}'$.  We
can then disregard correlations between subsequent piercing points
with time differences at least of the order Ehrenfest time.  Using
(\ref{equidistribution}), repeatedly invoking (\ref{mixing}) for the
product of $\delta$-functions in (\ref{fauxiliary}), and subsequently
integrating over $t_{\alpha j}'$ as in \ref{sec:contribution}, we obtain
\begin{eqnarray}
\label{wfancy}
&&\frac{1}{T}\left\langle\sum_\gamma|A_\gamma|^2\delta(T-T_\gamma)
\big[f\big]_\gamma\right\rangle_{\!\!\! \Delta T}\nonumber\\&&
= \frac{T(T-\sum_\alpha l_\alpha
  t_{\rm enc}^\alpha)^{L-1}} {(L-1)!\prod_\alpha t_{\rm
    enc}^\alpha\Omega^{L}}
\end{eqnarray}
which as expected coincides with $w_T(s,u)$ of Eq.  (\ref{wintegral}),
up to division by $\Omega^V$ and the ${\bf x}$-dependence of $t_{\rm
  enc}^\alpha({\bf x},s,u)$.

The $t_{\rm enc}$-independent terms in the multinomial expansion of
(\ref{wfancy}) yield the same contribution to the form factor as
before, since the divisor $\Omega^V$ is canceled by integration over ${\bf
  x}$. All other contributions can be neglected in the semiclassical
limit: they are either of a too low order in $T$ or proportional to
integrals of the form
\begin{equation*}
\int \frac{d\mu({\bf x})}{\Omega}\int_{-c}^cd^{l-1}sd^{l-1}u\frac{1}{t_{\rm
    enc}({\bf x},s,u)}{\rm e}^{{{\rm i}}\Delta S/\hbar}\,,
\end{equation*}
which we reveal as negligible similarly as (\ref{osc_integral}).  For
each contribution $I_J^\pm$, we transform from ${\bf x},s,u$ to
phase-space points ${\bf x}^{{\rm e}}$ and separations $s^{{\rm
    e}}_j,u^{{\rm e}}_j$ ($u^{{\rm e}}_J=\pm c$ fixed) inside a Poincar{\'e}
section ${\cal P}^{{\rm e}}$ in the encounter end, and the separation
$t_u$ between ${\cal P}$ and ${\cal P}^{{\rm e}}$. For general
hyperbolic dynamics, the stable and unstable coordinates are related
by $s_j^{{\rm e}}=\Lambda({\bf x},t_u)^{-1}s_j$ and $u_j^{{\rm e}}=\Lambda({\bf
  x},t_u)u_j$; see (\ref{linearize}).  The Jacobian\footnote{ In
  particular, we have $\frac{du_J}{dt_u} =-\Lambda({\bf
    x},t_u)^{-1}\frac{d\ln|\Lambda({\bf x},t_u)|}{dt_u}u_J^{{\rm e}}
  =-\Lambda({\bf x},t_u)^{-1}\chi({\bf x}^{{\rm e}})u_J^{{\rm e}}$ with
  $u_J^{{\rm e}}=\pm c$, where we used that $\Phi_{t_u}({\bf x})={\bf
    x}^{{\rm e}}$.  The factor $\Lambda({\bf x},t_u)^{-1}$ is compensated
  by the remaining transformations $u_j\to u_j^{{\rm e}} (j\neq J)$ and
  $s_j\to s_j^{{\rm e}}$.  } now reads $\chi({\bf x}^{{\rm e}})c$ with
the local stretching rate defined as $\chi(\Phi_t({\bf
  x}))=\frac{d\ln|\Lambda({\bf x},t)|}{dt}$ \cite{Gaspard}.  We thus obtain
\begin{eqnarray}
\label{osc_result_fancy}
 I_J^\pm&=&\int \frac{d\mu({\bf x}^{\rm e})}{\Omega}\chi({\bf x}^{{\rm e}})c
\int_{-c}^cd s_J^{{\rm e}}{\rm e}^{\pm {\rm i}s_J^{{\rm e}}c/\hbar}\nonumber\\&&\times
\prod_{j\neq J}\left(\int_{-c}^cd s_j^{{\rm e}} du_j^{{\rm e}}
{\rm e}^{{\rm i}s_j^{{\rm e}}u_j^{{\rm e}}/\hbar}\right)\nonumber\\
&& \times\frac{1}{t_{\rm enc}({\bf x}^{\rm e},s^{{\rm e}},u^{{\rm e}})}
\int_0^{t_{\rm enc}({\bf x}^{\rm e},s^{{\rm e}},u^{{\rm e}})}d t_u,\;
\end{eqnarray}
coinciding with (\ref{osc_result}) since the energy-shell average of
the local stretching rate yields the Lyapunov exponent of the system
$\lambda$ \cite{Gaspard}.

\begin{figure} 
\begin{center} 
  \includegraphics[scale=0.25]{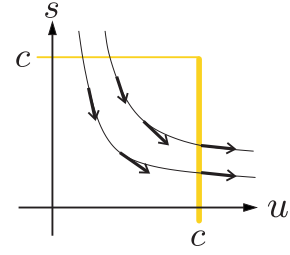}
\end{center}
\caption{ 
  Motion of piercing points through Poincar{\'e} section ${\cal P}$ of a
  2-encounter.  As ${\cal P}$ is shifted, unstable components grow and
  stable ones shrink, traveling on a hyperbola $\Delta S=s u$; arrows
  denote the direction of motion.  At end of the encounter, piercing
  points traverse the line $u=c$.  Negative $s,u$ are not shown.}
\label{fig:travel} 
\end{figure}

The Jacobian $\chi({\bf x}^{{\rm e}})c$ has an interesting physical
interpretation \cite{Higherdim}.  If we shift our Poincar{\'e} section
along the orbit, the piercing points travel, changing their unstable
coordinates with the velocity $\frac{d u_j}{d t}=\chi({\bf x}_t)u_j$;
see Fig.  \ref{fig:travel}.  The Jacobian $\chi({\bf x}^{{\rm e}})c$
thus gives the velocity in the end of the encounter region.
Restricting ourselves to ${\cal P}^{{\rm e}}$ and multiplying with the
above velocity, we simply measure the flux of piercings through the
line $u_J=\pm c$.  Since each point has to traverse that line, our
transformation indeed provides an alternative counting of piercings.
Note that the unstable coordinates temporarily shrink rather than grow
if the local stretching rate is negative, and thus may traverse the
line $u_j=\pm c$ several times.  Due to the asymptotic growth of
$|u_J|$, there is one more traversal with growing $|u_J|$ (and
positive contribution to (\ref{osc_result_fancy}) than with shrinking
$|u_J|$ (and negative contribution); hence only one contribution
remains effective.
 
\subsection{More than two freedoms}

\label{sec:multi} 
 
For dynamics with any number $f\geq 2$ of degrees of freedom, the
Poincar{\'e} section ${\cal P}$ at point ${\bf x}$ is spanned by $f-1$
pairs of stable and unstable directions ${\bf e}^s_k({\bf x})$, ${\bf
  e}^u_k({\bf x})$ ($k=1,2,\ldots,f-1$).  A displacement $\delta {\bf x}$
inside ${\cal P}$ may thus be decomposed as
\begin{equation*} 
  \delta{\bf x}=\sum_{k=1}^{f-1}\left(\hat{s}_k{\bf e}^s_k({\bf x})+\hat{u}_k{\bf e}^u_k({\bf x}) 
\right), 
\end{equation*}
compare (\ref{decompose}).  Each pair of directions comes with
separate stretching factors $\Lambda_k({\bf x},t)$ and Lyapunov exponent
$\lambda_k$.  The directions are mutually normalized as \cite{Higherdim}
\begin{eqnarray*}
&{\bf e}^u_k({\bf x})\land{\bf e}^s_l({\bf x})=\delta_{kl}&\\ 
&{\bf e}^u_k({\bf x})\land{\bf e}^u_l({\bf x})={\bf e}^s_k({\bf x})\land{\bf e}^s_l({\bf x})=0\,,
&\nonumber 
\end{eqnarray*} 
where ${\bf e}^u_k({\bf x})\land{\bf e}^s_k({\bf x})=1$ is a useful
convention, whereas all other relations follow from hyperbolicity.
 
Writing out the additional index $k$, the uniform piercing probability
(see \ref{sec:chaos}) reads $\frac{dt}{\Omega}\prod_{k=1}^{f-1}d\hat{s}_k
d\hat{u}_k$.  The encounters (defined by
$|\hat{s}_{jk}|,|\hat{u}_{jk}|<c$, for all $j$, $k$) have heads and
tails with durations
\begin{equation*} 
  t_u=\min_{j,k}\left\{\frac{1}{\lambda_k}\ln\frac{c}{|\hat{u}_{jk}|}\right\}, 
  \; 
  t_s=\min_{j,k}\left\{\frac{1}{\lambda_k}\ln\frac{c}{|\hat{s}_{jk}|}\right\}.
\end{equation*} 
(cp. \ref{tu}) and contribute to the action difference an amount given
by $\Delta S=\sum_{jk} s_{jk}u_{jk}$, with $s_{jk}$, $u_{jk}$ defined
similarly to \ref{sec:action_difference}.  The integral over $s_{\alpha
  jk}$, $u_{\alpha jk}$ in the second line of (\ref{contribution}) yields
$(2\pi\hbar)^{(L-V)(f-1)}$, which is just what we need since Heisenberg
time now reads $T_H=\frac{\Omega}{(2\pi\hbar)^{f-1}}$.  Given that the
encounter ends as soon as one unstable component $u_{JK}$ reaches $\pm
c$, the $\frac{1}{t_{\rm enc}}$-integral of Appendices
\ref{sec:integral} and \ref{sec:maths} is split into components
$I_{JK}^\pm$, with $\lambda$ replaced by $\lambda_K$, and $\chi({\bf x})$ by
$\chi_K({\bf x})$.

\section{Action correlations}

The semiclassical form factor (\ref{doublesum}) can be written in
terms of an ``action correlation function'' \cite{Argaman},
\begin{eqnarray}
P(y)&=&\left\langle \frac{1}{T}\sum_{\gamma,\gamma ^{\prime
    }} A_{\gamma
}A_{\gamma ^{\prime }}^*\,\delta \left(T-\frac{T_{\gamma }+T_{\gamma ^{\prime }}}{2}
\right)\right.\nonumber\\
&&\;\;\;\,\times\left.\delta \left(y-\frac{2\pi T(S_{\gamma}-S_{ \gamma ^{\prime
  }})}{\Omega } \right)\right\rangle\,,\nonumber\\
K(\tau )&=& \tau
\int\limits_{-\infty }^{\infty }P(y)\,{\rm e}^{{\rm i}y/\tau
}dy\,,\quad \tau>0\label{koffthrupy}\,.
\end{eqnarray}
Using the density of action differences $P_{\vec{v}}(\Delta S)$
(\ref{numorb}), we have evaluated the contributions to $P(y)$ which
arise from diagonal pairs and orbits $(\gamma,\gamma')$ differing by
reconnections in close self-encounters.  Collecting the terms relevant
for the form factor we obtain
\begin{equation} 
\label{action_goe}
P(y)=\!\begin{cases} \quad \delta(y) &\mbox{unitary}\\ 2\delta(y) 
  -\frac{\sin ^{2}y}{\left\vert y\right\vert } +\frac{1}{ \pi 
  }\frac{\sin 2y}{y}\ln \left\vert y\right\vert& \mbox{orthog}. 
\end{cases} 
\end{equation} 
Random-matrix theory predicts, through the inverse Fourier transform
of the above (\ref{koffthrupy}),
\begin{eqnarray*} 
P_{\rm GUE}(y)&=& 
\delta(y)-\frac{2}{\pi }\left( \frac{\sin
(y/2)}{y}\right) ^{2}  \\ 
P_{\rm GOE}(y)&=&2\delta(y)-\frac{4}{\pi }\left( \frac{\sin 
(y/2)}{y}\right) ^{2}\\ 
& &+\frac{2}{\pi y}\Big(\cos ^{2}y\,\mathrm{si\,}y-\cos 2y\,\mathrm{si\,} 
2y\\&&+\cos y\sin y\big[2\operatorname{Ci}(2y)-\operatorname{Ci}y\big]\Big) 
\end{eqnarray*}
with $\mathrm{si\,}y$ and $\operatorname{Ci}\,y$ the integral sine and
cosine \cite{Abramowitz}.  There appears to be, on first glance, a
contradiction between periodic-orbit theory and RMT. However, for both
symmetry classes the respective results differ by smooth functions of
the real variable $y$, smooth implying continuous derivatives of all
orders.  According to the Riemann-Lebesgue theorem, the respective
Fourier transforms (\ref{koffthrupy}) have identical $\tau$-expansions
of $K(\tau)$.  The even and odd parts of that expansion are respectively
caused by the logarithmic and modulus terms in (\ref{action_goe}).

\section{Encounter overlap}
 
\label{sec:overlap_appendix}

So far, we have confined ourselves to encounters whose {\it stretches
  are separated by intervening loops, i.e. do not overlap}.  To
justify this, we shall show that encounters without intervening loops
do not contribute to the form factor.  The overlap of stretches
involved in {\it different encounters} has already been treated in
\ref{sec:statistics}.  We will show that overlap of {\it antiparallel}
stretches (if not prohibited for dynamical reasons, as in the
Hadamard-Gutzwiller model) can be regarded as the reflection of a
single stretch at a hard wall, e.g. in billiards.

The case of overlapping {\it parallel} stretches is the most
complicated one.  As shown in \ref{sec:statistics}, two subsequent
parallel stretches are separated by a non-vanishing loop if the time
$t_{12}$ between the respective piercings fulfills
\begin{equation}
\label{min_parallel}
t_{12}>t_{\rm enc}\equiv t_s+t_u.
\end{equation}
On the other hand, we shall see that any encounter with $t_{12}<t_{\rm
  enc}$ is accompanied by {\it several other encounters}, all leading
to the {\it same partner orbit}; two of these encounters obey
(\ref{min_parallel}).  This means that blindly considering all
encounters and associating with each of them a partner orbit we would
count certain orbit pairs several times!  It also follows that
imposing the condition (\ref{min_parallel}) we do not loose any orbit
pair.  Moreover, to count each orbit pair once, we need a more
restrictive condition than (\ref{min_parallel}). It will be shown that
this condition is of the form
\begin{equation}
\label{strong_min_parallel}
t_{12}>t_{\rm enc}+t_{\rm fringe}\,,
\end{equation} where
$t_{\rm fringe}$ is a certain ``fringe duration".  It turns out that
this boundary leads to the same value of the form factor as
(\ref{min_parallel}), which is why the simpler condition
(\ref{min_parallel}) was used in the main part.

\subsection{Overlap of antiparallel encounter stretches}

\label{sec:retracers}

\begin{figure}
\begin{center}
  \includegraphics[scale=0.27]{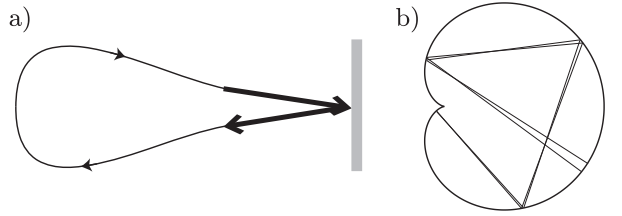}
\end{center}
\caption{Encounters with almost self-retracing reflections:
  a) orbit scheme in configuration space; b) example for cardioid
  billiard.}
\label{fig:retracer}
\end{figure}

First, let us consider two almost {\it mutually time-reversed
  encounter stretches} not separated by an intervening loop. Such a
scenario is only possible if the orbit undergoes a nearly
self-retracing reflection from a hard wall.  After the reflection, the
particle will for some time travel close to the pre-collision
trajectory, such that technically an antiparallel 2-encounter is
formed (Fig.  \ref{fig:retracer}), however with just one loop and two
``ports" attached.  As shown in \cite{Mueller}, no partner can be
connected to such an encounter; formally attempting to construct a
``partner" one obtains either the original orbit or its time-reversed.

The proof becomes surprisingly simple if we use symbolic dynamics.
Here, periodic orbits are fixed uniquely by sequences of symbols, e.g.
denoting in certain billiards the pieces of the boundary where the
orbit is reflected (see e.g. \cite{Baecker}); symbol sequences of {\it
  periodic} orbits are defined modulo cyclic permutations.  Even loops
or encounter stretches can be assigned a sequence of symbols, which
remains unchanged if the loop or stretch is slightly deflected.  Given
a Sieber/Richter pair, the orbit $\gamma$ must have a sequence $\gamma={\cal
  L}_1{\cal E}{\cal L}_2\bar{{\cal E}}$ , ${\cal L}_1$ and ${\cal
  L}_2$ denoting the two loops and ${\cal E}$ and $\overline{{\cal
    E}}$ two almost time-reversed encounter stretches
\cite{ThePaper,Mueller}. The symbol sequence $\overline{{\cal E}}$ is
obtained from ${\cal E}$ by ``time reversal"; in the above example we
simply have to revert the order of symbols.  The partner has one loop
inverted in time, and is thus described by the symbol sequence
$\gamma'={\cal L}_1{\cal E}\bar{{\cal L}_2}\bar{{\cal E}}$.

As announced, both the symbol sequences of the two partner orbits turn
out equal if the loop with symbol sequence ${\cal L}_2$ is absent.  A
generalization to systems without symbolic dynamics is given in
\cite{Mueller,Spehner,Turek,Higherdim}.

We thus see that almost self-retracing orbit pieces as in Fig.
\ref{fig:retracer} each have to be regarded as {\it one single
  stretch}, folded back into itself.  Clearly this carries over if
they form part of a larger $l$-encounter.

\subsection{Overlap of parallel encounter stretches}

\label{sec:por}

Now turning to the {\it parallel} case, we first would like to
establish an intuitive picture of ``overlapping'' parallel encounter
stretches.

\begin{figure}
\begin{center}
  \includegraphics[scale=0.15]{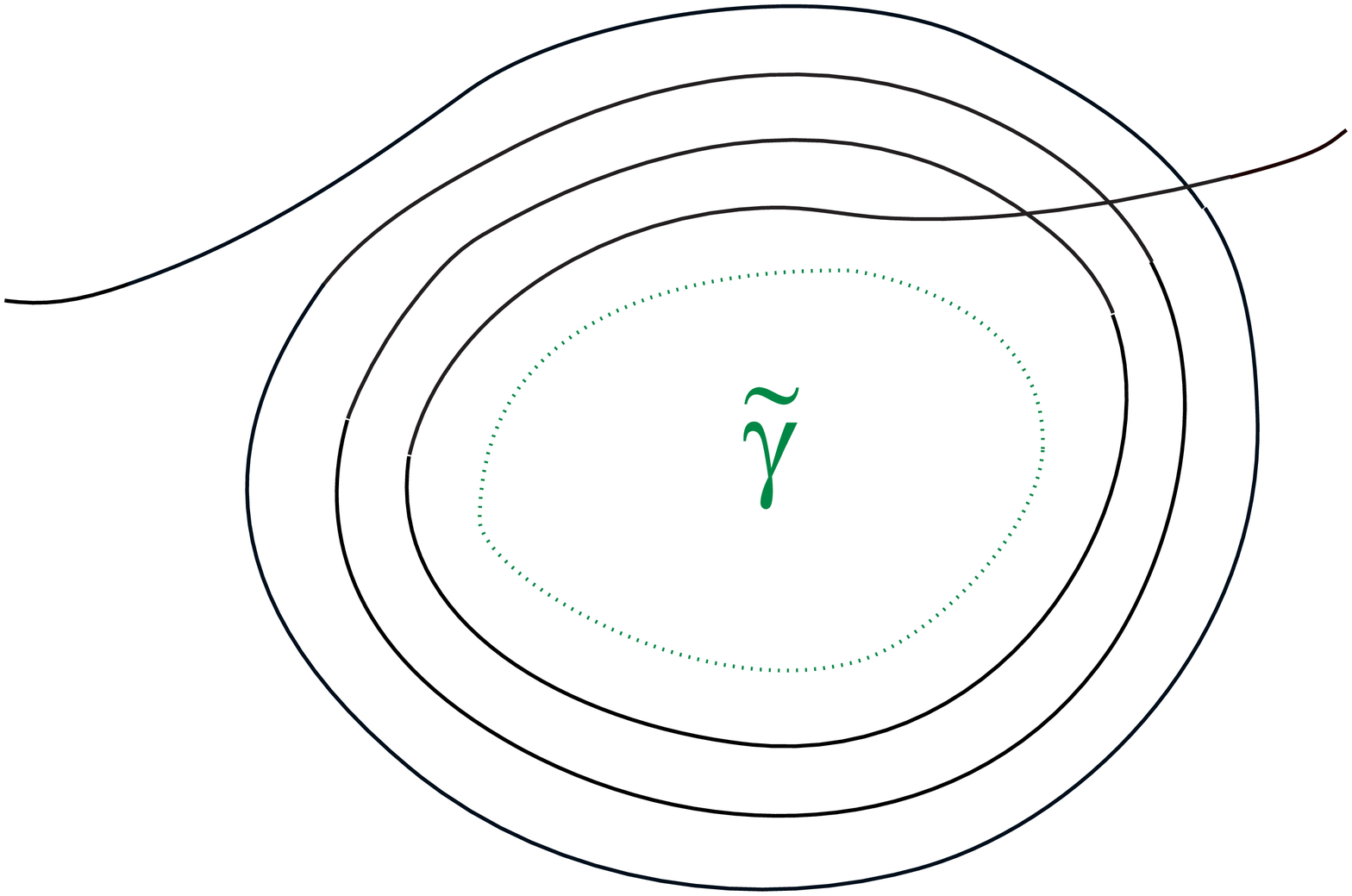}
\end{center}
\caption{Parallel 2-encounter with internal triple revolution about periodic
  orbit $\tilde\gamma$ }
\label{fig:por3}
\end{figure}

For that purpose it is worthwhile to momentarily come back to parallel
2-encounters (even though these give rise to pseudo-orbits as partner
orbits). The close parallel 2-encounters considered previously have
two external loops attached. We shall here get to know a qualitatively
different kind of ``parallel 2-encounter" with but a single external
loop.  Correspondingly, there is only a single intra-encounter orbit
stretch which remains ``close to itself'' until it exits into the
external loop. Such a 2-encounter looks like an $n$-fold revolution
close to some periodic orbit $\tilde{\gamma}$; the simplest case is $n=2$
but multiple revolutions with any integer $n$ are possible (see Fig.
\ref{fig:por3} for $n=3$).

\begin{figure}
\begin{center}
  \includegraphics[scale=0.148]{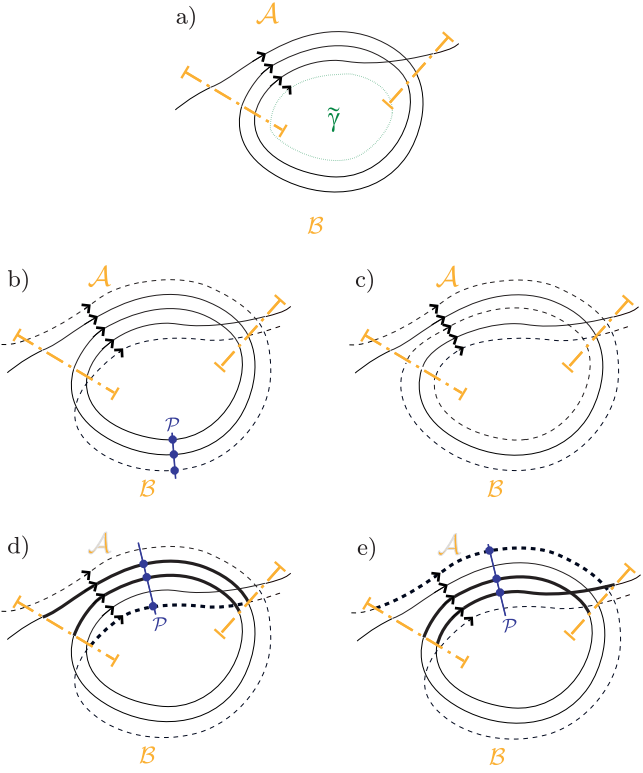}
\end{center}
\caption{
  Overlap of parallel encounter stretches: a) An orbit $\gamma$ comes
  close to a shorter orbit $\tilde{\gamma}$ (dotted line); the ``merged
  stretch" (full line) approaches $\tilde{\gamma}$ three times in the
  region ${\cal A}$ and twice in the region ${\cal B}$.  b) An
  ``isolated stretch" (dashed line) of $\gamma$ approaches ${\cal A}$
  twice and ${\cal B}$ once. Also depicted is a Poincar{\'e} section
  ${\cal P}$ inside the region ${\cal B}$.  c) The partner $\gamma'$ has
  one repetition of $\tilde{\gamma}$ shifted between the merged stretch
  and the isolated stretch.  d), e) Two shorter 3-encounters (depicted
  by thick lines) can be found in the region ${\cal A}$.  They pierce
  three times through a Poincar{\'e} section ${\cal P}$ inside ${\cal
    A}$.}
\label{fig:por}
\end{figure}

Long periodic orbits with such close self-encounters are
straightforward to construct, at least for systems with symbolic
dynamics. To see that, we employ Fig. \ref{fig:por}a which represents
$n=2$.  Two Poincar{\'e} sections mark beginning and end of the encounter
where the orbit $\gamma$ in question respectively enters from and departs
into the external loop. On the other hand, these sections cut the
encounter into two parts ${\cal A}$ and ${\cal B}$; a bundle of three
mutually close substretches traverses ${\cal A}$, while two appear in
${\cal B}$. The sequence of traversals of the two parts is ${\cal
  A}{\cal B}{\cal A}{\cal B}{\cal A}$. If the underlying system has
symbolic dynamics, ${\cal A}$ and ${\cal B}$ each stand for a string
of symbols and the foregoing sequence ${\cal A}{\cal B}{\cal A}{\cal
  B}{\cal A}$ becomes a symbolic representation of the full encounter
stretch while the small periodic orbit $\tilde\gamma$ by itself has the
code $\tilde\gamma={\cal A}{\cal B}$. With ${\mathcal L}$ the symbol
string for the external loop, the whole of $\gamma$ is represented by
$\gamma={\cal A}{\cal B}{\cal A}{\cal B}{\cal A}{\mathcal L}$.  We may
imagine that orbit to have arisen from another one with two external
loops (symbol sequence ${\mathcal E}{\mathcal L}'{\mathcal E}{\mathcal
  L}$ with ${\mathcal E}$ for two parallel encounter stretches and
${\mathcal L}',{\mathcal L}$ for the external loops), by shrinking
away the loop ${\mathcal L}'$ and thus letting the two encounter
stretches merge into one. The special structure ${\mathcal E}={\cal
  A}{\cal B}{\cal A}$ lets the merging stretches overlap in ${\cal A}$
and entails the orbit $\gamma={\cal A}{\cal B}{\cal A}{\cal B}{\cal
  A}{\mathcal L}$ in question, and that special structure obviously
restricts the length of the overlap ${\cal A}$ to less than half the
length of ${\cal E}$.  Otherwise, if the ``overlap'' extends over more
than half the length of $\mathcal E$ we have two pieces ${\mathcal
  E}=({\cal A}{\cal B})^{n-1}{\cal A}$ overlapping in $({\cal A}{\cal
  B})^{n-2}{\cal A}$ and merging to $({\cal A}{\cal B})^{n}{\cal A}$,
now $n>2$, which latter then represents a close 2-encounter with $n$
revolutions about the shorter periodic orbit $\tilde{\gamma}={\cal A}{\cal
  B}$.

Sticking to the double-revolution case of Fig.~\ref{fig:por}a one
might wonder why it is at all appropriate to speak of a 2-encounter,
given the closeness of three substretches in part ${\cal A}$. One
justification was in fact already mentioned: We may imagine the
encounter in question generated by shrinking away one external loop
from a standard 2-encounter.  Another motivation is that of the three
close substretches in part ${\cal A}$ ($n+1$ in the case of $n$
revolutions about $\tilde{\gamma}$) only two are independent; the third
(the $n-1$ others) are uniquely determined by two, calculable through
the linearized equations of motion.

Clearly, if one tries to reconnect (sub-)stretches of a parallel
2-encounter one does not get an acceptable partner orbit of $\gamma$ but a
pseudo-orbit.  However, the foregoing discussion carries over to
parallel stretches in arbitrary $l$-encounters. For instance, by
shrinking away an external loop from a parallel 3-encounter $\pc$ we
arrive at Fig. \ref{fig:por}b. Here the full line revolves twice about
some $\tilde\gamma$ and thus is the ``merger'' of two ``overlapping''
parallel encounter stretches as in Fig. \ref{fig:por}a, while the
dashed line revolves only once around $\tilde\gamma$.  The 3-encounter of
$\gamma$ in question thus includes a ``merged" stretch ${\cal A}{\cal
  B}{\cal A}{\cal B}{\cal A}$ and an ``isolated" stretch ${\cal
  E}={\cal A}{\cal B}{\cal A}$.

Now a partner orbit $\gamma'$ may arise wherein the dashed stretch has
pinched one revolution about $\tilde{\gamma}$ from the full one (see Fig.
\ref{fig:por}c), a fact we now turn to explain.  Recall that we can
construct a partner orbit by changing connections inside the
encounter.  To do so, we place a Poincar{\'e} section within the
encounter, say in region ${\cal B}$ as in Fig.  \ref{fig:por}b.  The
orbit $\gamma$ will pierce through that section in three points; the whole
of $\gamma$, the external loops included, is thus divided into three
parts.  To determine the partner $\gamma'$, we change the connections
between these parts, and thus their ordering.  The change of
connections can be interpreted as cutting out one of the parts and
reinserting it between the two other parts.  In the case depicted in
Fig.  \ref{fig:por}b, the merged stretch pierces through ${\cal P}$
twice, and the part of $\gamma$ between these piercings is one revolution
about $\tilde{\gamma}$. The third piercing belongs to the isolated
stretch.  The partner $\gamma'$ is thus obtained by cutting out one
revolution about $\tilde{\gamma}$ from the merged stretches and inserting
it inside the isolated one; see Fig.  \ref{fig:por}c.  To translate to
symbolic dynamics, we label the two remaining loops of $\gamma$ by ${\cal
  L}_1$ and ${\cal L}_2$ such that $\gamma={\cal L}_1({\cal A}{\cal
  B}{\cal A}{\cal B}{\cal A}){\cal L}_2({\cal A}{\cal B}{\cal A})$.
The partner orbit $\gamma'$ has one $\tilde{\gamma}={\cal A}{\cal B}$ shifted
between the two stretches, i.e.  $\gamma'={\cal L}_1({\cal A}{\cal B}{\cal
  A}){\cal L}_2({\cal A}{\cal B}{\cal A}{\cal B}{\cal A})$.

It is important to note that the encounter shown in Fig.
\ref{fig:por}b may be represented by three pieces ${\cal A}$, two from
the merged stretch and one from the isolated stretch. In other words,
it suffices to consider a shorter encounter made of three independent
traversals of the region ${\cal A}$.  For example, we may take an
encounter consisting of the first and second traversal of ${\cal A}$
by the merged stretch and the latest traversal by the isolated
stretch; these traversals are depicted in Fig. \ref{fig:por}d by thick
full and dotted lines.  We shall refer to the full 3-encounter and its
shorter representative as to the ``long'' and ``short'' one.

The choice of the three representative traversals in the short
encounter is not arbitrary.  First, we have to make sure that the
encounter is limited by the boundaries of ${\cal A}$.  For the above
example, we easily check that outside these boundaries, at least one
of the traversing stretches starts to depart from the others. We have
thus identified a shorter 3-encounter whose stretches {\it do not
  overlap}.

Second, we want our short encounter to give rise to the same partner
orbit as the long one.  To check this, let us place a Poincar{\'e} section
somewhere inside ${\cal A}$; see Fig.  \ref{fig:por}d.  Again, the two
piercings of the merged stretch are separated by one revolution of
$\tilde{\gamma}$, and there is one piercing by the isolated stretch.  Just
like for the larger encounter, the partner has one revolution of
$\tilde{\gamma}$ cut out of the merged stretch and reinserted into the
isolated one.

Both restrictions are fulfilled only for one further short
3-encounter, consisting of the second and third traversal by the
merged stretch and the first traversal by the isolated stretch; see
Fig.  \ref{fig:por}e.

For a first generalization we drop the above restriction on the length
of the ``overlapping part'' within the merged stretch of a parallel
3-encounter.  In general, all three encounter stretches must have the
form ${\cal E}=({\cal A}{\cal B})^{n-1}{\cal A}$, $n\geq 2$, where the
overlapping part is given by $({\cal A}{\cal B})^{n-2}{\cal A}$ (the
above case referred to $n=2$).  The merger of two of these reads
$({\cal A}{\cal B})^{n}{\cal A}$ and executes $n$ revolutions about
$\tilde\gamma={\cal A}{\cal B}$, whereas the isolated stretch remains as
${\cal E}=({\cal A}{\cal B})^{n-1}{\cal A}$ and revolves only $n-1$
times.  Next, we would like to generalize to parallel $l$-encounters
with larger $l$. If two of the stretches involved merge into one, this
leaves $l-2$ isolated stretches with symbol sequences as above. As
before, the partner orbit $\gamma'$ has one revolution about
$\tilde{\gamma}={\cal A}{\cal B}$ cut out of the merged stretch and
inserted into one of the isolated stretches. In addition, the ordering
of the intervening loops may be changed.  Finally, given time-reversal
invariance, the isolated stretches may also have symbol sequences
time-reversed w.r.t.  $({\cal A}{\cal B})^{n-1}{\cal A}$.

For all three generalizations just mentioned, we can identify a
shorter encounter without overlap, consisting of $l$ traversals of the
region ${\cal A}$, and that shorter representative suffices to get all
partner orbits $\gamma'$.  To do so, we divide the stretches of the larger
encounter in two groups. For each stretch of the first group, the
earliest traversal of ${\cal A}$ will be included in the shorter
encounter.  For each stretch of the second group, we choose the latest
traversal of ${\cal A}$.  For the ``merged" stretch also the second
traversal of ${\cal A}$ needs to be included (or the next-to-last, if
the merged stretch forms part of the second group).  These $l$
traversals were chosen such that before the beginning of ${\cal A}$,
the traversals of the first group deviate from the others, while the
traversals of the second group do so after the end of ${\cal A}$.  One
further short encounter is built out of the latest traversals by the
first group, and the earliest traversals by the second group.  As
before, we disregard the long encounter in favor of the shorter ones.
Note that the division of stretches in two groups is not arbitrary;
the groups must be chosen to guarantee that the shorter encounters
give rise to the same partner as the larger one. (See
\cite{MuellerPhD} for details, and \cite{HeuslerPhD,MuellerPhD} for an
alternative description of overlap in parallel 3-encounters.)

To sum up, each orbit pair generated by overlapping parallel stretches
within some encounter can be thought of as originating from any of
{\it two} shorter encounters without overlapping stretches. One of
these two shorter encounters will, in the following subsection, be
excluded and a one-to-one correspondence between orbit pairs and
encounters will be reestablished, in effect by the condition
(\ref{strong_min_parallel}) for the separation of piercings. As
regards the form factor that condition will turn out semiclassically
equivalent to the weaker condition (\ref{min_parallel}), i.e. the
requirement of non-vanishing loops we had worked with in the main part
of this paper.

\subsection{Fringes}

\label{sec:fringes}

\begin{figure}
\begin{center}
  \includegraphics[scale=0.19]{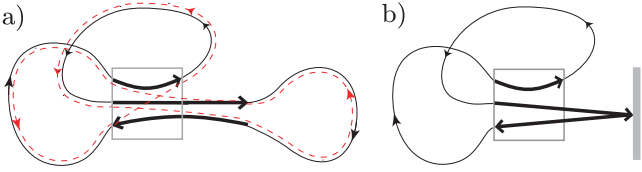}
\end{center}
\caption{
  3-encounter (in the box) with a ``fringe" where only {\it two}
  antiparallel stretches remain close. Stretches are separated by a
  loop in a) but not in b); an almost self-retracing reflection arises
  in the latter case. A partner orbit (dashed line) is obtained only
  in case a).}
\label{fig:fringe}
\end{figure}

Recall that we define an $l$-encounter as a region inside a periodic
orbit in which $l$ orbit stretches come close up to time reversal.
Attached to the sides of the encounter are ``fringes'' in which only
some of the $l$ stretches remain close while others have already gone
astray, as shown in Fig. \ref{fig:fringe}a.  We shall now show that
these fringes do not affect the spectral form factor.

As a first example, let us assume that two {\it antiparallel}
stretches remain close after the end of the encounter, as in Fig.
\ref{fig:fringe}a. If there is no intervening loop (see Fig.
\ref{fig:fringe}b), the orbit has to undergo an almost self-retracing
reflection, just like the one studied in Appendix \ref{sec:retracers}.
Since no connections can be switched between the two stretches
involved we have to disregard such encounters.

On the other hand, if the ``fringe" stretches are separated by an
external loop, a partner orbit is obtained as usual, by reshuffling
connections inside the encounter, see Fig. \ref{fig:fringe}a.  To make
sure that there is an intervening loop, we impose minimal separations
between piercing points, similar to those used for excluding overlap
between {\it encounters} in \ref{sec:statistics}.  After two
antiparallel stretches (labelled by $j$ and $j+1$) have pierced
through a Poincar{\'e} section ${\cal P}$ with unstable coordinates
$\hat{u}_{\alpha j}$ and $\hat{u}_{\alpha,j+1}$, they will remain close for a
time $\frac{1}{\lambda}\ln\frac{c}{|\hat{u}_{\alpha j}-\hat{u}_{\alpha,j+1}|}$.
This duration contains both the time $t_u$ (\ref{tu}) till the end of
the $\alpha$-th encounter, and an additional time span {\it after} the end
of the encounter, which gives the duration of the fringe.  Using the
unstable coordinates $\hat{u}_{\alpha j}^{\rm e}$ of piercing through a
section in the end of the encounter (depending on $\hat{u}_{\alpha j}$ via
$\hat{u}_{\alpha j}^{\rm e}=\hat{u}_{\alpha j}{\rm e}^{\lambda t_u}$), this
additional time may be written as
$\frac{1}{\lambda}\ln\frac{c}{|\hat{u}_{\alpha j}^{\rm e}-\hat{u}_{\alpha,j+1}^{\rm
    e}|}$.

Hence, the minimal time difference $2t_u$ between piercings demanded
in \ref{sec:statistics} has to be incremented by a time $t_{\rm
  fringe}=\frac{2}{\lambda}\ln\frac{c}{|\hat{u}_{\alpha j}^{\rm e}-\hat{u}_{\alpha,
    j+1}^{\rm e}|}$, depending on $\hat{u}$ (or, equivalently, the
coordinates $u$ defined in \ref{sec:action_difference}) only via
$\hat{u}^{\rm e}$ (or, equivalently, $u^{\rm e}$).  Similarly, the
minimal separations related to stable coordinates have to be
incremented by an amount purely depending on the stable coordinates
$s^{\rm b}$ in the beginning of the encounter.  The contribution of
each encounter to the total sum of minimal separations $t_{\rm excl}$
can therefore be written in the form
\begin{equation}
\label{texcl}
t_{\rm excl}^\alpha=l_\alpha t_{\rm enc}^\alpha+\Delta
  t_s^\alpha(s^{{\rm b}}) +\Delta t_u^\alpha(u^{{\rm e}}) \end{equation}
with $\Delta t_s$ and $\Delta t_u$ respective functions of $s^{\rm b}$ and
$u^{\rm e}$ only (whose explicit form will not be needed in the
following).  The numerator in the density of phase-space separations
$w_T(s,u)$ (\ref{density}) has to be modified accordingly.

By reasoning as in Subsection \ref{sec:contribution}, we see that only
those terms of the multinomial expansion of the numerator in
$w_T(s,u)$ contribute which involve a product of all $t_{\rm
  excl}^\alpha$.  They now have to be written as (compare (\ref{wcontr}))
\begin{eqnarray}
  \frac{w_T^{\rm contr}(s,u)}{L}&=&h(\vec{v})
  \left(\frac{T}{\Omega}\right)^{L-V} \prod_\alpha\frac{t_{\rm excl}^\alpha}{l_\alpha
    t_{\rm enc}^\alpha}\nonumber \\ &=& h(\vec{v})
  \left(\frac{T}{\Omega}\right)^{L-V}\prod_\alpha\left[1+\frac{\Delta t_s^\alpha+\Delta
      t_u^\alpha}{l_\alpha t_{\rm enc}^\alpha}\right],\nonumber
\end{eqnarray}
and contribute to the form factor as (compare (\ref{contribution}))
\begin{eqnarray}
\label{contribution_fringes}
&&\kappa \tau h(\vec{v}) \left(\frac{T}{\Omega}\right)^{L-V}\prod_\alpha\left[\int
  d^{L-V}s\;d^{L-V}u\right.
  \nonumber\\
  &&\left.\left(1+\frac{\Delta t_s^\alpha(s^{{\rm b}})+\Delta
      t_u^\alpha(u^{{\rm e}})}{l_\alpha t_{\rm enc}^\alpha}\right){\rm
    e}^{\frac{\rm i}{\hbar}\sum_{j}s_{\alpha j}u_{\alpha j}}\right]\,.
\end{eqnarray}
The integral in (\ref{contribution_fringes}) coincides with the one in
(\ref{contribution}) up to the fringe corrections $\propto\frac{\Delta
  t_s^\alpha}{t_{\rm enc}^\alpha}$ and $\propto\frac{\Delta t_u^\alpha}{t_{\rm enc}^\alpha}$,
respectively.  By reasoning similar to Appendix \ref{sec:integral},
one easily shows that the resulting integrals vanish in the
semiclassical limit.  For the integrand $\frac{\Delta t_u^\alpha}{t_{\rm
    enc}^\alpha}$, the rapidly oscillating integral over $s_J^{\rm e}$
inside (\ref{osc_result}) is not modified; for $\frac{\Delta
  t_s^\alpha}{t_{\rm enc}^\alpha}$ we have to consider a surface of section in
the beginning rather than in the end of the encounter. For
antiparallel orbit stretches, fringe corrections to the form factor
are hence revealed as negligible semiclassically.

Let us now consider fringes of {\it parallel} encounter stretches.  In
the absence of intervening loops, there may be {\it different
  encounters} yielding the {\it same partner orbit}, like the two
short encounters of Appendix \ref{sec:por}, depicted in Fig.
\ref{fig:por}d and e.  One easily sees that these short encounters do
not have their fringes separated by loops.  For instance, in Fig.
\ref{fig:por}d, the two full thick stretches remain close after the
end of the encounter region ${\cal A}$, thus forming a fringe.  They
are still close when the orbit reenters ${\cal A}$, such that there is
no external loop.

In order to count the resulting partner $\gamma'$ exactly once, we choose
to take into account only the encounter in Fig. \ref{fig:por}e, and
disregard the one in Fig. \ref{fig:por}d. To do so, we again have to
increase the minimal distances between piercing points; see
(\ref{strong_min_parallel}).  As above, the increment $t_{\rm fringe}$
may depend only on $s^{\rm b}$ and $u^{\rm e}$, given that these
quantities determine how long each two stretches will stay close
inside the fringe region.  More precisely, one can even show that
$t_{\rm fringe}$ is given by a sum of two terms depending on $s^{\rm
  b}$ and $u^{\rm e}$, respectively.  (For instance, for the first two
piercings of a 3-encounter, $t_{\rm fringe}=
\frac{1}{\lambda}\ln\frac{c}{|\hat{u}_{\alpha 1}^{\rm e}-\hat{u}_{\alpha 2}^{\rm
    e}|} +\frac{1}{\lambda}\ln\frac{c}{|\hat{s}_{\alpha 2}^{\rm b}-\hat{s}_{\alpha
    3}^{\rm b}|}$; see \cite{MuellerPhD} for the derivation and a
generalization to arbitrary $l$-encounters.)  Hence, the sum of all
minimal distances will still be of the form (\ref{texcl}); the
increment will thus not contribute to $K(\tau)$ in the semiclassical
limit.

The present treatment immediately generalizes to $f>2$, by writing out
the additional index $k$, and to non-homogeneously hyperbolic systems
by keeping the dependence on ${\bf x}$. In particular, $\Delta t_s$ will
depend both on $s^{\rm b}$ and the phase-space points ${\bf x}^{\rm
  b}$ in the beginning of the encounter, and $\Delta t_u$ on $u^{\rm e}$
and ${\bf x}^{\rm e}$ in the end of the encounter.

\section{The nonlinear $\sigma$-model}

\label{sec:sigma_appendix}

\subsection{Replica trick and perturbation expansion}

We here sketch the derivation of the integral representation
(\ref{rs}) for the Wigner/Dyson ensembles.  For simplicity, we
concentrate on the GUE.  For the GOE see Subsection
\ref{sec:sigma_orthogonal} below. We start from the ``replica
representation'' of the Green function $G(z)\equiv (z-H)^{-1}$,
\begin{eqnarray} 
    \label{eq:tr_det} 
    {\rm tr}\,G(z)&=&\partial_z {\,\rm tr}\ln(G(z)^{-1})=\partial_z \ln ({\rm 
    det\,}G(z)^{-1})\nonumber\\ 
    &=&- \lim_{r\to 0 
    }{1\over r}\partial_z\det G(z)^r\,. 
\end{eqnarray} 
We take the Hamiltonian $H$ as an $N$-dimensional Hermitian random
matrix whose Gaussian statistics is defined by the two moments
\begin{eqnarray} 
     &  & \overline{H_{\mu\nu}}=0,  \nonumber\\ 
     &  & \overline{H_{\mu\nu}\,H_{\nu'\mu'}}=\frac{\lambda^2}{N} 
     \delta_{\mu\mu'}\delta_{\nu\nu'}\label{eq:H_dist}\,; 
\end{eqnarray} 
these imply Wigner's semicircle for the mean level density as
$\overline{\rho}(E)=(N/\pi\lambda^2)\sqrt{\lambda^2-(E/2)^2} \,\Theta(\lambda^2-(E/2)^2)$
and thus the mean spacing $\pi\lambda/N$ at $E=0$.
 
The $r$th power of the determinant in (\ref{eq:tr_det}) can be written
as the Gaussian integral (summation convention)
\begin{equation} 
    \det(G(z))^r=\int d\psi\,\exp(\pm {\rm i} 
    \psi^{a\dagger}(z-H)\psi^a)\,; 
    \label{eq:det_gaussian} 
\end{equation} 
here $\psi^a=\{\psi^a_\mu\}$, $a=1,\ldots, r$ are $N$--dimensional complex
vectors of integration variables; the measure
$d\psi\propto\prod_{\mu=1}^N\prod_{a=1}^r d{\rm Re} \psi_\mu^a d{\rm Im}\psi_\mu^a$
comprises a normalization constant of the form $({\rm const})^r\sim 1$;
here and in the remainder of this Appendix we let the symbol $\sim$ mean
equality in the limit $r\to0$; the positive/negative sign in the
exponent applies to the case ${\rm Im\,}z >0$/${\rm Im\,}z < 0$.
 
We proceed to the retarded and advanced Green functions ${\rm
  tr}\,G_{\pm}(E)= {\rm tr}\,G(E\pm{\rm i}\delta)$ with real $E$ and $\delta\downarrow 0$.
Their ensemble averaged product
\begin{equation} 
C(\epsilon)=\overline{ {(\rm tr}\,G_+(E+\epsilon/2)) 
    ({\rm tr}\,G_-(E-\epsilon/2))} 
    \label{eq:Cdef} 
\end{equation}
leads to the spectral correlator and the form factor according to
(\ref{complex_correlator}). Applying the replica trick
(\ref{eq:tr_det}) and the integral representation
(\ref{eq:det_gaussian}) twice, i.e. both for $G_+$ and $G_-$, we get
to a {\it bosonic replica variant of the sigma model}: the correlation
function $C(\epsilon)$ appears as a two--fold derivative of a generating
function,
\begin{eqnarray} 
 C(\epsilon)&=&\lim_{r\to0}\frac{1}{r^2} 
\left[\frac{\partial^2  Z(\hat \epsilon)}{\partial \epsilon_ 
+\partial\epsilon_-}\right]_{\epsilon_{\pm
 =\pm \frac{\epsilon}{2}}}\,, 
    \label{eq:CvsZ_def}\\ 
\mathcal{Z}(\hat \epsilon)&=& 
\int d\psi\, 
\overline{\exp({\rm i}\psi^{\dagger}\Lambda(\hat \epsilon+E-H)\psi)}\,, 
    \label{eq:C_gaussian}\\ 
\hat{\epsilon}&=&\mathrm{\,diag\,}(\epsilon_++{\rm i}\delta, 
\epsilon_--{\rm i}\delta)\,, 
\quad\delta\downarrow 0\,,\label{diag_eps}\nonumber\\ 
\Lambda&=&\mathrm{\,diag\,}(1,-1)\label{Lambda}\,.\nonumber 
\end{eqnarray} 
Compared to (\ref{eq:det_gaussian}) the number of integrations is
doubled in the generating function $\mathcal{Z}$: a doublet
$\psi\equiv(\psi^1,\psi^2)$ comprises $rN$--dimensional vectors $\psi^{1}$ and
$\psi^2$ needed to represent the retarded and the advanced Green
functions by Gaussian integrals. The diagonal matrices $\hat{\epsilon},\Lambda$
act in the newly introduced advanced/retarded space like
$\Lambda{\psi^1\choose\psi^2}= {\psi^1\choose-\psi^2}$, and like unity in the
replica space.  Note that $\Lambda$ and the imaginary part of $\hat{\epsilon}$
together secure convergence of the Gaussian integrals in $\mathcal{Z}$
as did the plus/minus sign in (\ref{eq:det_gaussian}). The replica
index $a$ is suppressed in the compact notation for $\mathcal{Z}$.
 
The GUE average is now easily done in (\ref{eq:C_gaussian}) since the
random Hamiltonian appears linearly in the exponent,
\begin{equation}\nonumber 
\mathcal{Z}(\hat \epsilon)=\!\int \!d\psi\, 
\exp\!\left(\!{\rm i} \psi^{\dagger}\Lambda(\hat 
  \epsilon+E)\psi-\frac{\lambda^2}{2N} {\rm tr\,}A^2\!\right) 
\end{equation} 
with the $2r\times 2r$ matrix $A=\{A^{\alpha\beta}=
\sum_{\mu=1}^N\psi_\mu^{\alpha}\psi_\mu^{\beta\ast}\Lambda^{\beta\beta}\}$; the index $\alpha=(p,a)$
comprises the replica index $a=1\ldots r$ and an index $p=1,2$
discriminating between ``advanced'' and ``retarded''.  We next
decouple the term $\propto {\rm tr}\,A^2$ quartic in the integration
variables $\psi$ by a Hubbard--Stratonovich transformation.  To this
end, we multiply $\mathcal{Z}$ by the unit--normalized Gaussian
integral $ 1= \mathcal{N}\int dQ \,\exp\left(\frac{N}{2}{\rm
    tr}\,Q^2\right) $, where $Q=\{Q^{\alpha\beta}\}$ is a $2r$--dimensional
anti-Hermitian matrix and $\int dQ$ denotes integration over all its
independent matrix elements; the latter comprise $(2r)^2$ independent
real and imaginary parts such that the normalization constant is of
the form ${\mathcal N}= ({\rm const})^{r^2}\sim 1$. The shift $Q\to Q -
\lambda A/N$ leads to
\begin{eqnarray}
        \mathcal{Z}(\hat \epsilon)
      \! \! &\sim&\! \!\!\!\!\int\!\! d\psi\!\int\!\! dQ\exp
\!
\left({\textstyle{\frac{N}{2}}}{\rm tr}\,Q^2+{\rm i}
\psi^{\dagger}\Lambda(\hat \epsilon+E+ {\rm i}\lambda Q)\psi\right) 
\nonumber\\ 
        &\sim& \!\!\!\!\int \!\!dQ \,\exp\left({\textstyle{\frac{N}{2}}}{\rm
tr}\,Q^2-N{\rm\, tr\,ln}(\hat \epsilon+E+ {\rm i}\lambda Q)\right)\;\;\;\;\;\; \nonumber
\end{eqnarray} 
where in the second step we have performed the Gaussian integration
over $\psi$. The large parameter $N$ in the foregoing exponential allows
for a stationary-phase approximation of the $Q$--integral.  Variation
of the exponent, the so-called action, w.r.t. the independent matrix
elements of $Q$ yields the saddle-point equation
\begin{equation} 
     Q=\frac{{\rm i}\lambda}{\hat{\epsilon}+E+{\rm i}\lambda Q}\,. 
     \label{eq:pastur_eq} 
\end{equation} 
In order to shorten the saddle-point analysis we confine ourselves to
$E=0$; the final result is valid throughout Wigner's semicircle. The
saddle-point equation then invites solution to zeroth order in
$\hat{\epsilon}$ since the scale for the energy offset variables is the mean
level spacing, $\epsilon_\pm= \mathcal{O}(N^{-1})$; we can even dispatch the
infinitesimal imaginary part $\mathrm{Im}(\hat\epsilon)={\rm i}\Lambda\delta$ which
is only needed to ensure absolute convergence of all integrals. A
simple matrix--diagonal saddle point thus arises, $Q\simeq \Lambda$. (Strictly
speaking, there are $2^{2r}$ diagonal solutions with diagonal entries
$\pm 1$.  However, the solutions different from $\Lambda=
\mathrm{\,diag\,}(1,-1)$, have to be discarded since they cannot be
reached without crossing the cuts of the logarithm in the action
\cite{Weidenmueller}).
 
A continuous manifold of more general (non--diagonal yet compatible
with the cut structure of the logarithm) solutions of the saddle-point
equation can now be obtained by the conjugation $\Lambda \to T \Lambda T^{-1}\equiv
Q_{\rm s}$, where $T\in {\rm U}(r,r)$ is a $(2r)$--dimensional
pseudo-unitary matrix. (The pseudounitarity condition, $T\Lambda T^\dagger=\Lambda$,
is required to make the subsequent integration over all configurations
$Q$ convergent \cite{SymmSpaces}.)  This observation identifies the
non--compact symmetric space ${\rm U}(r,r)/{\rm U}(r)\times{\rm U}(r)$ as
the saddle-point manifold. (Transformations $T\in {\rm U}(r)\times{\rm
  U}(r)$ commute with $\Lambda$ and, therefore, do not affect the diagonal
saddle.) It may be a noteworthy property of the saddle-point manifold
in question that the matrices $Q$ thereon are no longer anti-Hermitian
\cite{SymmSpaces}.
 
We next substitute the saddle configurations $Q_{\rm s} =T\Lambda T^{-1}$
into the action and expand to order $\hat \epsilon$,
\begin{eqnarray} 
        \mathcal{Z}(\hat \epsilon)&\sim& \int dQ_{\rm s} \,\exp 
      \left(-N{\rm\, tr\,}\left[\hat \epsilon(E+ {\rm i}\lambda 
      Q_{\rm s} )^{-1}\right]\right)\nonumber\\ 
      &\stackrel{\protect(\ref{eq:pastur_eq})}{\sim}&\int dQ_{\rm s} \,\exp 
      \left({{\rm i}N\over \lambda}{\rm\, tr\,}\left[\hat \epsilon 
       Q_{\rm s}\right]\right)\nonumber\\ 
      &\sim&\int d Q_{\rm s} \,\exp
      \left({{\rm i}(\epsilon_+-\epsilon_-+{\rm i}2\delta)\pi 
      \bar \rho \over 2}{\rm\, tr\,} 
      \left[\Lambda 
       Q_{\rm s}\right]\right)\nonumber\\ 
      &\sim& \int d Q_{\rm s}\, \exp{\left(\frac{{\rm i}s^+}{2}
      {\rm tr}\Lambda  Q_{\rm s}\right)}\,, 
      \label{finZ} 
\end{eqnarray} 
where now $\int d Q_{\rm s}$ demands integration over the saddle-point
manifold \cite{Weidenmueller}; in the foregoing calculation we used
that ${\rm tr}( Q_{\rm s}^2 )={\rm tr}(\Lambda^2)=r\sim 0$ and that ${\rm
  tr}( Q_{\rm s}\hat \epsilon) = (\epsilon_++\epsilon_-){\rm tr}( Q_{\rm s})/2 + (\epsilon_+-
\epsilon_-+2{\rm i}\delta) {\rm tr}( Q_{\rm s} \Lambda)/2= (\epsilon_+- \epsilon_-+2{\rm
  i}\delta){\rm tr}( Q_{\rm s} \Lambda)/2$; finally, in the last step we
introduced the scaled offset $s^+=(\epsilon_+ - \epsilon_-+2{\rm i}\delta)\pi\bar \rho$.
In order to evaluate the matrix integral above we need an explicit
parametrization of the $Q$--matrices on the saddle-point manifold.
Taylor made to perturbative calculations is the so-called rational
parametrization\footnote{Other parametrizations allow to conveniently
  do the $Q_{\rm s}$-integral exactly, to get the non-oscillatory part
  of $R(s)$ in closed form instead of the asymptotic $1/s^+$-series we
  are after.}
\begin{equation}\label{parametrization} 
 Q_{\rm s}=(1-W)\Lambda (1-W)^{-1}\,;\; 
W=\frac{1}{\sqrt{s^+}}\left({0\atop B^\dagger}{B\atop 0}\right) 
\end{equation} 
where the scaling factor $1/\sqrt{s^{+}}$ makes for notational
convenience.  The main merit of that representation is that the
invariant measure $d Q_{\rm s}$ is effectively (i.e. in view of the
eventual limit $r\to 0$) the flat measure $(s^+)^{-r^2}\prod_{i,j}^{1\ldots
  r}d{\rm Re}B_{ij}d{\rm Im}B_{ij}\equiv (s^+)^{-r^2}dB$ for $r\times r$
matrices $B$. Inserting the parametrization (\ref{parametrization}) in
the generating function (\ref{finZ}) we expand as $(1-W)^{-1}=
\sum_{n=0}^\infty W^n$ to get
\begin{equation}
\label{finfinZ}
 \mathcal{Z}(\hat \epsilon)\sim(s^+)^{-r^2}\, {\rm e}^{{\rm i}s^+r}
\int {dB} \,{\rm e}^{(2{{\rm i}}/\kappa ) \sum_{l=1}^{\infty}
         (s^+)^{1-l}{\rm tr}(B {B^\dagger})^l}\,,
\end{equation}
where the factor $\kappa=1$ has been sneaked in for later use.
 
We had shown in \ref{sec:sigmamodel_formfactor} that the foregoing
$B$-integral yields $\{{\rm const}^{r^2}+
r^2f_2(s^+)+r^3f_3(s^+)+{\mathcal O}(r^4)\}$ with $f_i(s^+)$
polynomials in $1/s^+$. Therefore, nonzero contributions to the
correlator $C(\epsilon)$ according to (\ref{eq:CvsZ_def}) arise only when
both derivatives $\partial^2/ \partial\epsilon_+\partial\epsilon_-$ act together on each of the
three factors on the r.~h.~s. of (\ref{finfinZ}).  It is easy to see
that the contribution from $\partial^2{\rm e}^{{\rm i}s^+r}/ \partial\epsilon_+\partial\epsilon_-$
yields the disconnected part $\bar{\rho}^2$ such that upon simply
setting ${\rm e}^{{\rm i}s^+r}\to1$ and invoking (\ref{eq:CvsZ_def}) we
get (the non-oscillatory contributions to) the connected part $C_{\rm
  conn}(\epsilon)$ instead of $C(\epsilon)$.  The connected correlation function
defined in (\ref{complex_correlator}) is thus obtained as
$R=\frac{1}{2\pi^2\overline{\rho}^2}\mbox{Re}\,C_{\rm conn}
=-\frac{1}{2}\mbox{Re}\frac{\partial^2}{\partial {s^+}^2}\big(Z|_{{\rm e}^{{\rm
      i}s^+r}\to 1}\big)$ which leads us to (\ref{rs}) for the unitary
class, with $\kappa=1$.

\subsection{Generalization to the orthogonal class}
\label{sec:sigma_orthogonal} 
 
For the GOE, Eq.~(\ref{eq:H_dist}) generalizes to
\begin{eqnarray*} 
     &  & \langle H_{\mu\nu}\rangle=0,  \nonumber\\ 
     &  & \langle H_{\mu\nu}\,H_{\nu'\mu'}\rangle=\frac{\lambda^2}{N}
     (\delta_{\mu\mu'}\delta_{\nu\nu'}+ 
     \delta_{\mu\nu'}\delta_{\nu\mu'})%\label{eq:H_dist_orth}
\,. 
\end{eqnarray*} 
As a consequence, the averaged action will contain the sum of two
contributions quartic in the integration variables, $\psi^\dagger_\mu \Lambda
\psi_\nu\left(\psi^{\dagger}_\nu \Lambda \psi_\mu+\psi^{\dagger}_\mu \Lambda \psi_\nu\right)= 2\bar \Psi_\mu
\Lambda \Psi_\nu \bar \Psi_\nu \Lambda \Psi_\nu$, where we have defined
\begin{equation} 
    \bar \Psi\equiv \frac{1}{\sqrt 2}(\psi^\dagger,\psi^T),\qquad 
    \Psi=\frac{1}{\sqrt 2}\left( 
    \begin{array}{c} 
        \psi\cr \psi^\ast 
    \end{array}\right). 
    \label{eq:Psi_def} 
\end{equation} 
Notice that the $4r$--component vectors $\Psi$ and $\bar \Psi$ are
connected by the symmetry relation
\begin{equation} 
    \bar \Psi = \Psi^T 
    \sigma_{1}, 
    \label{eq:Psi_sym} 
\end{equation} 
where the Pauli matrix $\sigma_{1}$ acts in the two--component space of
Eq.~(\ref{eq:Psi_def}).  To decouple the quartic $\Psi$--term in the
action, we introduce a $Q$--matrix subject to the constraint $Q=\sigma_{1}
Q^T \sigma_{1}$.  Noting that $\psi^\dagger(\hat \epsilon + E)\psi=\bar \Psi(\hat \epsilon +
E)\Psi$ and proceeding as in the unitary case we get
\begin{eqnarray*} 
\mathcal{Z}(\hat \epsilon)&=&\!\! \int\!\! d \Psi\!\!\int\!\!
      dQ\exp\!\left(\!{\textstyle{\frac{N^2}{4}}}
      {\rm tr}(Q^2)+{\rm i} 
    \bar\Psi\Lambda(\hat \epsilon+E+ {\rm i}\lambda Q)\Psi\!\right) \\ 
&=& \!\!\int\!\! dQ \,\exp\left({\textstyle{\frac{N^2}{4}}}{\rm 
tr}(Q^2)-{\textstyle{{N\over 2}}}{\rm\, tr\,ln}(\hat \epsilon+E+ {\rm i}\lambda Q)\right),
\end{eqnarray*} 
where we made use of the symmetry of $Q$. A stationary-phase approach
yields the diagonal saddle $Q=\Lambda$ and its non-diagonal generalization
$ Q_{\rm s}=T \Lambda T^{-1}$, where $T\in {\rm O}(2r,2r)$. (Compatibility
with the symmetries of $Q$ requires $T^T=\sigma_{1} T^{-1} \sigma_{1}$, hence
the restriction to the (pseudo)orthogonal group.)  This identifies
$\mathrm{O}(2r,2r)/\mathrm{O}(2r)\times \mathrm{O}(2r)$ as the saddle-point
manifold.
 
As in the unitary case, we represent the $ Q_{\rm s}$--matrices in the
rational parametrization (\ref{parametrization}). Presently,
time--reversal invariance implies $T^T =\sigma_{1} T^{-1} \sigma_{1}$ and, as
the symmetry of the integration variables $B$,
\begin{equation} 
  \label{sigma_time_reversal}
   B^\dagger = -\sigma_1 B^T \sigma_1\,. 
\end{equation} 
Simplifying the action to leading order in $\hat \epsilon$ and expanding in
$W$, we recover Eq.  (\ref{rs}), now with $\kappa=2$.
 
\subsection{Contraction rules}
\label{sec:contraction-rules} 
 
To compute the matrix integral perturbatively in an expansion in
$1/s^+$, one expands the exponentiated action in powers of traces
${\rm tr}[(B B^\dagger)^{l}]$, $l~\!\!\!\!\geq~\!\!\!\!2$. This leads to a
series of terms of the structure $\langle ({\rm tr}[(B B^\dagger)^{2}])^{v_2}
({\rm tr}[(B B^\dagger)^{3}])^{v_3}\dots\rangle$, where $v_l$ are integers and
the averaging over the quadratic action, $\langle \dots \rangle$, is defined
by~(\ref{sigma_quadratic_av}). Each term contributing to the series is
then computed by Wick's theorem, i.e. as a sum over all non--vanishing
pair contractions~(\ref{sigma_con}a-d) of $B$--matrices.  We now
briefly outline the derivation of the basic contraction rules. In an
index notation, the quadratic action reads $S^{(2)}[B,B^\dagger] = {({\rm
    i}/\kappa) }\sum_{\alpha\alpha'} |B_{\alpha\alpha'}|^2$.  This implies the prototypical
contraction rule
\begin{equation*} \langle B_{\alpha\alpha'} {B^\dagger}_{\beta'\beta}\rangle =
  -{1\over 2{\rm i}}\delta_{\alpha\beta} \delta_{\alpha'\beta'}\,, 
\end{equation*} 
the $\kappa$--independence of which follows from the fact that in the
orthogonal case, $\kappa=2$, only one half of the matrix elements
$B_{\alpha\alpha'}$ are independent integration variables --- a consequence of
the time reversal relation (\ref{sigma_time_reversal}).  When
expressed as a sum over independent integration variables only, the
coefficient of the action effectively doubles, i.e. in either case
$\kappa=1,2$ the integration over matrix components obtains a factor ${\rm
  i}/2$.  Using this relation we obtain
\begin{eqnarray}
  \big< {\rm tr}  \contraction[2ex]{}{  {\mathbf  B}}{(\;X)}
 {      {\mathbf B} } { B}
         X \ {\rm tr} { {  B^\dagger}} Y \big> &=&
\big< %%preamble 
  \contraction[2ex]{}{  {\mathbf  B}}{(\; X_{\beta\gamma})} {  {\mathbf B} } {%Positioners 
  B_{\alpha\beta}%%First contracted 
  } 
         X_{\beta\alpha}%%Insertion 
          { {B^\dagger_{\gamma\delta}%%Second contracted
         }} 
         Y_{\delta\gamma} \big> \nonumber\\%final part 
& = &-{1\over 2{\rm i}}\langle X_{\beta
\alpha} Y_{\alpha\beta}\rangle\nonumber\\
& =& -{1\over 2{\rm i}} \langle{\rm
tr}(X \,Y)\rangle,\nonumber
\end{eqnarray}
which is contraction rule (\ref{sigma_con}a). Rule (\ref{sigma_con}b)
is proven in the same manner.  Rule (\ref{sigma_con}d) for the GOE
results from
\begin{eqnarray}
  \langle{\rm tr}\contraction[2ex]{}{{B}}{}{(X){B}}{B}X{{ B}} Y\rangle 
  &=&   \langle {\rm tr}(
  (\sigma_1 X)  %%preamble
  \contraction[2ex]{}{B}{(X \sigma_1)\,(\sigma_1\,\,}
  {  } %%Positioners
  {B}%%First contracted
        (Y \sigma_1)\,(\sigma_1 %%Insertion
          { {B%%Second contracted
         }}
          \sigma_1))\rangle \nonumber\\ 
  &=&%final part
  \langle 
  (\sigma_1 X)_{\alpha\beta}%%preamble 
  \contraction[2ex]{}{  {\mathbf  B}}{((Y \,\;\sigma_1)_{\gamma \delta} (\sigma_1 )\!\!} {  {\mathbf B} } {%Positioners
  B_{\beta\gamma}%%First contracted 
  } 
        (Y \sigma_1)_{\gamma \delta} (\sigma_1 %%Insertion 
           { {  B%%Second contracted
         }}
         \sigma_1)_{\delta \alpha}\rangle \nonumber\\%final part 
  &\stackrel{\protect~(\ref{sigma_time_reversal})}{=}&
   - \langle 
  (\sigma_1 X)_{\alpha\beta} %%preamble
  \contraction[2ex]{}{  {\mathbf  B}}{((Y \sigma_1)_{\gamma \delta})} {\;  {\mathbf B} } {%Positioners
  B_{\beta\gamma}%%First contracted 
  } 
       (Y \sigma_1)_{\gamma \delta}%%Insertion
          { {B^\dagger_{\alpha\delta}%%Second contracted 
         }} 
        \big> \nonumber\\ %final part 
  &=&+{1\over 2{\rm i}}\langle
(\sigma_1 X)_{\alpha\beta}  ( \sigma_1 Y^T)_{\beta\alpha}\rangle\nonumber\\ 
  &=&+{1\over 2{\rm i}}
\langle{\rm tr} ( X  ( \sigma_1 Y^T \sigma_1))\rangle\nonumber\\
&=&-{1\over 2{\rm i}}\langle{\rm
tr}(X\,Y^\dagger)\rangle\,,\nonumber
\end{eqnarray}
where in the last step we used that $Y$ contains an odd number of
matrices $B$ and $B^\dagger$ (i.e. that the time reversal relation
(\ref{sigma_time_reversal}) implies $\sigma_1 Y^T \sigma_1= - Y^\dagger$.) The
proof of rule (\ref{sigma_con}c) proceeds along the same lines.


\begin{thebibliography}{99}
  
\bibitem{BGS} O. Bohigas, M. J. Giannoni, and C. Schmit, Phys. Rev.
  Lett.  {\bf 52}, 1 (1984); G. Casati, F. Valz-Gris, and I. Guarneri,
  Lett.  Nuovo Cim. {\bf 28}, 279 (1980); M. V. Berry, Ann. Phys.
  (N.Y.) {\bf 131}, 163 (1981).
  
\bibitem{Stoeckmann} H.-J. St{\"o}ckmann, {\it Quantum Chaos: An
    Introduction} (Cambridge University Press, Cambridge, England,
  1999).
  
\bibitem{Haake} F. Haake, {\it Quantum Signatures of Chaos} (Springer,
  Berlin, 2nd ed 2001).
  
\bibitem{Mehta} M. L. Mehta, {\it Random Matrices and the Statistical
    Theory of Spectra} (Academic, New York, 2nd ed 1991).
  
\bibitem{Berry} M. V. Berry, Proc. R. Soc. Lond., Ser. A {\bf 400},
  229 (1985).
  
\bibitem{Argaman} N. Argaman, F.-M. Dittes, E. Doron, J. P. Keating,
  A. Yu.  Kitaev, M. Sieber, and U. Smilansky, Phys. Rev. Lett. {\bf
    71}, 4326 (1993).
  
\bibitem{SR} M. Sieber and K. Richter, Physica Scripta {\bf T90}, 128
  (2001); M. Sieber, J. Phys. A {\bf 35}, L613 (2002).
  
\bibitem{Gutzwiller} M. Gutzwiller, {\it Chaos in Classical and
    Quantum Mechanics} (Springer, New York, 1990).
  
\bibitem{Cohen} D. Cohen, H. Primack, and U. Smilansky, Ann. Phys.
  {\bf 264}, 108 (1998).
  
\bibitem{Primack} H. Primack and U. Smilansky, Phys. Rep. {\bf 327}, 1
  (2000).
  
\bibitem{Smilansky} U. Smilansky and B. Verdene, J. Phys. A {\bf 36},
  3525 (2003).
  
\bibitem{Aleiner} I. L. Aleiner and A. I. Larkin, Phys. Rev. B {\bf
    54}, 14423 (1996).
  
\bibitem{Disorder} R. A. Smith, I. V. Lerner, and B. L. Altshuler,
  Phys. Rev. B {\bf 58}, 10343 (1998); R. S. Whitney, I. V. Lerner,
  and R.  A. Smith, Waves Random Media {\bf 9}, 179 (1999).
  
\bibitem{Mueller} S. M{\"u}ller, Eur. Phys. J. B {\bf 34}, 305 (2003);
  Diplomarbeit, Universit{\"a}t Essen (2001).
  
\bibitem{Spehner} D. Spehner, J. Phys. A {\bf 36}, 7269 (2003).
  
\bibitem{Turek} M. Turek and K. Richter, J. Phys. A {\bf 36}, L455
  (2003).
  
\bibitem{Braun} P. A. Braun, F. Haake, and S. Heusler, J. Phys. A {\bf
    35}, 1381 (2002).
  
\bibitem{Higherdim} M. Turek, D. Spehner, S. M{\"u}ller, and K. Richter,
  Phys.  Rev. E {\bf 71}, 016210 (2005).
  
\bibitem{Tau3} S. Heusler, S. M{\"u}ller, P. Braun, and F. Haake, J. Phys.
  A {\bf 37}, L31 (2004).
  
\bibitem{Letter} S. M{\"u}ller, S. Heusler, P. Braun, F. Haake, and A.
  Altland, Phys. Rev. Lett. {\bf 93}, 014103 (2004).
  
\bibitem{Heusler} S. Heusler, J. Phys. A {\bf 34}, L483 (2001).
  
\bibitem{BolteHarrison} J. Bolte and J. Harrison, J. Phys. A {\bf 36},
  2747 (2003); J. Phys. A {\bf 36}, L433 (2003).
\bibitem{Berkolaiko} G. Berkolaiko, H. Schanz, and R. S. Whitney,
  Phys.  Rev. Lett. {\bf 88}, 104101 (2002); J. Phys. A {\bf 36}, 8373
  (2003); G. Berkolaiko, Waves Random Media {\bf 14}, S7 (2004).
  
\bibitem{GnutzAlt} S. Gnutzmann and A. Altland, Phys. Rev. Lett. {\bf
    93}, 194101 (2004).

%later
  
\bibitem{Gaspard} P. Gaspard, {\it Chaos, scattering, and statistical
    mechanics} (Cambridge University Press, Cambridge, England, 1998).
  
\bibitem{Equidistribution} W. Parry and M. Pollicott, Ast{\'e}risque {\bf
    187}, 1 (1990).
  
\bibitem{HOdA} J. H.  Hannay and A. M. Ozorio de Almeida, J. Phys. A
  {\bf 17}, 3429 (1984).
  
\bibitem{Robbins} J. A. Foxman and J. M. Robbins, J. Phys. A {\bf 30},
  8187 (1997).
  
\bibitem{HeuslerPhD} S. Heusler, PhD thesis, Universit{\"a}t
  Duisburg-Essen (2004).
  
\bibitem{MuellerPhD} S. M{\"u}ller, PhD thesis, Universit{\"a}t
  Duisburg-Essen, in preparation.
  
\bibitem{Permutations} E. P. Wigner, {\it Group Theory and its
    Application to the Quantum Mechanics of Atomic Spectra} (Academic,
  New York, 1971).
  
\bibitem{JMueller} J{\"u}rgen M{\"u}ller, private communication.
  
\bibitem{Keppeler} J. Bolte and S. Keppeler, J. Phys. A {\bf 32}, 8863
  (1999); S. Keppeler, {\it Spinning Particles - Semiclassics and
    Spectral Statistics} (Springer, Berlin, 2003).
  
\bibitem{Sigma} F. Wegner, Z. Physik B {\bf 35}, 207 (1979).
  
\bibitem{Efetov} K. Efetov, {\it Supersymmetry in disorder and chaos}
  (Cambridge University Press, Cambridge, England, 1997).
  
\bibitem{Bogomolny} E. Bogomolny and C. Schmit, J. Phys. A {\bf 37},
  4501 (2004) and references therein.
  
\bibitem{Transport} K. Richter and M. Sieber, Phys. Rev. Lett. {\bf
    89}, 206801 (2002); H. Schanz, M. Puhlmann, and T. Geisel, Phys.
  Rev. Lett. {\bf 91}, 134101 (2003); M. Puhlmann, H. Schanz, T.
  Kottos, and T. Geisel, Europhys. Lett. {\bf 69}, 313 (2005); O.
  Zaitsev, D. Frustaglia, and K. Richter, Phys. Rev. Lett. {\bf 94},
  026809 (2005); R. S.  Whitney and P. Jacquod, {\tt
    arXiv:cond-mat/0408487} (2004) and to be published in Phys. Rev.
  Lett.
  
\bibitem{Tenfold} M. R. Zirnbauer, J. Math. Phys. {\bf 37}, 4986
  (1996); J. J. M. Verbaarschot and I. Zahed, Phys. Rev. Lett. {\bf
    70}, 3852 (1993); A. Altland and M. R. Zirnbauer, Phys. Rev. B
  {\bf 55}, 1142 (1997).
  
\bibitem{GnutzSeif} S. Gnutzmann, B. Seif, F. v. Oppen, and M. R.
  Zirnbauer, Phys. Rev. E {\bf 67}, 046225 (2003); S. Gnutzmann and B.
  Seif, Phys. Rev. E {\bf 69}, 056219 (2004); Phys. Rev. E {\bf 69},
  056220 (2004).
  
\bibitem{Jan} J. M{\"u}ller and A.~Altland, J. Phys. A. {\bf 38}, 3097
  (2005).
  
\bibitem{Nagao} T. Nagao and K. Saito, Phys. Lett. A {\bf 311}, 353
  (2003).
  
\bibitem{Abramowitz} M. Abramowitz and I. A. Stegun, {\it Handbook of
    Mathematical Functions} (Dover, New York, 1970).
  
\bibitem{Baecker} A. B{\"a}cker and N. Chernov, Nonlinearity \textbf{11},
  79 (1998).
  
\bibitem{ThePaper} P. A. Braun, S. Heusler, S. M{\"u}ller, and F. Haake,
  Eur.  Phys. J. B {\bf 30}, 189 (2003).
  
\bibitem{Weidenmueller} J. J. M. Verbaarschot, H. A. Weidenm{\"u}ller, and
  M. R. Zirnbauer, Phys. Rep. {\bf 129}, 367 (1985).
  
\bibitem{SymmSpaces} M. R. Zirnbauer, J. Math. Phys. {\bf 38}, 2007
  (1997).

\end{thebibliography}
\end{document}